\documentclass[preprint,12pt]{article}

\newcommand{\pD}[2]{\frac{\partial #1}{\partial #2}}

\newcommand{\D}[2]{\frac{d #1}{d #2}}

\newcommand{\eval}[1]{\bigg{|}_{#1}}

\newcommand{\intcell}{\int^{x_R(k,t)}_{x_L(k,t)}\!dx\,}
\newcommand{\intcellt}[1]{\int^{x_R(k,#1)}_{x_L(k,#1)}\!dx\,}
\newcommand{\Sn}{$S_n$\,}

\newcommand{\sigmat}{\sigma_\mathrm{t}}
\newcommand{\sigmas}{\sigma_\mathrm{s}}
\newcommand{\psiu}{\psi_\mathrm{u}}
\newcommand{\psic}{\psi_\mathrm{c}}
\newcommand{\phiu}{\phi_\mathrm{u}}

\usepackage{amssymb}
\usepackage{amsmath}
\usepackage{graphicx}
\usepackage{ulem}

\usepackage{geometry}
\usepackage{afterpage}
\usepackage{caption}
\usepackage{subcaption}
\usepackage{color}
\usepackage{float}
\usepackage{placeins}

\begin{document}




\title{Accurate solutions to time dependent transport problems with a moving mesh and exact uncollided source treatments}

\author{William Bennett\\
Ryan G. McClarren}
\maketitle
\begin{abstract}
    
    
    To find benchmark quality solutions to time dependent \Sn transport problems, we develop a numerical method in a Discontinuous Galerkin (DG) framework that utilizes time dependent cell edges, called a moving mesh, and an uncollided source treatment. The moving mesh and uncollided source treatment is devised to circumvent discontinuities in the solution and better realize the potential of the DG method to converge on smooth problems. The resulting method achieves spectral convergence on smooth problems. When applied to problems with discontinuity-inducing sources, our combined method returns a significantly better solution than the standard DG method. On smooth problems, we observe spectral convergence even in problems with wave fronts. In problems where the angular flux is inherently non-smooth, we do not observe high-order of convergence when compared with static meshes, but there is a quantitative reduction in error of nearly three orders of magnitude. 
    
\end{abstract}









\section{Introduction}
Implementations of the Discontinuous Galerkin method for spatial discretization to solve the time dependent \Sn neutral particle transport equation must reckon with the inevitable nonsmoothness due to the finite number of wave speeds akin to contact discontinuities in fluid equations. Either the method will fail to achieve the high-order convergence, or  computational scientists must find a way to deal with the inevitable discontinuities that arise. 

This is not a trivial undertaking. Different discontinuities are possible in the solution and the first derivative in the angular flux and in the solution and derivative of the scalar flux. Moreover, these discontinuities are time dependent. Ganapol's plane pulse benchmark solution \cite{ganapol}, ubiquitous in the transport literature, is a good example of the behavior of nonsmooth regions in transport solutions. The so called uncollided flux, which refers to particles that have not collided with other particles since being emitted from the source, is more structured and nonsmooth. In this case, the uncollided angular flux is a decaying delta function in each angular direction traveling away from the initial pulse in the direction that it was emitted. The time decay is a result of scattering, which has an overall smoothing effect on the solution. The scalar flux solution for this problem is smooth everywhere except for a travelling wavefront moving out from the origin at the particle speed. 

Other source configurations also exhibit the behaviors shown by the Ganapol problem: the uncollided flux is the cause of discontinuities, scattering smooths the solution over time, and the angular flux discontinuities can coalesce into discontinuities in the scalar flux. Since the Ganapol problem is a Green's function for the time dependent transport equation, uncollided solutions for other sources can be understood as superpositions of this uncollided solution. This leads to another insight, if the uncollided solution is known, then the location of any discontinuities can be determined at any time.
Also, we see from this problem that the uncollided flux in any configuration will be the most ``unsmooth'' when compared to the flux from particles that have experienced collisions. Therefore, finding the uncollided flux has a twofold advantage. First, it gives the location of discontinuities and second, it gives the least smooth part of the solution. Knowing the location of discontinuities presents the opportunity to resolve them with mesh edges and knowing the uncollided solution presents the possibility to significantly smooth the system being solved. This is the motivation for our method: a moving mesh to resolve discontinuities and an uncollided source treatment to reduce nonsmoothness in the system being solved. 

It is necessary to note that, as described in the plane pulse example, the discontinuities in the angular flux are responsible for less than optimal convergence of a DG method. Tracking and resolving each one of these discontinuities would require a different mesh defined for each angle in the \Sn discretization. Since this is unrealistic, especially for solutions with large numbers of angles, we choose to only attempt to resolve nonsmoothness in the scalar flux and abandon the hubristic quest for a spectrally convergent method on fundamentally nonsmooth problems. The method we propose of time dependent cell edges is similar to methods used in the field of Computational Fluid Dynamics (CFD). For example, \cite{amsallem2012nonlinear} solves an optimization problem to find the location of discontinuities and \cite{henrick2006simulations} uses shock fitting to achieve high order accuracy solutions. The principles in those works are the same: align non-smooth features with the mesh so that highly accurate and efficient solutions can be obtained.

Since our moving mesh is only meant to make the problem smoother inside mesh elements, and not to resolve all discontinuities, we employ an uncollided source treatment to make the problem smoother still. It has already been mentioned that the solution can be split into collided and uncollided parts. This decomposition of the solution can be extended to defining the flux based on collisions. There is a  uncollided flux (i.e., the zeroth collided flux), a first collided flux, a second, and so on. This is called Multiple Flux Decomposition and is used by numerical methods to distribute computational resources more efficiently (see \cite{Alcouffe_1990, Hauck_2013, Walters_2017}). Our implementation involves using the analytic solution for the uncollided flux as a source term to solve for the remaining collided flux and adding the uncollided and collided solutions at the final step. We adopt the uncollided flux solutions from \cite{bennett2022benchmarks}.

In order to verify the effectiveness of the moving mesh and uncollided source methods in solving smooth and nonsmooth problems, we chose a set of representative sources with varying degrees of solution nonsmoothness. Recently, we presented calculations for these sources in \cite{bennett2022benchmarks}. Also, to quantify the effectiveness of each method on its own and in combination, we implement four different methods. The first has both an uncollided source and a moving mesh. One is a standard DG implementation. The final two test each method individually: a static mesh with an uncollided source and a moving mesh with a standard source treatment. Conveniently, the latter three methods can be easily implemented with small modifications to the system that we derive to find the moving mesh, uncollided solutions case. 

Given that there are semi-analytic results for the problems we present here, it may seem that computing highly accurate numerical solutions to these problems is to carry coal to Newcastle (that is perform a pointless action). Our ultimate goal is to develop benchmark quality solutions to nonlinear radiative transfer without utilizing linearizations (as in the specific form of the heat capacity widely used \cite{SU19971035, BINGJING1996337, P1S2, MCCLARREN2011119,pomraning1969extension,GANAPOL1983311} or resorting to the equilibrium diffusion limit as the Marshak wave \cite{petschek1960penetration, marshak, bennett2021self, Lane_2013, verma, hammer, Heizler_2010}). By demonstrating that our approach is sound on problems of linear particle transport, we build a foundation for confidence in solutions to nonlinear problems, where there are no known semi-analytic solutions,  in future work.

The uncollided solution method is presented in Section \ref{sec:uncollided}. The following section contains a derivation of our DG method with moving mesh edges. The implementation, a short description of our convergence analysis methods, and a discussion of the results follow in Sections \ref{sec:implement}, \ref{sec:error}, and \ref{sec:results} respectively.

\section{The transport model and uncollided solution}\label{sec:uncollided}

We begin with the single-speed neutral particle transport equation in slab geometry with isotropic scattering \cite{ganapol}:
\begin{equation}\label{eq:1dtransport}
    \left(\pD{}{t} + \mu\pD{}{x} + 1 \right)\psi = \frac{c}{2}\phi + \frac{1}{2}S(x,t, \mu).
\end{equation} Here,  $\psi(x,t,\mu)$ is the angular flux of particles, and $\phi(x,t) = \int^1_{-1} \!d\mu'\,\psi(x,t,\mu')$ is the scalar flux given by the integral over the comment angle. The cosine of the angle between the polar angle of the direction of travel and the $x$-axis is represented by $\mu \in [-1,1]$. $S$ is a source and $c$ is the scattering ratio, $c=\sigmas/\sigmat$ where $\sigmas$ is the scattering cross section and $\sigmat$ is the total absorption plus scattering cross section. The coordinates $x$ and $t$ are measured in units of mean free path lengths and mean free times, respectively.


We represent the angular flux as the sum of an uncollided and collided angular flux, $\psi \equiv \psiu + \psic $. This allows us to obtain from Eq.~\eqref{eq:1dtransport} an expression for the the particles that have not collided after being emitted by the source (i.e., the \textit{uncollided} flux, $\psiu$),
\begin{equation}\label{eq:uncollided}
    \left(\pD{}{t} + \mu\pD{}{x} + 1 \right)\psiu =  \frac{1}{2}S(x,t, \mu).
\end{equation}
It is possible to solve Eq.~\eqref{eq:uncollided} analytically and recover an expression for the uncollided flux in many cases. This is done by integrating the Green's solution provided by Ganapol \cite{ganapol} over an arbitrary source. That uncollided Green's function solution, $\psiu = \frac{1}{2}\exp(-t)\delta\left(\mu - \frac{x}{t}\right)$, is a  piecewise constant function that has a non-zero part expanding out from the origin at speed $\mu$.
We adopt the uncollided solutions presented in \cite{bennett2022benchmarks} for our chosen source configurations: a plane pulse, a square pulse, a Gaussian pulse, a square source, and a Gaussian source. 

With an expression for the uncollided scalar flux, the equation for the collided flux has the same form as the original transport equation ,
\begin{align}\label{eq:collided}
    \left(\pD{}{t} + \mu\pD{}{x} + 1 \right)\psi_c(x,t,\mu) &=   \frac{c}{2}\int^1_{-1}\!d\mu'\left( \psi_c(x,t,\mu') + \psi_u(x,t,\mu')\right) \nonumber \\&=   \frac{c}{2}\int^1_{-1}\!d\mu' \psi_c(x,t,\mu') + S_u(x,t).
\end{align}
Here the uncollided source is given by
\begin{equation}
    S_u(x,t) = \frac{c}{2}\int^1_{-1}\!d\mu' \psi_u(x,t,\mu') \\=
    \frac{c}{2} \phi_u(x,t).
\end{equation}
To find the full solution, it is necessary to sum the uncollided and collided solutions once the collided solution has been calculated. 

Since Eq.~\eqref{eq:1dtransport} and Eq.~\eqref{eq:collided} have the same form, a numerical method that handles arbitrary sources can easily be applied to both. One of the objectives of this work is to demonstrate the efficiency improvement of analytically representing the more discontinuous, anisotropic uncollided flux and using that to obtain benchmark-quality solutions for the total scalar flux.


\section{Moving mesh DG spatial discretization}\label{sec:DG}
The second ingredient in our benchmark solutions is the use of a moving mesh. We develop a Discontinuous Galerkin (DG) scheme to solve equations of the form Eq.~\eqref{eq:1dtransport}. First, we use the method of discrete ordinates to approximate Eq.~\eqref{eq:1dtransport} as a system of coupled partial differential equations, 
\begin{equation}\label{eq:sn}
     \left(\pD{}{t} + \mu_l\pD{}{x} + 1 \right)\psi^l = \frac{c}{2}\sum_{l'=1}^{N} w_{l'} \psi^{l'} + \frac{1}{2}S(x,t,\mu) \qquad \mathrm{for}\quad l = 1 \dots N,
\end{equation}
where the scalar flux is a weighted sum of the angular flux, $\phi \approx  \sum_{l'=1}^Nw_{l'}\psi^{l'}$. The number of discrete directions the angular flux is evaluated at is $N$; $\mu_l$ is a discrete angle and $\psi^l(x,t)$ is the angular flux in the direction $\mu_l$; the quadrature weights are $w_l$. We use standard Gauss-Lobatto quadrature rules.

We define our solution domain on a mesh of $K$ non-overlapping cells, each with time dependent edges $x_L(k,t)$ and $x_R(k,t)$. The left edge of the $k^{\mathrm{th}}$ cell is always at the same position as the right edge of the  $(k-1)^{\mathrm{st}}$ cell, and so on. We define a new variable $z$ to map to each cell to [-1,1],
\begin{equation*}
    z(k,t) \equiv \frac{x_L(k,t)+x_R(k,t)-2x}{x_L(k,t)-x_R(k,t)}, \qquad k=1\dots K.
\end{equation*}
We define an orthonormal basis function in $z$ for each cell, 
\begin{equation}
    B_{i,k}(z) = \frac{\sqrt{2i +1}}{\sqrt{x_R(k,t)-x_L(k,t)}}P_i(z),
\end{equation}
where $P_i$ is the $i^{th}$ Legendre Polynomial. Now we may define the weak solution on each cell for every angle as a sum of basis functions with time dependent coefficients, 
\begin{equation}\label{eq:solution}
    \psi^l(x,t) \approx \sum_{j=0}^M B_{j,k}(z) \,u^l_{k,j}.
\end{equation}
We obtain the weak formulation of Eq.~\eqref{eq:sn} by multiplying by a basis function and integrating over cell $k$, 
\begin{multline}\label{eq:IandII}
    \underbrace{\intcell B_{i,k}(z) \pD{\psi^l}{t}}_{\mathrm{I}} + \underbrace{\mu_l \intcell B_{i,k}(z)  \pD{\psi^l}{x}}_{\mathrm{II}} + \intcell B_{i,k}(z)  \psi^l = \\
    \intcell B_{i,k}(z)  \left(\frac{c}{2}\sum_{l'=1}^N w_{l'} \psi^{l'} + \frac{1}{2}S(x,t,\mu)\right) .
\end{multline}
In term II, integration by parts is used to push the derivative on to the basis function as is typical with the DG method. Since the integration domain in term I is time dependent, we invoke a special case of the Reynolds Transport Theorem \cite{marsden2003vector} to write the total derivative as
\begin{multline}\label{eq:reynolds}
     \D{}{t}\intcell\psi^l(x,t)B_{i,k}(z) =\\ \intcell\left(\pD{\psi^l}{t}B_{i,k}(z)+\psi^l\pD{B_{i,k}(z)}{t}\right) \\ + \D{x_R(k,t)}{t} \psi^l(x_R,t)B_{i,k}(z=1)  -\D{x_L(k,t)}{t}\psi^l(x_L,t)B_{i,k}(z=-1).
\end{multline}
Solving Eq.~\eqref{eq:reynolds} for term I and substituting into Eq.~\eqref{eq:IandII} with term II integrated by parts, we obtain
\begin{multline}\label{eq:timedep}
    \D{}{t}\intcell \psi^l B_{i,k}(z)    +  \\\underbrace{\D{x_L(k,t)}{t}{\psi^l}(x_L,t)B_{i,k}(z=-1)-\D{x_R(k,t)}{t} \psi^l(x_R,t)B_{i,k}(z=1)}_{III}\\ -\intcell\boldsymbol{\psic}^l\D{B_{i,k}(x,t)}{t}  + \underbrace{\mu_l{\psi^l} B_{i,k}(z)\eval{x_L(k,t)}^{x_R(k,t)}}_{IV}\\ -\mu_l 
    \intcell \boldsymbol{\psi^l}^l \D{B_{i,k}(z)}{x} +  \intcell\boldsymbol{\psi^l}^l B_{i,k}(z) \\=  \intcell B_{i,k}(z) \left(\frac{c}{2}\sum_{l'=1}^N w_{l'} \psi^{l'} + \frac{1}{2}S(x,t,\mu) \right).
\end{multline}
III and IV both involve evaluating the solution at the edges of the cell and are combined to create a numerical flux (not to be confused with the scalar flux or the angular flux) term that governs the flow of information based on the  speed of the particles relative to the mesh  ($\boldsymbol{LU}^{\mathrm{surf}}$). Then, substituting Eq.~\eqref{eq:solution} into Eq.~\eqref{eq:timedep} and exploiting the orthonormality of the chosen basis functions to simplify the mass matrix and the scalar flux term, we obtain
\begin{equation}\label{eq:vectorEQ}
    \D{\boldsymbol{U}_l}{t} - \boldsymbol{\underline{\underline{G}}U}_l+ \left(\boldsymbol{LU}_l\right)^{(\mathrm{surf})} - \mu_l\boldsymbol{\underline{\underline{L}}U}_l + \boldsymbol{U}_l =\frac{c}{2} \sum_{l'=1}^N
  w_{l'}\boldsymbol{U}_{l'} + \frac{1}{2}\boldsymbol{Q},
\end{equation}
where the time dependent solution vector is
 $$\boldsymbol{U}_{l,k} = [u^l_{k,0},u^l_{k,1},...,u^l_{k,M}]^T,$$  where $M+1$ is the number of basis functions. We also define
\begin{equation}
    L_{i,j} = \int_{x_L}^{x_R}\!\,dx\, B_{j,k}(z)\,\D{B_{i,k}(z)}{x},
\end{equation}
\begin{equation}
    G_{i,j} = \int^{x_R}_{x_L}\!dx\,B_{j,k}(z)\,\D{B_i(z)}{t},
\end{equation}
\begin{equation}\label{eq:S}
        Q_i = \int^{x_R}_{x_L}\!dx\,B_{i,k}(z)\,S(x,t,\mu),
\end{equation}
and
\begin{equation}
    \left(LU\right)^{\mathrm{surf}}_i =  
    \left(\mu_l- \D{x_R}{t}\right)B_{i,k}(z=1)\boldsymbol{\psi^{l+}}-\left(\mu_l-\D{x_L}{t}\right)B_{i,k}(z=-1)\boldsymbol{\psi^{l-}}.
\end{equation}
$\boldsymbol{\psi^{l+}}$ and $\boldsymbol{\psi^{l-}}$ are found by evaluating Eq.~\eqref{eq:solution} with an upwinding scheme relative to the mesh motion at the right and left cell edges respectively.

The initial condition is found from 
\begin{equation}\label{eq:icbcmms}
     u_{k,j}^l = \intcellt{0} B_{i,k}(z) \psi(x,t=0,\mu_l).
\end{equation}
Equation~\eqref{eq:vectorEQ} is a system of coupled ordinary differential equations for the solution in an given cell that requires a time integration algorithm to update. To capture particles traveling with the wavefront, a Gauss-Lobatto quadrature scheme that includes the endpoints [-1,1] was used to calculate the angles and the weights \cite{1960ratr.book.....C}. These we calculated with the Python package \texttt{quadpy} \cite{quadpy}.
This scheme easily handles arbitrary sources, which is useful since the uncollided solutions we use as source terms are usually complicated functions of space and time. To apply this scheme on a moving mesh, it is necessary to specify the velocities of the cell edges. We detail our approach to this in  Section \ref{sec:implement}.
\subsection{$M=1$ Example equations}
To illustrate the method in a more tangible way, we show the method for  $M=1$. This choice makes Eq.~\eqref{eq:vectorEQ} for an angle, $l$ and cell, $k$, have the form
\begin{multline}\label{eq:vector_M1}
    \D{}{t} \left(
\begin{array}{c}
 u^l_{k,0} \\
 \\
 u^l_{k,1} \\
\end{array}
\right) -
\underbrace{\left(
\begin{array}{cc}
 -\frac{\D{x_R}{t}-\D{x_L}{t}}{2 (x_R
(t)-x_L(t))} & 0 \\
 \frac{\sqrt{3} \left(\D{x_L}{t}+\D{x_R}{t}\right)}{x_L(t)-x_R(t)} &
   \frac{3 \left(\D{x_R}{t}-\D{x_L}{t}\right)}{2 (x_L(t)-x_R(t))} \\ 
\end{array}
\right)}_{\boldsymbol{G}}  \boldsymbol{\cdot} \left(
\begin{array}{c}
 u^l_{k,0} \\
 \\
 u^l_{k,1} \\
\end{array}
\right) 
-\mu_l \underbrace{\left(
\begin{array}{cc}
 0 & 0 \\
 -\frac{2 \sqrt{3}}{x_L(t)-x_R(t)} & 0 \\
\end{array}
\right)}_{\boldsymbol{L}} \boldsymbol{\cdot} \left(
\begin{array}{c}
 u^l_{k,0} \\
 \\
 u^l_{k,1} \\
\end{array}
\right)  \\ \\
+ \underbrace{\left(
\begin{array}{c}
 u^l_{k,0} \\
 \\
 u^l_{k,1} \\
\end{array}
\right)}_{\boldsymbol{U}^l}    + \underbrace{\left(
\begin{array}{c}
 \left(\mu_l- \D{x_R}{t}\right)\frac{1}{\sqrt{x_R(t)-x_L(t)}}\boldsymbol{\psi^{l+}} - \left(\mu_l-\D{x_L}{t}\right)\frac{1}{\sqrt{x_R(t)-x_L(t)}}\boldsymbol{\psi^{l-}} \\
 \\
  \left(\mu_l- \D{x_R}{t}\right)\frac{2}{\sqrt{x_R(t)-x_L(t)}}\boldsymbol{\psi^{l+}} + \left(\mu_l-\D{x_L}{t}\right)\frac{2}{\sqrt{x_R(t)-x_L(t)}}\boldsymbol{\psi^{l-}} \\
\end{array}
\right)}_{\boldsymbol{LU}^{\mathrm{surf}}}  \\ \\ 
= \frac{c}{2} \left(w_0 \left(
\begin{array}{c}
 u^0_{k,0} \\
 \\
 u^l_{k,1} \\
\end{array}
\right)  + w_1 \left(
\begin{array}{c}
 u^1_{k,0} \\
 \\
 u^1_{k,1} \\
\end{array}
\right) + \dots + w_N \left(
\begin{array}{c}
 u^N_{k,0} \\
 \\
 u^N_{k,1} \\
\end{array}
\right)\right)
 + \underbrace{\left(
\begin{array}{c}
 \int^{x_R}_{x_L}\!dx\,B_{0,k}(z)\,S(x,t,\mu) \\
 \\
 \int^{x_R}_{x_L}\!dx\,B_{1,k}(z)\,S(x,t,\mu) \\
\end{array}
\right)}_{\boldsymbol{Q}}.
\end{multline}
The terms $(\mu_l - \D{x_R}{t})$ and $(\mu_l - \D{x_L}{t})$ give the particle velocity relative to the right and left mesh edges respectively and determine where the solution is evaluated in the upwinding scheme. If $(\mu_l - \D{x_R}{t}) > 0$, 
\begin{equation}
    \boldsymbol{\psi^{l+}} = \sum_{j=0}^M B_{j,k}(z)\,u_{k,j}^l.
\end{equation}
If the relative velocity is negative, the right edge solution is evaluated in the next cell,
\begin{equation}\label{eq:right_right}
    \boldsymbol{\psi^{l+}} = \sum_{j=0}^M B_{j,k+1}(z)\,u_{k+1,j}^l.
\end{equation}

For the solution at the left edge, if $(\mu_l - \D{x_L}{t}) > 0$, the solution from the previous cell is used.
\begin{equation}\label{eq:left_left}
    \boldsymbol{\psi^{l-}} = \sum_{j=0}^M B_{j,k-1}(z)\,u_{k-1,j}^l.
\end{equation}
For a positive relative velocity,
\begin{equation}
    \boldsymbol{\psi^{l-}} = \sum_{j=0}^M B_{j,k}(z)\,u_{k,j}^l.
\end{equation}
When dealing with edges at the end of the domain, $k=1$ or $k=K$, the boundary condition is required to evaluate Eq.~\eqref{eq:right_right} and Eq.~\eqref{eq:left_left}. For the infinite medium problems we explored, the boundary is zero with the exception of the MMS problem (Section \ref{sec:MMS}).

\section{Implementation}\label{sec:implement}
To verify the effectiveness of our moving mesh DG scheme with an uncollided source treatment, we implement the method in a code written in Python with Numba \cite{Lam_2015} to solve transport problems with six representative sources: a Method of Manufactured Solutions (MMS) source, a Gaussian pulse and source, a square pulse and source, and a plane pulse. For error quantification, which is addressed in the next Section, the full solution for the MMS source is known from the problem setup, and we adopt the full solutions from \cite{bennett2022benchmarks} to calculate the accuracy of our method in the remaining five source configurations. All of our solutions are available on Github\footnote{{www.github.com/wbennett39/moving\textunderscore mesh\textunderscore radiative\textunderscore transfer}}.

Since we intend to compare the uncollided source treatment and the moving mesh individually and in unison, it is necessary to develop a code that readily switches between four different methods: (1) a moving mesh using an uncollided source, (2) a moving mesh with out an uncollided source,  a static mesh with (3) and without (4) the uncollided source. Each of these cases  requires solving the system of ODE's given by Eq.~\eqref{eq:vectorEQ} with different functions for the source term and the time dependent mesh edges.

In this section, the methods we used to describe the moving and static meshes and the source treatments are presented in a general way. The subsequent results section includes the particular implementations for each source. Also in this section is a discussion of the time integration method. 
\subsection{Moving mesh}
While our method admits mesh motion that is any function of time that does not result in mesh edges crossing or zero width cells, we have chosen to restrict our investigation to one simple method for moving the cells. This simple method requires the mesh to be subdivided into an even number of cells. The initial mesh spans a finite width of $2x_0$, centered on zero. The edges move with a constant velocity away from the origin that is dependent on their initial location. If the edges are defined as a vector of location values where $x_0$ is the leftmost cell edge, $x_1$ is the right edge of that cell or the left edge of the adjacent cell, etc., then
\begin{equation}
    \boldsymbol{X}(t) = \left[x_0(t), x_1(t),...,x_{K}(t)\right],
\end{equation}
and the location of the edges can be found with,
\begin{equation}\label{eq:mesh_edges}
    x_k(t) = x_k(0) + vt\frac{x_k(0)}{x_{K}(0)},
\end{equation}
where $x_{k}(0)$ is the initial location of a given edge and $v$, the particle velocity, is unity. 

If the initial widths are chosen to span a finite source, Eq.~\eqref{eq:mesh_edges} moves the outermost edges at speed one, matching the solution wavefront. Since a cell interface will be initialized at $x=0$, that interface will never move. For our static mesh calculations, we simply span a chosen width with evenly-spaced cells and set $v=0$. 

\subsection{Source treatment}
The pulsed sources, i.e., the Gaussian pulse, the square pulse, and the plane pulse, are equivalent to initial conditions for Eq.~\eqref{eq:1dtransport}. For these cases in the standard source treatments that do not employ the method of uncollided solutions, the source $S$ in Eq.~\eqref{eq:S} is set to zero and the initial condition is found by letting $\psi(x,t=0) = S(x,t=0)/2$ and inserting into Eq.~\eqref{eq:icbcmms}. For the uncollided source treatment of the pulsed sources, the initial condition is set to zero and the uncollided solution is used in Eq.~\eqref{eq:S} to find the source in the weak formulation, $\boldsymbol{Q}$. Similarly for the source cases (Gaussian source, square source), $\boldsymbol{Q}$ is found by integrating either the uncollided solution or the source term, depending on the method used.  

\subsection{Time integration}
To solve the system of coupled ordinary differential equations from Eq.~\eqref{eq:vectorEQ}, we employed an $8^{\mathrm{th}}$ order, explicit Runge-Kudda algorithm, \texttt{DOP853} \cite{1993}, as implemented in \texttt{scipy} \cite{Virtanen_2020}. The relative tolerance parameter was set to $5\times10^{-13}$ and the absolute tolerance to $10^{-12}$.

\section{Error characterization}\label{sec:error}
We judge the effectiveness of our scheme by characterizing the convergence of the solution on a test problem. We use the root mean squared error (RMSE) of the computed scalar flux as our error metric,
\begin{equation}
    \mathrm{RMSE} = \sqrt{\sum^N_i\frac{|\phi_i - \hat{\phi_i}|^2}{N}},
\end{equation}
where $\phi_i$ is the calculated scalar flux at a given node, $\hat{\phi_i}$ is the corresponding benchmark solution, and $N$ is the total number of nodes in the computational solution. To characterize the convergence, we solve a benchmark problem and increase the degrees of freedom. For a problem that is algebraically convergent, holding the number of basis functions constant and increasing the cell divisions leads to an error behavior that limits, as $K \rightarrow \infty$, to the form
\begin{equation}\label{eq:algebraic}
    \mathrm{RMSE} = C \,K^{-A},
\end{equation}
where $C$ is the $y$ intercept, and the constant $A$ is the rate of convergence. If $A$ is $2$, the method is said to converge at second order. The curve in Eq.~\eqref{eq:algebraic} is a straight line on a graph where both axes have a logarithmic scale. 

In characterizing problems where the method shows algebraic convergence, the intercept is significant. Two separate methods may have the same algebraic convergence order, but wildly different errors when tried on a problem if one method has a smaller intercept value. Therefore, we use our data for the errors as a function of the number of cells, $K$ to estimate the values of $A$ and $C$ based on Eq.~\eqref{eq:algebraic}.

For problems that demonstrate geometric spectral convergence, the error can be modeled as 
\begin{equation}\label{eq:spectral}
    \mathrm{RMSE} = C \,\exp(-c_1 M ),
\end{equation}
where $M$ is the highest polynomial order of the basis and $C$ and $c_1$ are constants that could depend on the number of cells used in the problem. This curve is a straight line on a logarithmic-linear scale. 
For spectral problems, the coefficient $C$ is less consequential as our results show that the error is exceedingly small for modest values of $M$. 
\section{Results}\label{sec:results}
\begin{table}[]
\caption{Parameters for each test case in section \ref{sec:results}}.
\centering
\resizebox{\columnwidth}{!}{
\begin{tabular}{|l|l|l|l|l|l|l|l|l|}
\hline
Section             & Problem         & Initial Condition                                                   & Source                                                                               & Uncollided source & $c$      & $x_0$            & $t_0$ & $\sigma$         \\ \hline
\ref{sec:MMS}       & MMS             & $ \psi(x,t=0,\mu)= \frac{e^{-x^2/2}}{2}\Theta(t-|x| + x_0)$         & $S(x,t,\mu)= -\frac{e^{-\frac{x^2}{2}} (\mu,(t+1) x+1)}{(t+1)^2}\Theta(x-|t| + x_0)$ &-                 & 1.0      & $0.1$   & -     & -                \\
\ref{sec:gauss_ic}  & Gaussian pulse  & $\psi(x,t=0,\mu)=\frac{1}{2}\exp\left(\frac{-x^2}{\sigma^2}\right)$ & -                                                                                    &  Eq.~\eqref{eq:gaussian_pulse_uncollided}                  & 1.0      & -                & -     & 0.5              \\
\ref{sec:gauss_s}   & Gaussian source & $\psi(x,t=0,\mu) = 0$                                               & $S(x,t) = \exp\left({\frac{-x^2}{\sigma^2}}\right)\,\Theta(t_0 - t)$                 &Eq.~\eqref{eq:gaussian_source_uncollided_1}                 & 1.0      & -                & $5.0$ & 0.5              \\
\ref{sec:pl_ic}     & Plane pulse     & $\psi(x,t=0,\mu) =\frac{1}{2}\delta(x)\delta(t)$                    & -                                                                                    &    Eq.~\eqref{eq:uncollided_plane_IC}              & 1.0      & 0.5              & -     & -                \\
\ref{sec:sq_ic}     & Square pulse    & $\psi(x,t=0,\mu) =\frac{1}{2}\Theta\left(x_0-|x|\right)$            & -                                                                                    & Eq.~\eqref{eq:sq_ic_cases}                  & 1.0      & 0.5              & -     & -                \\
\ref{sec:square_s}  & Square Source   & $\psi(x,t=0,\mu) = 0$                                               & $S(x,t) = \Theta(x_0-|x|)\Theta(t_0-t)$                                              & Eq.~\eqref{eq:sq_s_uncollided}                 & 1.0      & 0.5              & $5.0$ & -                \\
\ref{sec:c_not_one} & Square pulse    & $\psi(x,t=0,\mu) =\frac{1}{2}\Theta\left(x_0-|x|\right)$            & -                                                                                    & Eq.~\eqref{eq:sq_ic_cases}                 & 0.8, 1.2 & $0.625$, $0.417$ & -     & -                \\
\ref{sec:c_not_one} & Gaussian pulse  & $\psi(x,t=0,\mu)=\frac{1}{2}\exp\left(\frac{-x^2}{\sigma^2}\right)$ & -                                                                                    & Eq.~\eqref{eq:gaussian_pulse_uncollided}                 & 0.8, 1.2 & -                & -     & $0.625$, $0.417$ \\ \hline
\end{tabular}
}
\end{table}
\subsection{MMS}\label{sec:MMS}
\begin{figure}
     \centering
     \begin{subfigure}[b]{0.3\textwidth}
         \centering
         \includegraphics[width=\textwidth]{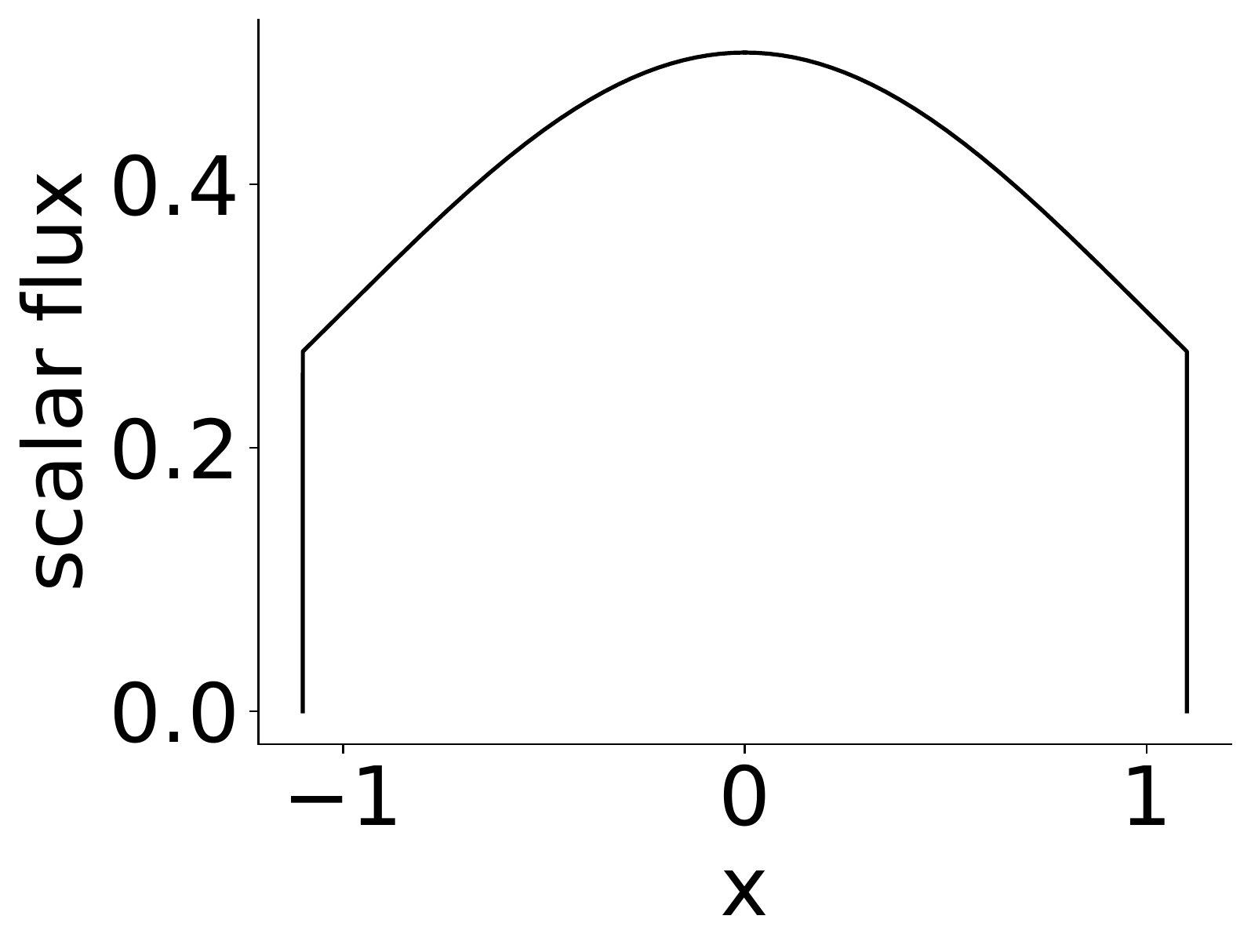}
         \caption{$t=1$}
         \label{fig:MMS_t1}
     \end{subfigure}
     \hfill
     \begin{subfigure}[b]{0.3\textwidth}
         \centering
         \includegraphics[width=\textwidth]{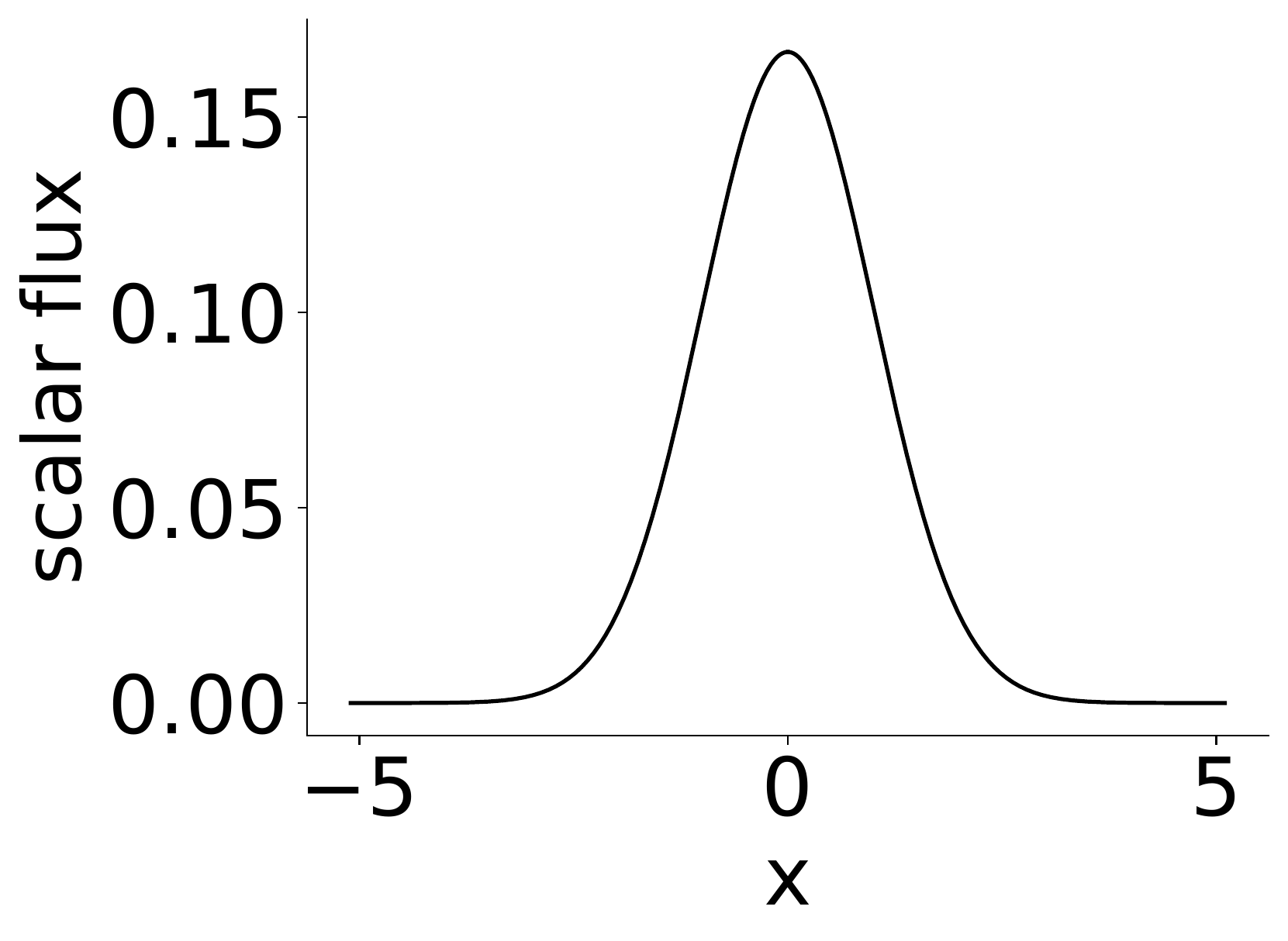}
         \caption{$t=5$}
         \label{fig:MMS_t5}
     \end{subfigure}
     \hfill
     \begin{subfigure}[b]{0.3\textwidth}
         \centering
         \includegraphics[width=\textwidth]{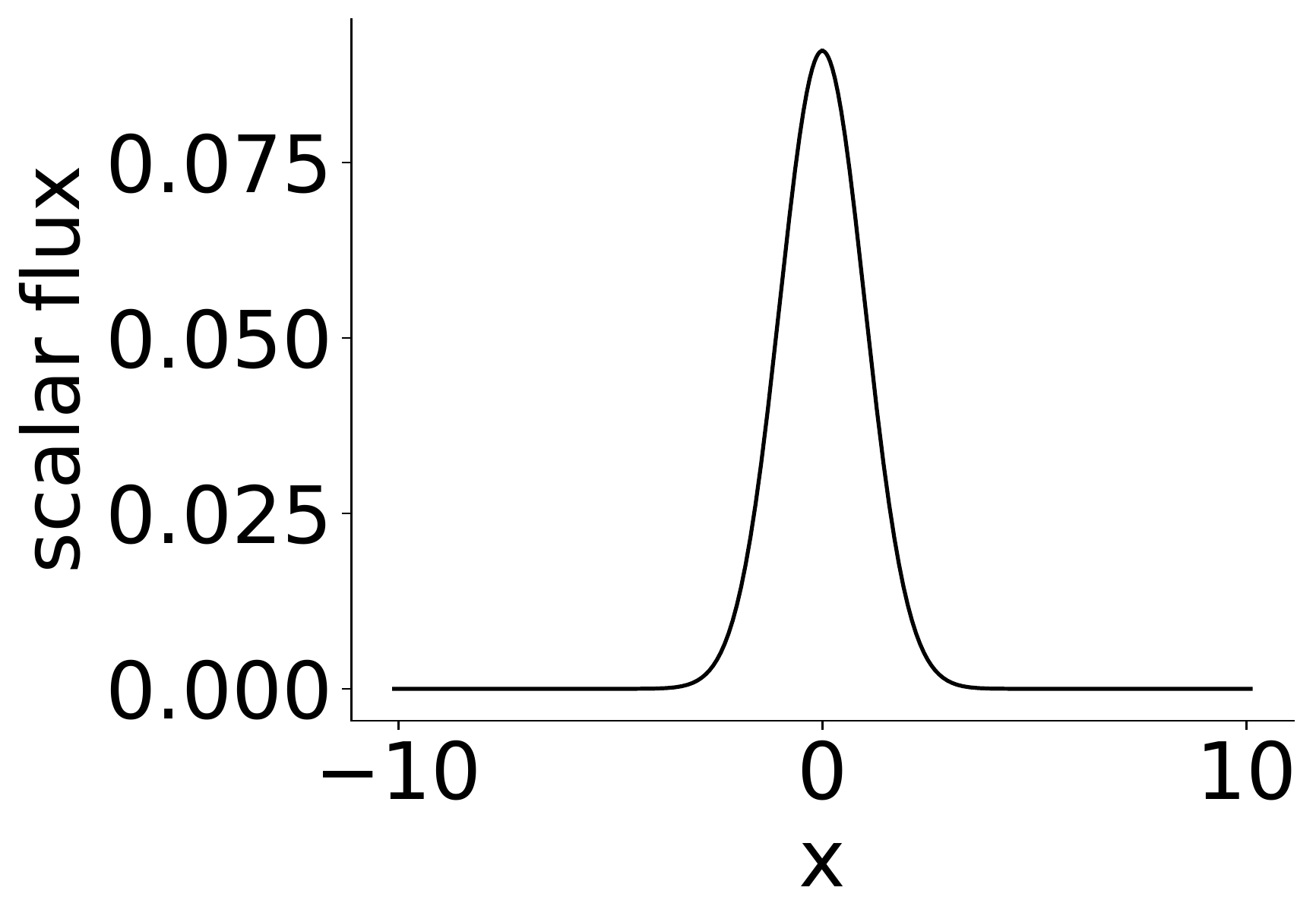}
         \caption{$t=10$}
         \label{fig:MMS_t10}
     \end{subfigure}
        \caption{MMS analytic solution, $\phi$, with $x_0=0.1$ at times $t=1, 5,$ and $10$.}
        \label{fig:MMS}
\end{figure}

In the Method of Manufactured solutions, a solution is chosen and inserted into the governing equations to solve for a source term \cite{lingus1971analytical,knupp2002verification, mcclarren2008manufactured}. This source is then used in a numerical implementation to converge to the already known solution. Here, we specify a solution that will mimic the behavior of a plane pulse source, where a discontinuous wavefront smooths into a solution with Gaussian characteristics. The solution is:
\begin{equation}\label{eq:psimms}
    \psi_{\mathrm{MMS}}(x,\mu,t) = \frac{e^{-x^2/2}}{2(1+t)}\Theta(t-|x| + x_0).
\end{equation}
where $\Theta$ is a step function. Notice that $\psi_{\mathrm{MMS}}$ does not depend on $\mu$ so that the quadrature order will not affect the numerical solution. For three times, $t=1$, $t=5$, and $t=10$ with $x_0=0.1$, the solution is plotted in Figure \ref{fig:MMS}.  The source term that yields this manufactured solution is 
\begin{equation}\label{eq:mmssource}
    S_{\mathrm{MMS}} = -\frac{e^{-\frac{x^2}{2}} (\mu  (t+1) x+1)}{(t+1)^2}\Theta(x-|t| + x_0).
\end{equation}
The source for Eq.~\eqref{eq:vectorEQ} is found by inserting Eq.~\eqref{eq:mmssource} into Eq.~\eqref{eq:S}.

It is important to note that the wavefront in this case is different than the wavefronts that appear in the solutions for the finite width sources (plane pulse, square pulse and source). In this MMS problem, the wavefront is not a feature of the finite wavespeed in the governing equation, but is instead imposed by the step function. Therefore, it is necessary in this problem, and this problem only, to impose a boundary condition at the wavefront, $x = \pm t \pm x_0$. The value of the solution vector at the edges is found by integrating
\begin{equation}\label{eq:bcmms}
     u_{k,j}^l = \intcell B_{i,k}(z) \psi_{\mathrm{MMS}}(x,t,\mu_l),
\end{equation}
at the wavefront. The initial condition is found from Eq.~\eqref{eq:icbcmms}. 

The moving mesh is ideal for enforcing this boundary condition since a cell edge can always be matched to the travelling wavefront. While it is possible to enforce the boundary condition, Eq.~\eqref{eq:bcmms}, with a static mesh, we implement this MMS problem not to compare the different schemes but to test the rate of convergence of the moving mesh method on a problem with a known smooth solution. We also do not attempt to find the uncollided solution in this configuration. 
The moving mesh cell edges are governed by Eq.~\eqref{eq:mesh_edges}.

\begin{figure}
    \centering
    \includegraphics{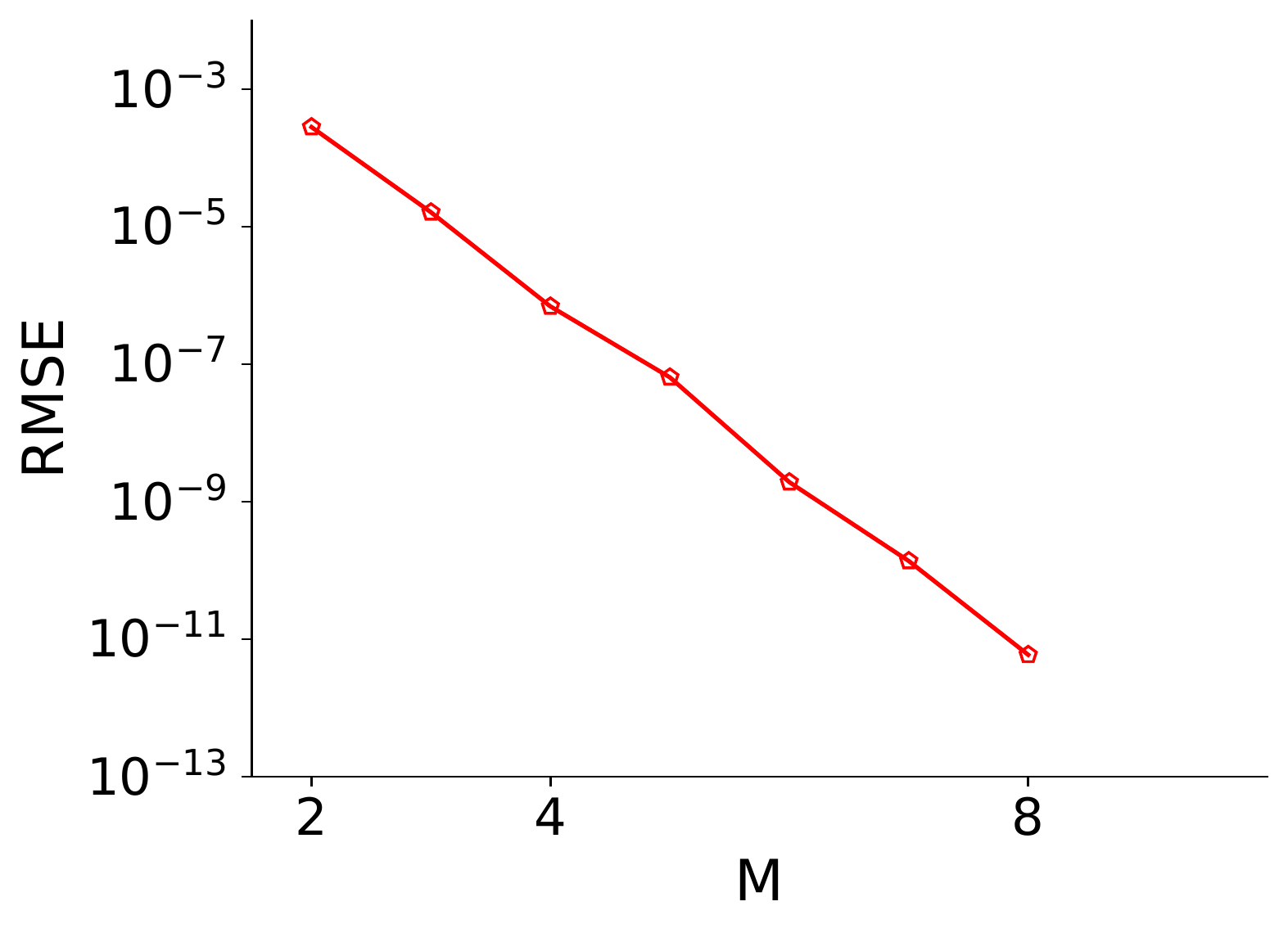}
    \caption{MMS problem convergence results on a logarithmic-linear scale with increasing number of basis functions for $t=1$ and $4$ cells in the moving mesh. The uncollided source treatment is not used in this problem. Using Eq.~\eqref{eq:spectral}, the decay rate in $M$ is $c_1 \approx 1.3$.}
    \label{fig:MMS_RMSE_MS}
\end{figure}

The MMS problem is the only problem we consider that we are assured the angular flux solution is a smooth function. The Gaussian source and pulse considered later have smooth source functions, but there is no analytic expression for the full solution. Figure \ref{fig:MMS_RMSE_MS} shows linear convergence on a log-linear scale which, as explained in Section \ref{sec:error}, indicates spectral convergence (Eq.~\eqref{eq:spectral}).

\begin{figure}
    \centering
    \includegraphics{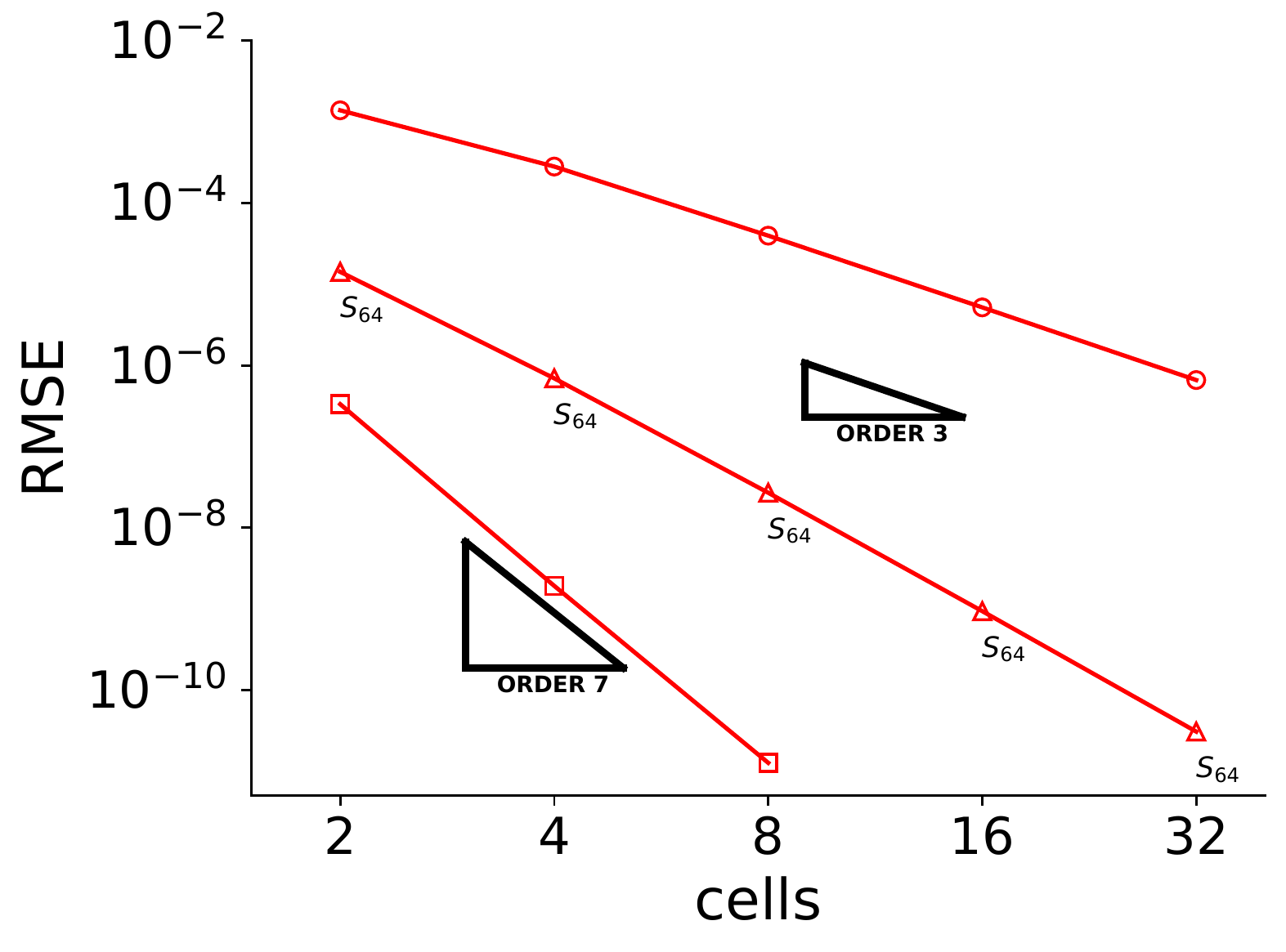}
    \caption{MMS problem convergence results  for a moving mesh, standard source treatment} on a logarithmic scale at $t=1$ for $M=2$ (circles), $M=4$ (triangles), and $M=6$ (squares).
    \label{fig:MMS_RMSE}
\end{figure}

If we fix $M$, we expect the error to converge at an algebraic rate equal to the number of basis functions $(M+1)$. In Figure \ref{fig:MMS_RMSE} we observe that this is indeed the case for the MMS problem. Also, this figure shows the phenomena which will be repeated in subsequent tests of a lowering of the $y$ intercept with increasing number of basis functions. This intercept is $C$ in Eq.~\eqref{eq:algebraic}. Although the MMS source is anisotropic, this solution required only $S_{32}$ to converge to machine precision. The MMS solution results do not necessarily prove the merit of the moving mesh since the wavefront is created in this problem with a boundary condition and says nothing about uncollided source treatments. However, the MMS test verifies that this method can converge at optimal rates when applied to a smooth problem.
\afterpage{\clearpage}
\subsection{Gaussian pulse}\label{sec:gauss_ic}
\begin{figure}
     \centering
     \begin{subfigure}[b]{0.3\textwidth}
         \centering
         \includegraphics[width=\textwidth]{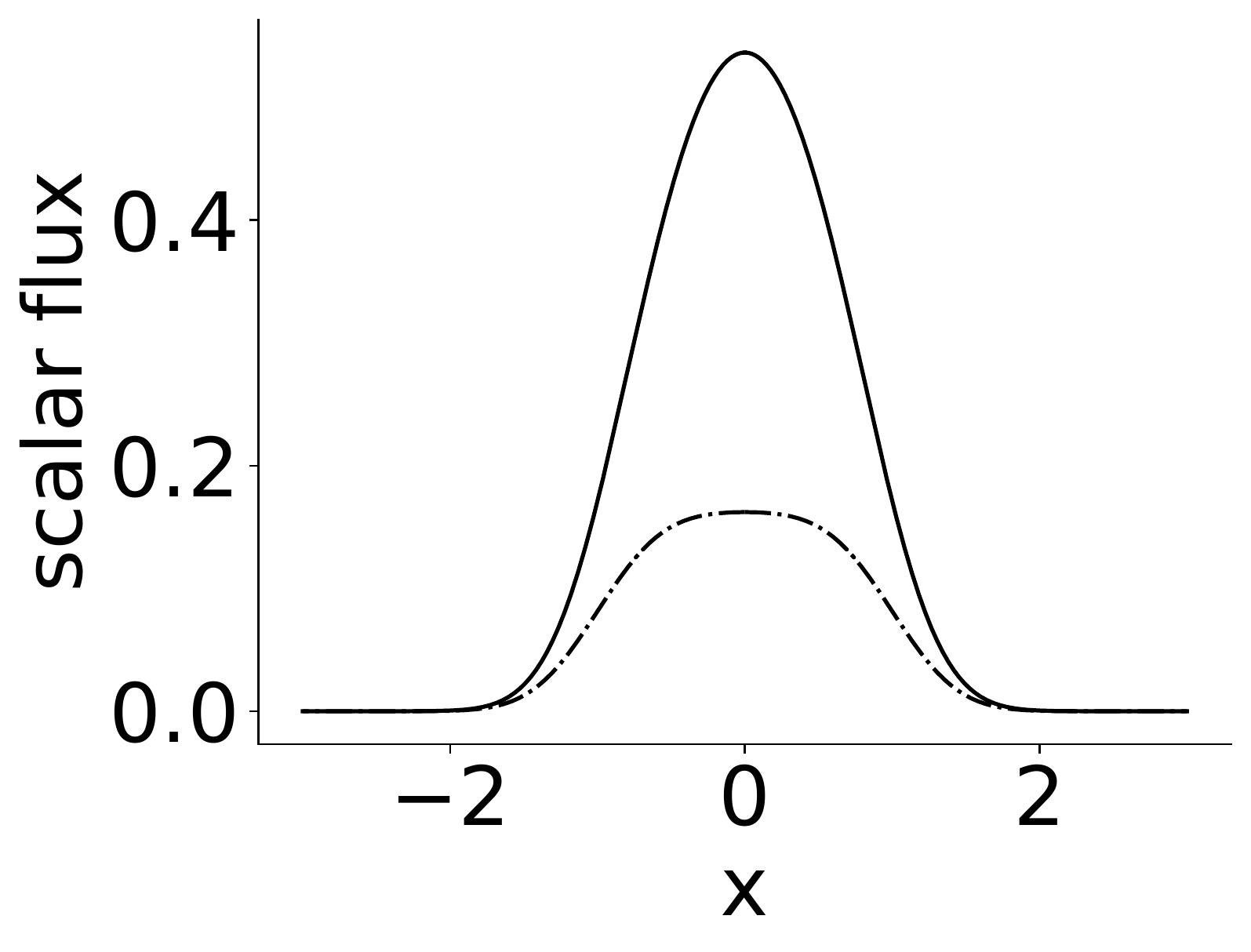}
         \caption{$t=1$}
         \label{fig:gs_IC_1}
     \end{subfigure}
     \hfill
     \begin{subfigure}[b]{0.3\textwidth}
         \centering
         \includegraphics[width=\textwidth]{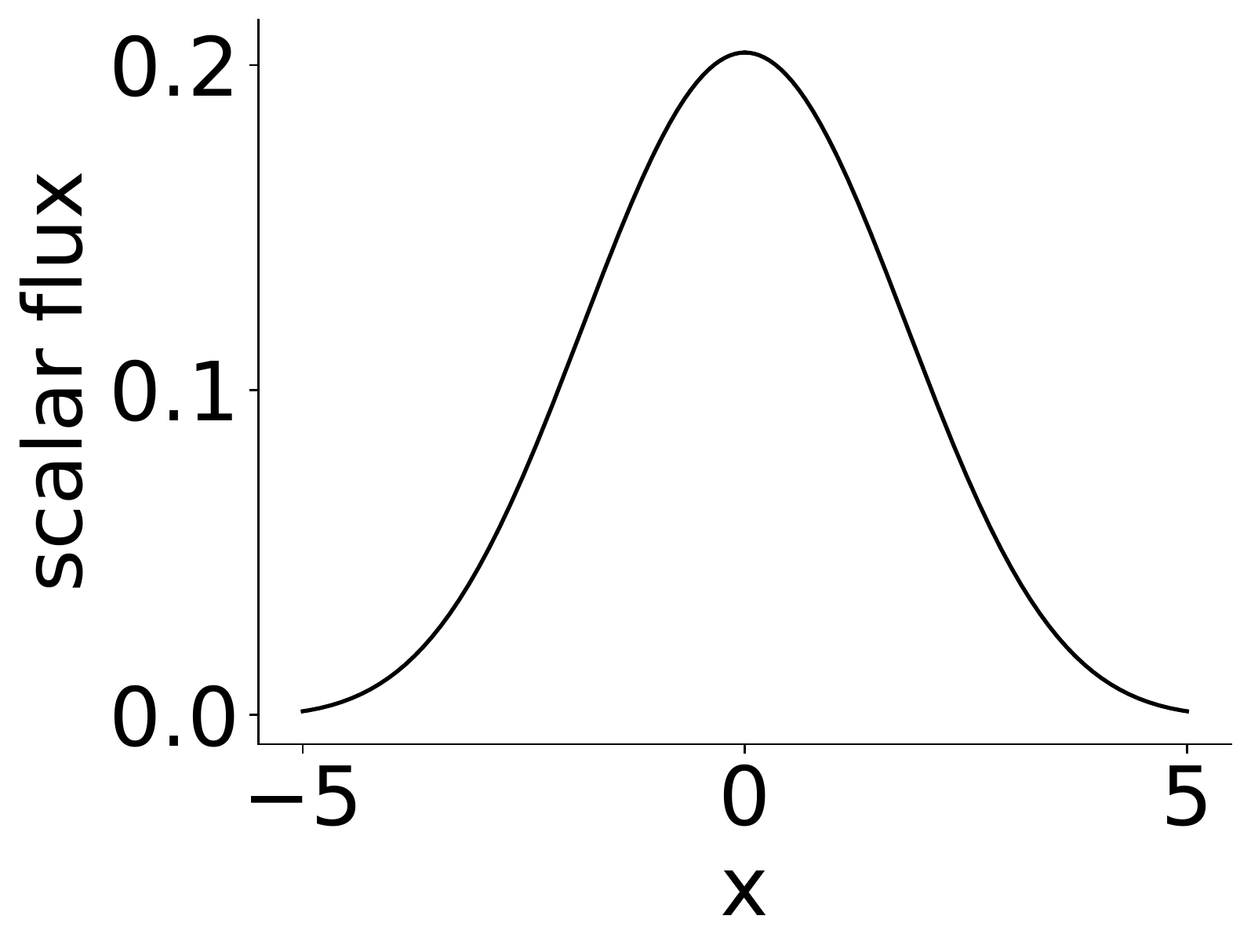}
         \caption{$t=5$}
         \label{fig:gs_IC_5}
     \end{subfigure}
     \hfill
     \begin{subfigure}[b]{0.3\textwidth}
         \centering
         \includegraphics[width=\textwidth]{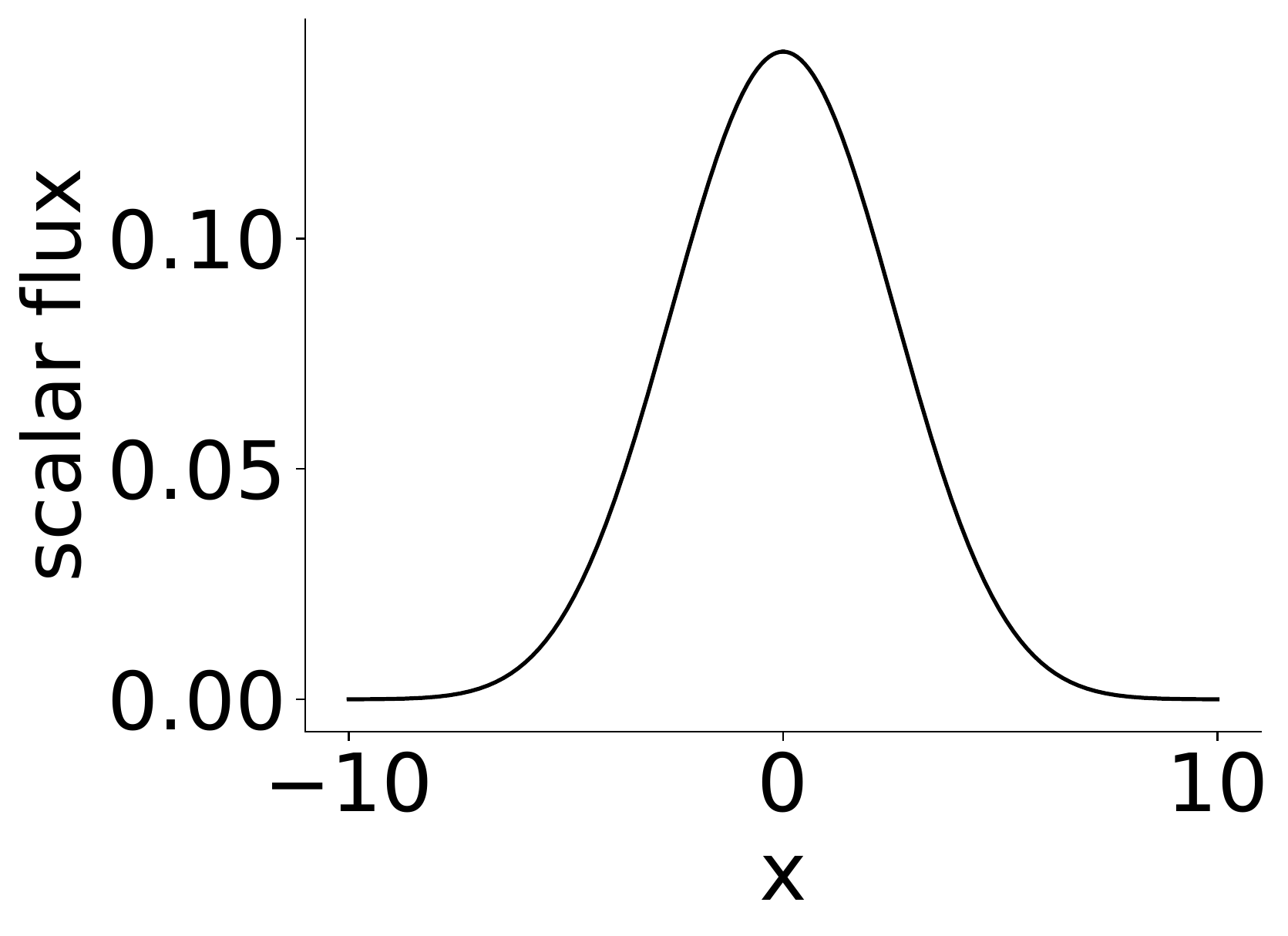}
         \caption{$t=10$}
         \label{fig:gs_IC_10}
     \end{subfigure}
        \caption{Gaussian pulse semi-analytic solution, $\phi$ (solid) and $\phiu$ (dashed) with $\sigma = 0.5$ and $c=1$. The uncollided solution is not shown for times where it is negligible.} 
        \label{fig:gs_IC}
\end{figure}

For the next problem we consider a Gaussian pulse of the form,
\begin{equation}\label{eq:gp}
    S_{\mathrm{gp}}(x,t) = \exp\left({\frac{-x^2}{\sigma^2}}\right)\,\delta(t),
\end{equation}
where $\sigma$, not to be confused with a cross section, is the standard deviation. The solution for $\sigma = 0.5$ is plotted in Figure \ref{fig:gs_IC}. The uncollided solution is significant at early times, but has decayed to approximately zero by $t=5$. Unlike the finite width sources examined later, the uncollided solution for this source is a smooth function.

The solution for the uncollided scalar flux in this configuration is \cite{bennett2022benchmarks},
\begin{equation}\label{eq:gaussian_pulse_uncollided}
    \phi_u^\mathrm{gs}(x,t) =  \sigma\,\sqrt{\pi }\, e^{-t} \,\frac{\text{erf}\left( \frac{t-x}{\sigma}\right)+\text{erf}\left(\frac{t+x}{\sigma}\right)}{4 t}.
\end{equation}
For the uncollided case, Eq.~\eqref{eq:gaussian_pulse_uncollided} is treated as $S$ in Eq.~\eqref{eq:S} and the initial condition is zero everywhere. For the methods that do not use an uncollided source, the initial condition is found by inserting $\psi(x,t=0) = \frac{1}{2} \exp\left({\frac{-x^2}{\sigma^2}}\right) $ into Eq.~\eqref{eq:icbcmms} and setting $S(x,t) = 0$.

\begin{figure}
    \centering
    \includegraphics{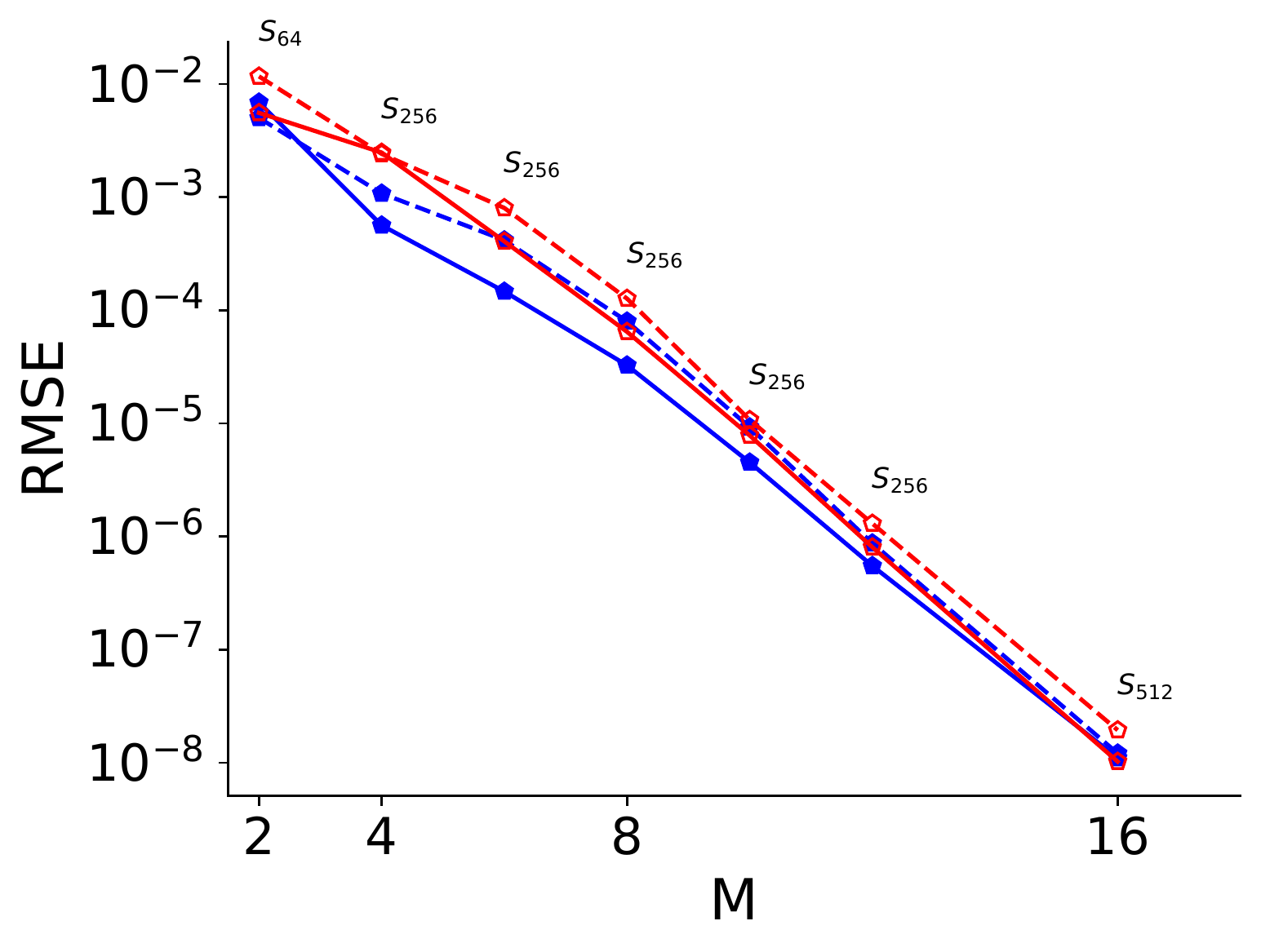}
    \caption{Gaussian pulse convergence with logarithmic- linear scaling at $t=1$ increasing number of basis functions with $4$ cells. Blue lines indicate the uncollided solution is used, red that no uncollided source is used.  Dashed lines are for a static mesh and solid lines are for the moving mesh.  Using Eq.~\eqref{eq:spectral}, the estimated decay rate in $M$ is $c_1 \approx 1.0$.}
    \label{fig:gauss_IC_spectral}
\end{figure}

While the initial condition does have an infinite width, setting the initial mesh width ($x_0$) to be the location where the initial condition is below a specified small tolerance in Eq.~\eqref{eq:mesh_edges} is sufficient to achieve accurate results. For example, the initial condition is less than $1\times 10^{-16}$ at $x=3.1$ with the standard deviation, $\sigma=0.5$. We use the same $x_0$ for static mesh calculations and initialize the mesh to span the space $[-t_{\mathrm{final}} - x_0, t_{\mathrm{final}} + x_0]$.
The Gaussian pulse and Gaussian source are a more realistic test than the MMS problem and still have the guarantee of a smooth source and smooth uncollided solution. All combinations of moving or static mesh and uncollided source or standard source that we applied to this problem achieved spectral convergence, shown in Figure \ref{fig:gauss_IC_spectral}. The uncollided, moving mesh method performed the best, consistently achieving lower error than the other methods. 

For the Gaussian pulse, only $S_{256}$ is required to achieve accuracies of  $\mathrm{RMSE}\approx10^{-6}$. After that point, more angles are required to ensure that the angular discretization error is lower than the errors in discretizing space or time. 

\begin{figure}
    \centering
    \begin{subfigure}[b]{0.48\textwidth}
    \includegraphics[width=\textwidth]{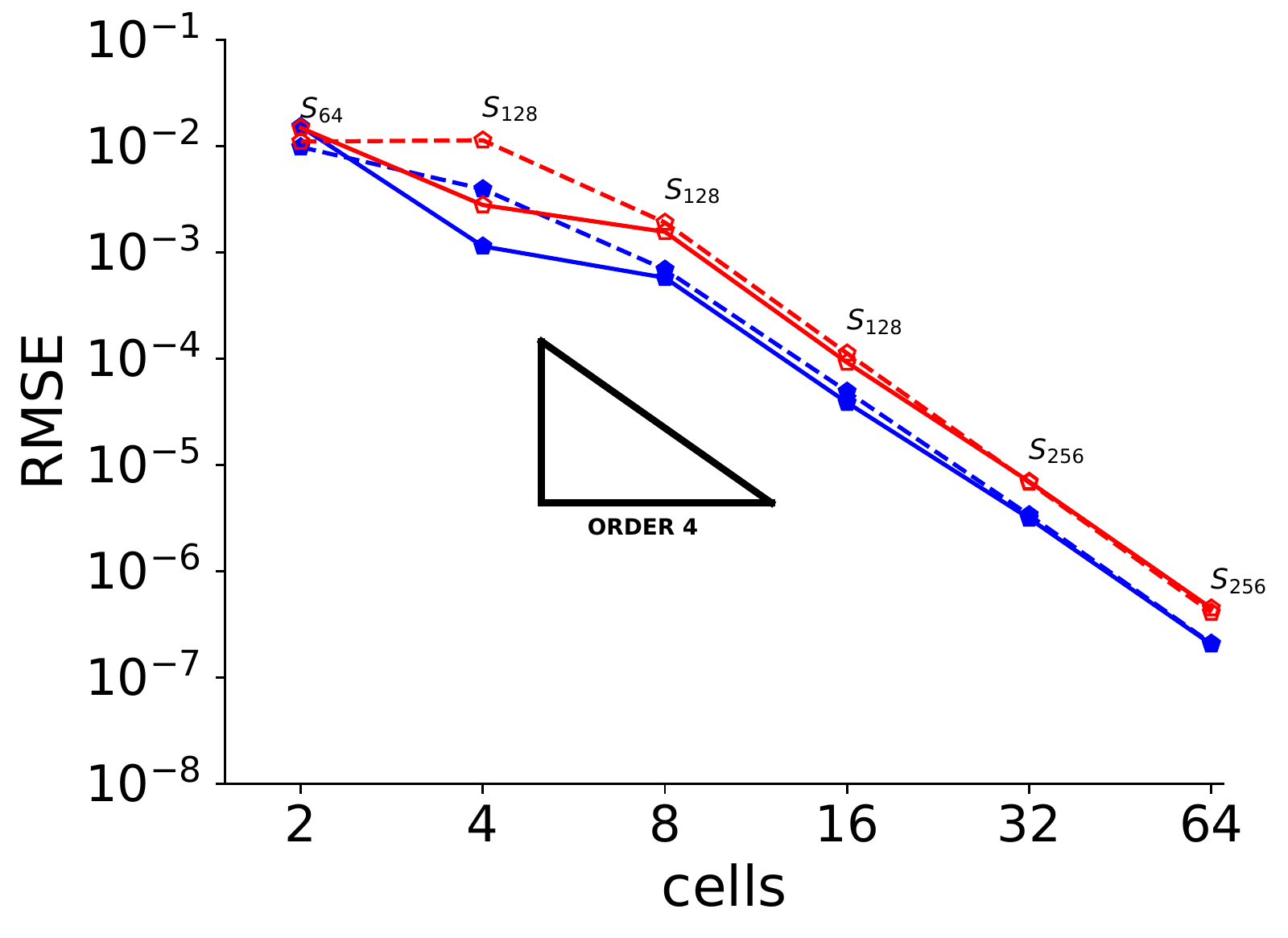}
    \caption{$M=3$}
    \label{subfig:gauss_M3_rms}
    \end{subfigure}
    \centering
    \begin{subfigure}[b]{0.48\textwidth}
        \includegraphics[width=\textwidth]{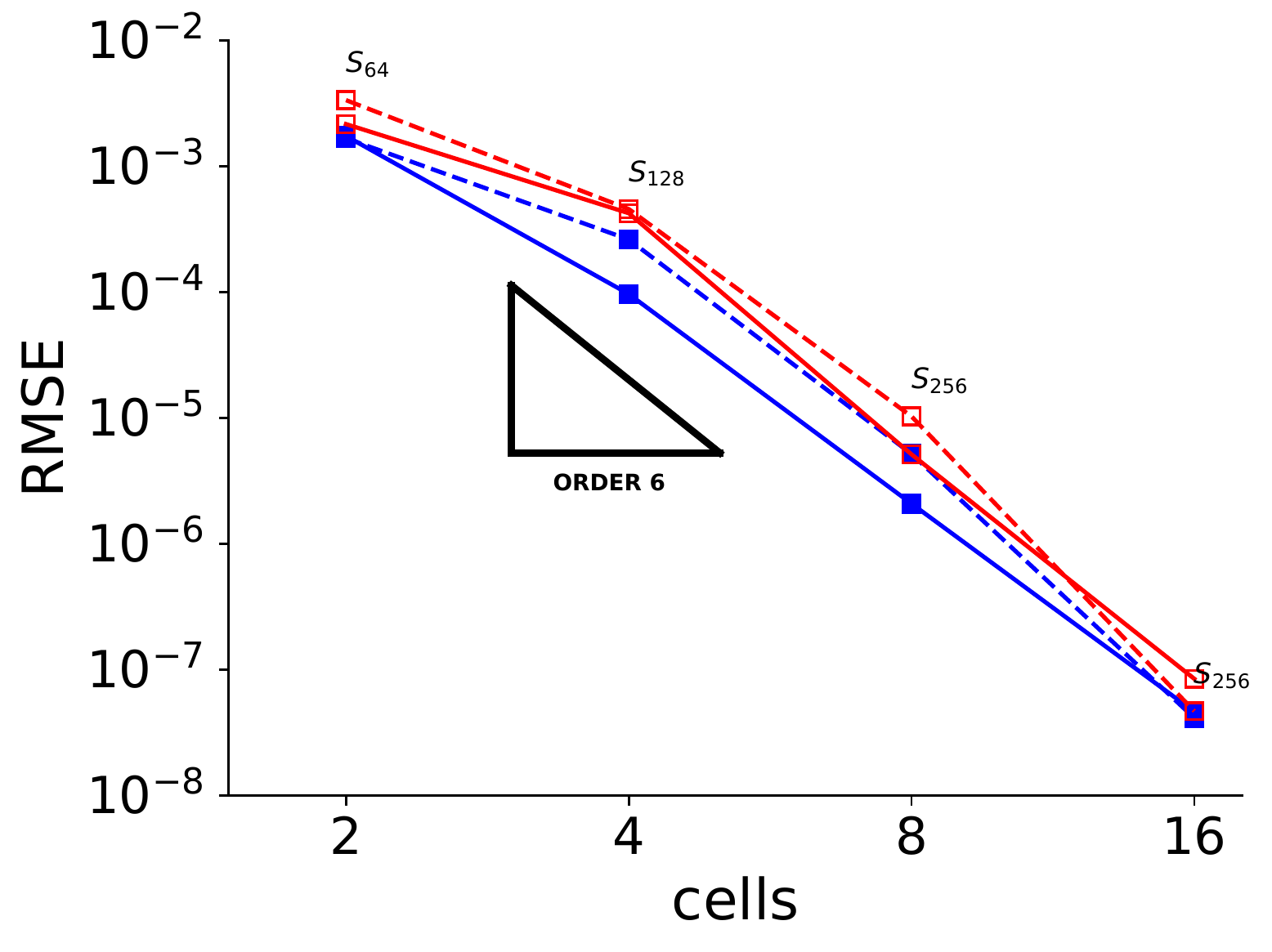}
        \caption{$M=6$}
        \label{subfig:gauss_M6_rms}
    \end{subfigure}
    \caption{Gaussian pulse convergence results on a logarithmic scale with $c=1$, $\sigma= 0.5$ at $t=1$. Blue lines indicate the uncollided solution is used, red that no uncollided source is used.  Dashed lines are for a static mesh and solid lines are for the moving mesh.}
    \label{fig:gauss_IC_RMS}
\end{figure}

Figure \ref{subfig:gauss_M6_rms} demonstrates that each method can achieve sixth-order convergence by holding $M=6$ and increasing the number of mesh subdivisions, $K$. Since our methods are spectrally convergent, with seven basis functions ($M=6$) one would expect the convergence to approach seventh-order as $K\rightarrow\infty$. This convergence test is limited by the accuracy of the benchmark solution before $K$ is sufficiently large to achieve seventh order convergence. However, Figure \ref{subfig:gauss_M3_rms} achieves fourth-order convergence with four basis functions ($M=3$) before reaching the accuracy limit of the benchmark. 

\afterpage{\clearpage}
\subsection{Gaussian source}\label{sec:gauss_s}
\begin{figure}
     \centering
     \begin{subfigure}[b]{0.3\textwidth}
         \centering
         \includegraphics[width=\textwidth]{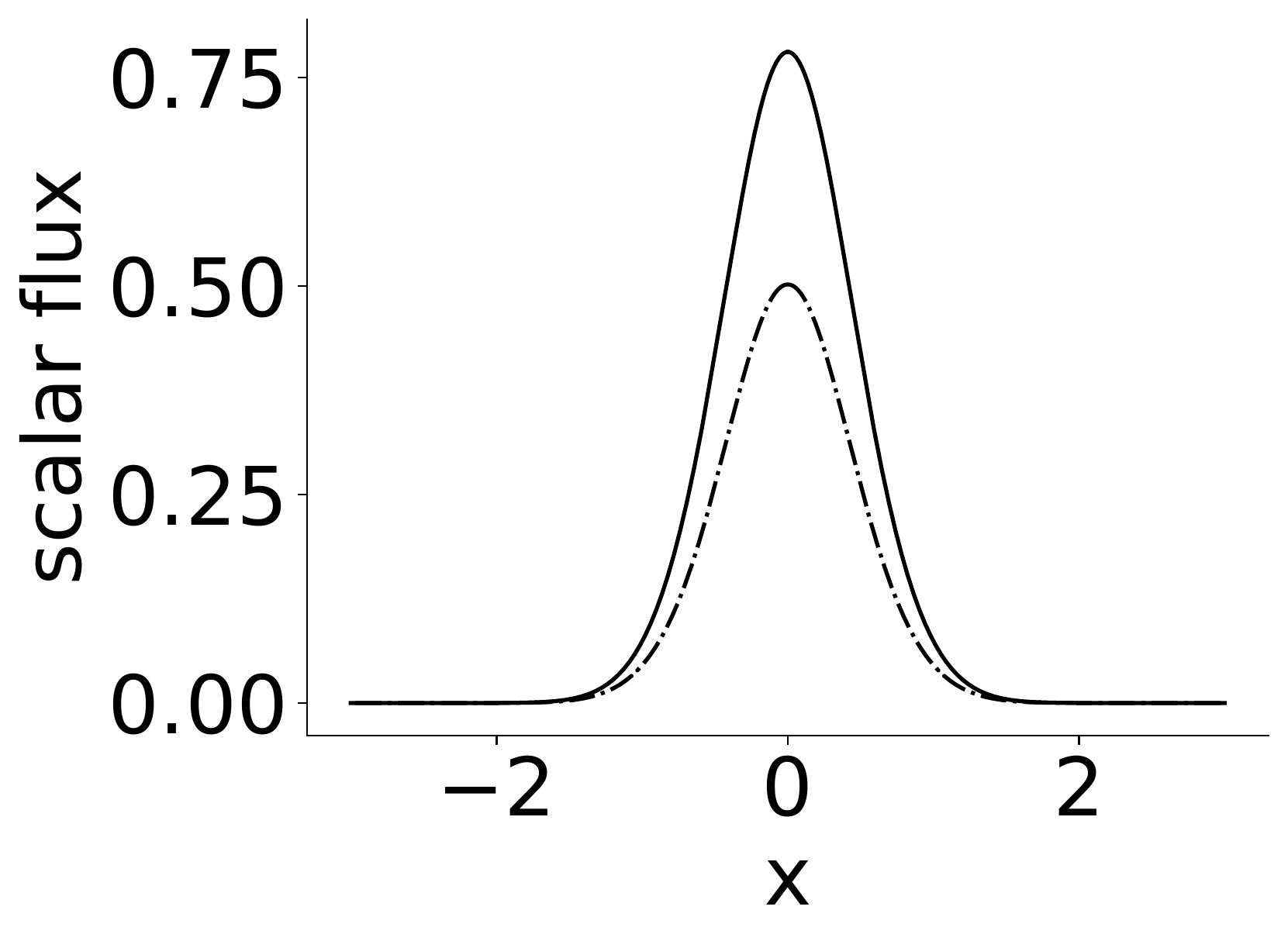}
         \caption{$t=1$}
         \label{fig:gs_s_1}
     \end{subfigure}
     \hfill
     \begin{subfigure}[b]{0.3\textwidth}
         \centering
         \includegraphics[width=\textwidth]{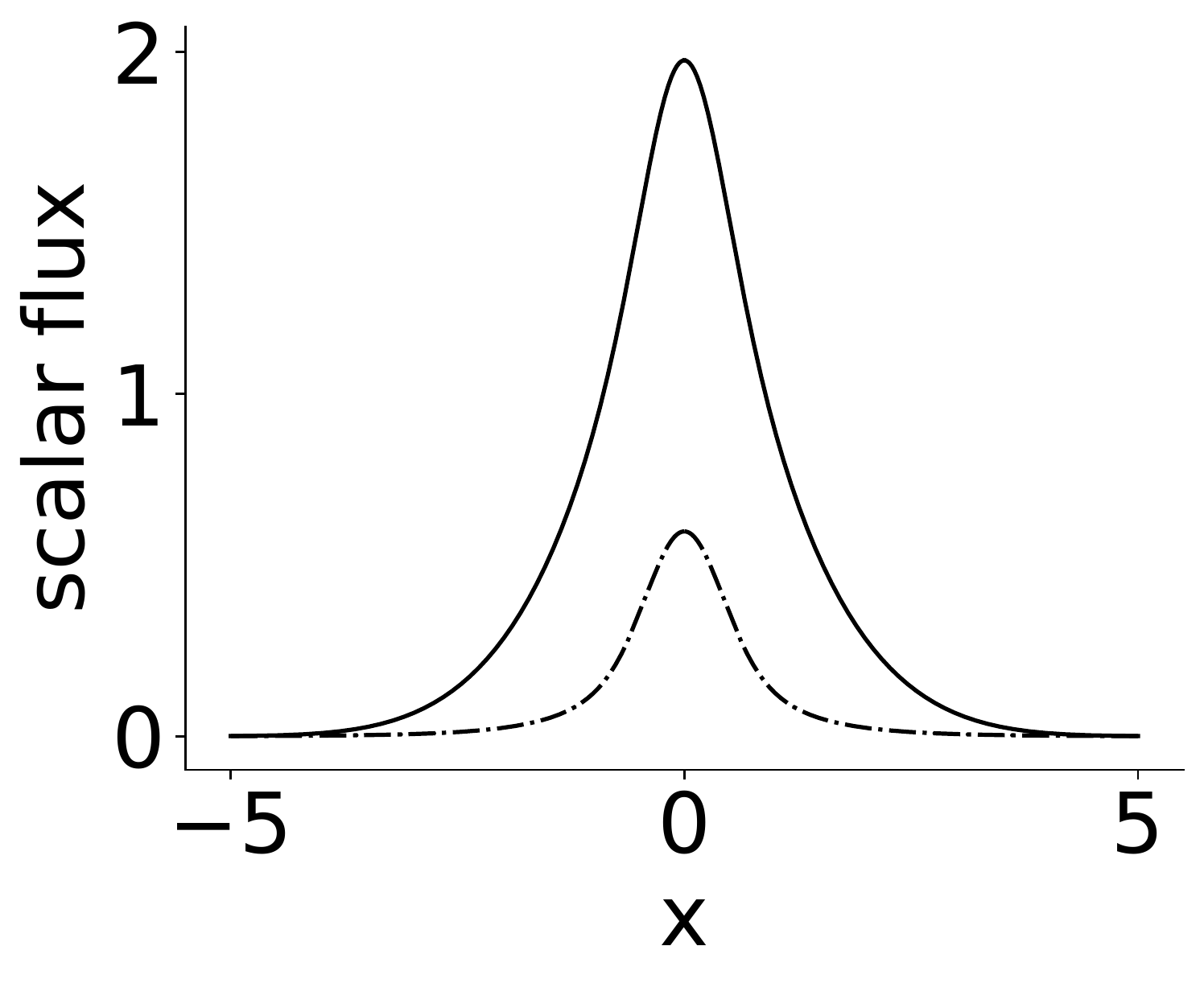}
         \caption{$t=5$}
         \label{fig:gs_s_5}
     \end{subfigure}
     \hfill
     \begin{subfigure}[b]{0.3\textwidth}
         \centering
         \includegraphics[width=\textwidth]{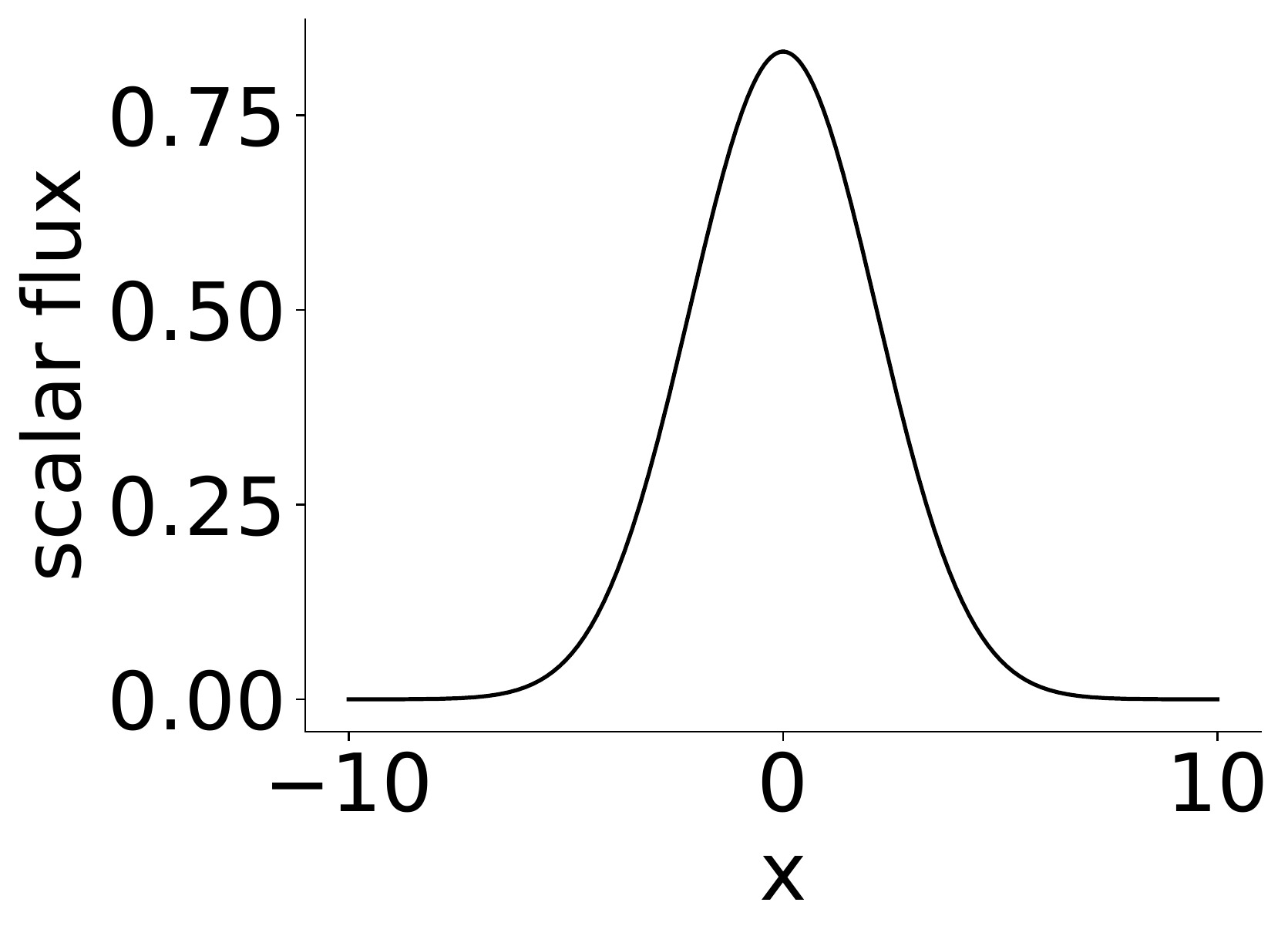}
         \caption{$t=10$}
         \label{fig:gs_s_10}
     \end{subfigure}
        \caption{Gaussian source semi-analytic solution, $\phi$ (solid) and $\phiu$ (dashed) with $\sigma = 0.5$, $t_0= 5$, and $c=1$. The uncollided flux is not shown for times where it is negligible.}
        \label{fig:gs_s}
\end{figure}

A Gaussian source turned on at $t=0$ and left on until $t=t_0$ is a superposition of Gaussian pulses,
\begin{equation}\label{eq:gauss_s}
    S_{\mathrm{gs}}(x,t) = \exp\left({\frac{-x^2}{\sigma^2}}\right)\,\Theta(t_0 - t).
\end{equation}
Figure \ref{fig:gs_s} shows this solution with  $\sigma=0.5$ and $t_0 = 5$. Notice that the uncollided solution is still significant at $t=5$. Like the Gaussian pulse, the uncollided angular flux and the collided angular flux are smooth functions. 

Reference \cite{bennett2022benchmarks} gives the uncollided scalar flux for this source as a convolution of the Gaussian pulse uncollided flux,
\begin{equation}\label{eq:gaussian_source_uncollided_1}
    \phi_u^\mathrm{gs}(x,t) =  \int_0^{\mathrm{min}(t,t_0)}\!d\tau\,\sigma\,\sqrt{\pi }\, e^{-(t-\tau)} \,\frac{\text{erf}\left( \frac{t-\tau-x}{\sigma}\right)+\text{erf}\left(\frac{t-\tau+x}{\sigma}\right)}{4 (t-\tau)}.
\end{equation}
For the two solution methods that use an uncollided source treatment, Eq.~\eqref{eq:gaussian_source_uncollided_1} is integrated in Eq.~\eqref{eq:S}. For the two cases that do not employ the uncollided source, $S$ in Eq.~\eqref{eq:S} is Eq.~\eqref{eq:gauss_s}. The angular flux is initialized to be zero everywhere for both source treatments. 

The moving mesh and static mesh are treated in the exact same way as Section \ref{sec:gauss_ic}, where the initial width is set to be the initial width of a Gaussian pulse with the same standard deviation.

\begin{figure}
    \centering
    \includegraphics{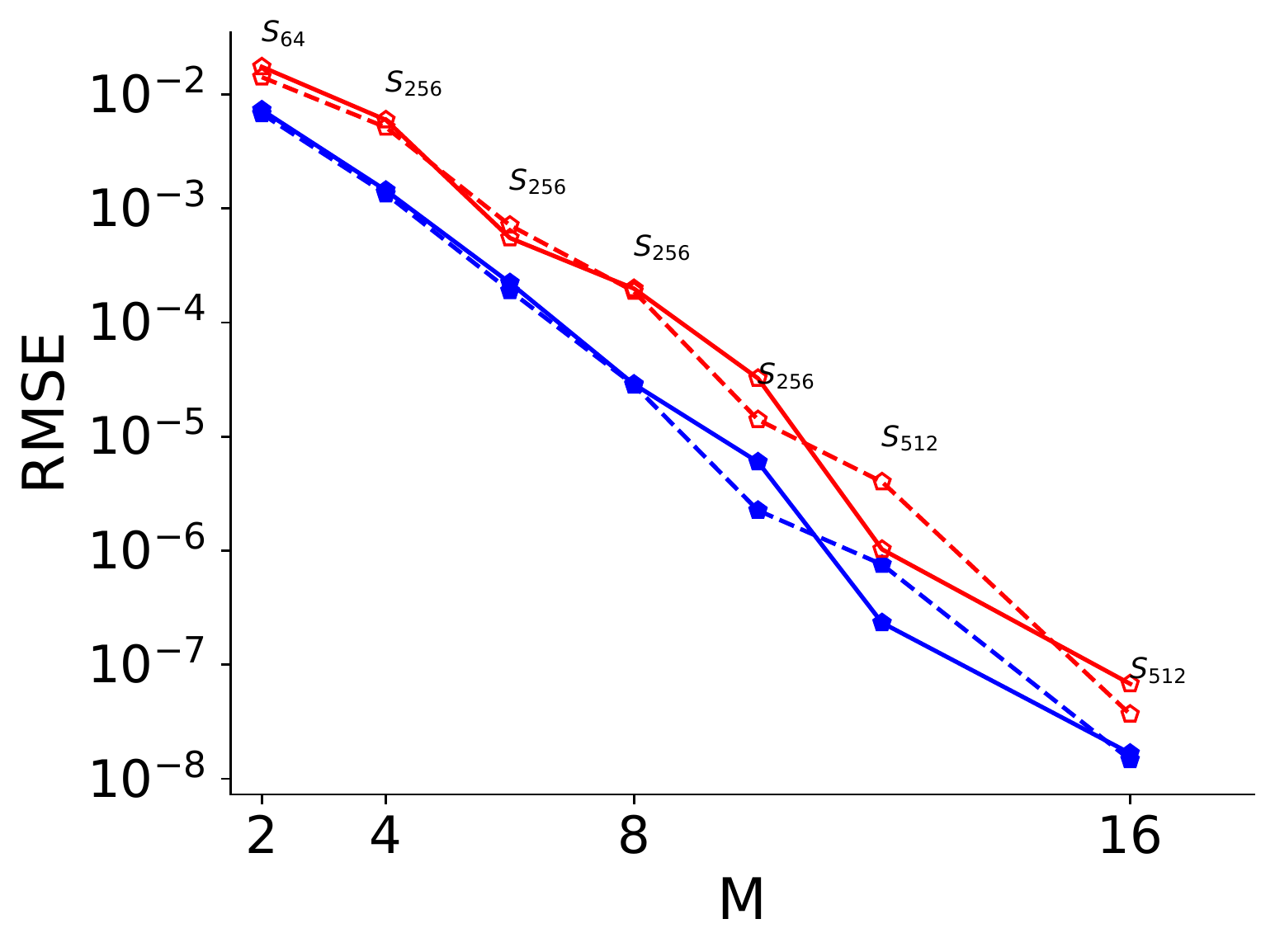}
    \caption{Gaussian source convergence on a logarithmic-linear scale at $t=1$ increasing number of basis functions with $4$ cells. Blue lines indicate the uncollided solution is used, red that no uncollided source is used.  Dashed lines are for a static mesh and solid lines are for the moving mesh.  Using Eq.~\eqref{eq:spectral}, the decay rate in $M$ is estimate to be $c_1 \approx 0.95$  }
    \label{fig:gauss_s_spectral}
\end{figure}

Our solutions for the Gaussian source achieve similar levels of accuracy as in the Gaussian pulse. Figure \ref{fig:gauss_s_spectral} shows spectral convergence for all four methods. Compared with the Gaussian pulse (Figure \ref{fig:gauss_IC_spectral}), these results show a more drastic difference in the intercepts between the methods that use the uncollided solution and those that do not. This could be because, as shown in Figure \ref{fig:gs_s}, the uncollided solution is a significant portion of the solution at later times. 

\begin{figure}
    \centering
    \begin{subfigure}[b]{0.48\textwidth}
    \includegraphics[width=\textwidth]{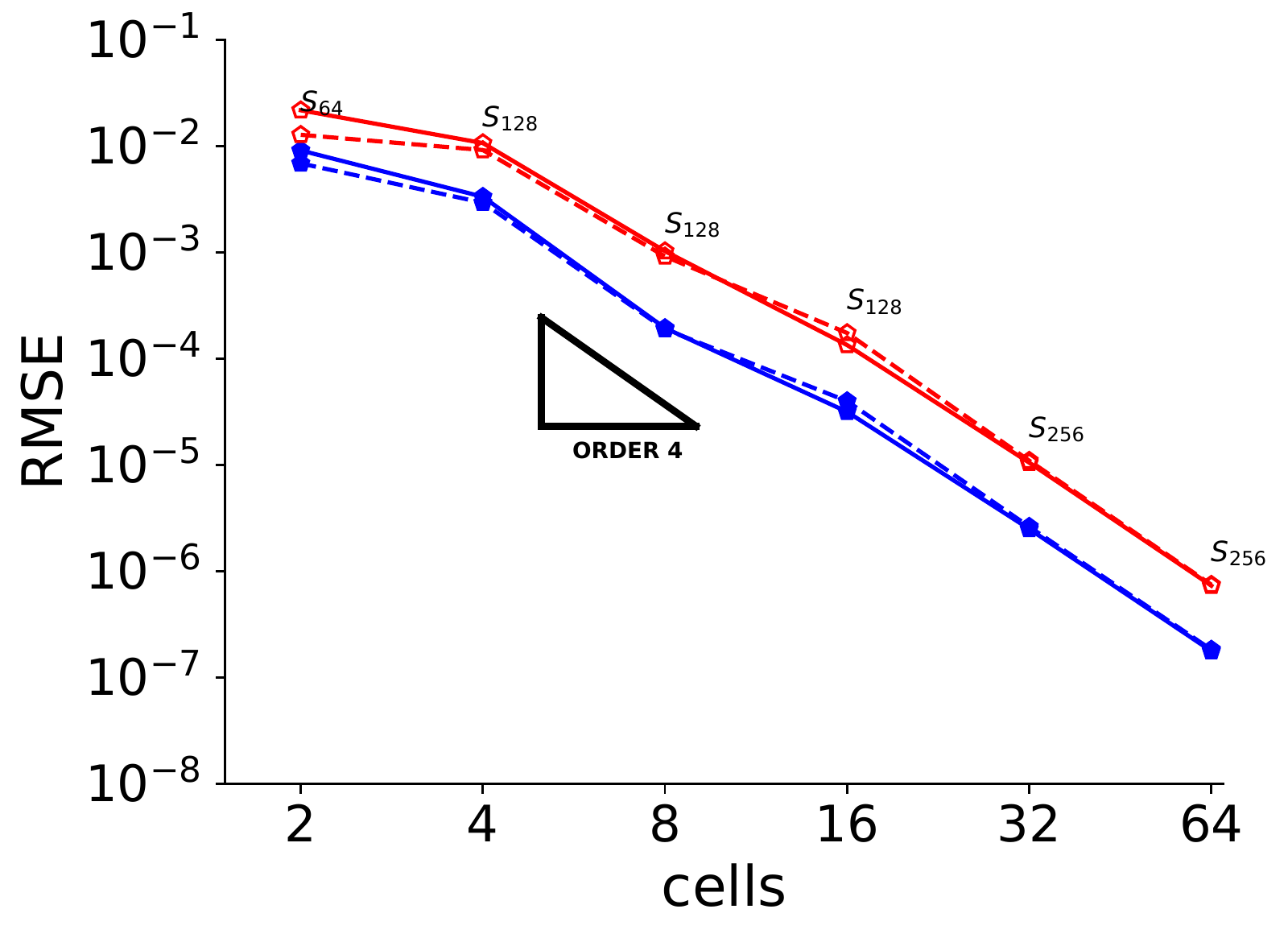}
    \caption{$M=3$}
    \label{subfig:gauss_s_M3_rms}
    \end{subfigure}
    \centering
    \begin{subfigure}[b]{0.48\textwidth}
        \includegraphics[width=\textwidth]{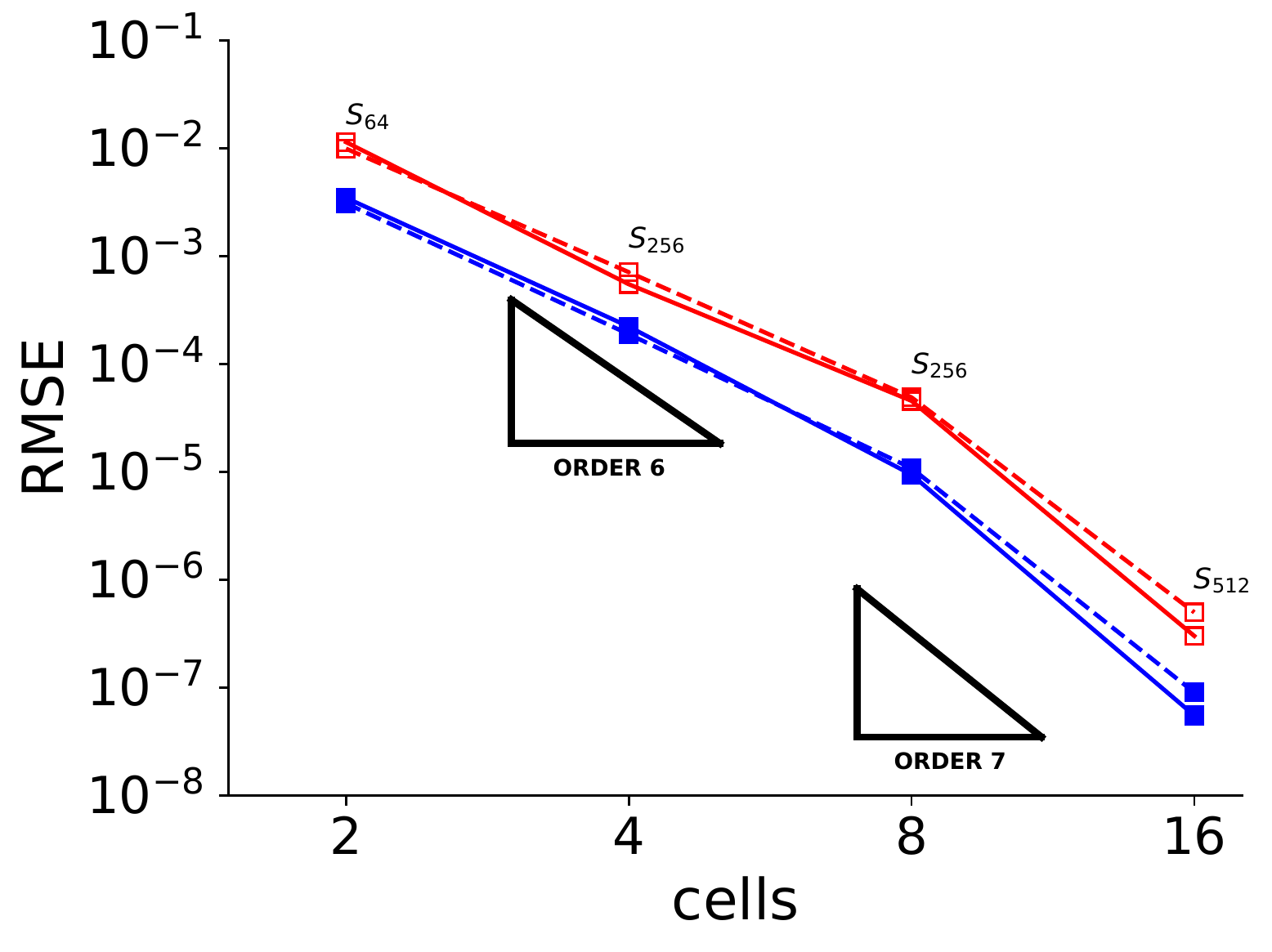}
        \caption{$M=6$}
        \label{subfig:gauss_s_M6_rms}
    \end{subfigure}
    \caption{Gaussian source convergence results on a logarithmic scale with $c=1$, $\sigma= 0.5$ at $t=1$. Blue lines indicate the uncollided solution is used, red that no uncollided source is used.  Dashed lines are for a static mesh and solid lines are for the moving mesh.}
    \label{fig:gauss_s_RMS}
\end{figure}

Figure \ref{subfig:gauss_s_M6_rms} shows these methods achieving almost sixth order convergence with $M=6$ and improving to seventh order for $K=16$ before the error approaches the number of accurate digits for the benchmark solution. Figure \ref{subfig:gauss_s_M3_rms} better illustrates how each method can achieve an order of convergence equal to the number of basis functions ($M+1$) as $K\rightarrow\infty$ with fourth-order convergence with $M=3$. These solutions have an angular error requiring $S_{512}$ to achieve $\mathrm{RMSE}\approx 10^{-7}$.

\afterpage{\clearpage}
\subsection{Plane pulse}\label{sec:pl_ic}
\begin{figure}
     \centering
     \begin{subfigure}[b]{0.3\textwidth}
         \centering
         \includegraphics[width=\textwidth]{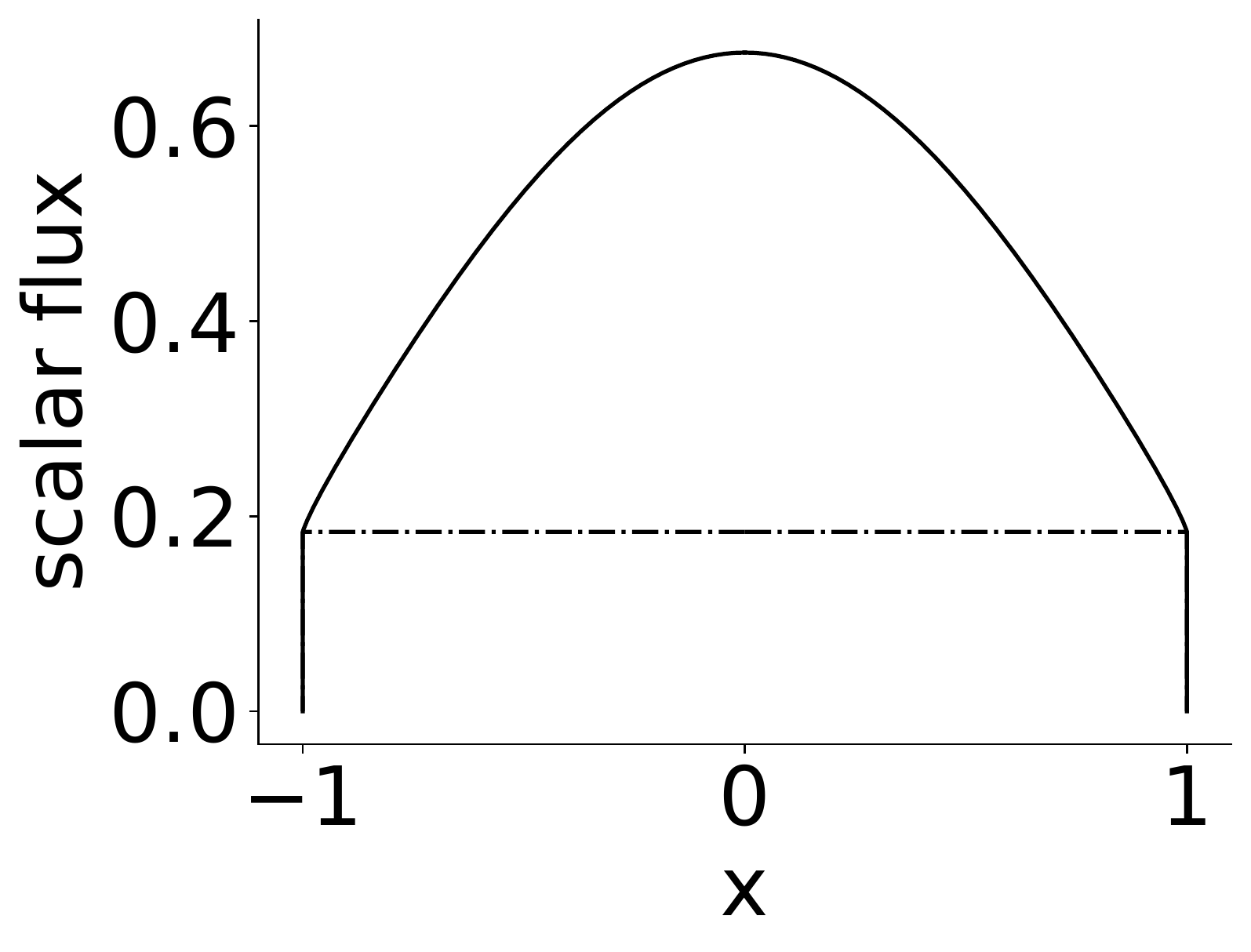}
         \caption{$t=1$}
         \label{fig:pl_IC_1}
     \end{subfigure}
     \hfill
     \begin{subfigure}[b]{0.3\textwidth}
         \centering
         \includegraphics[width=\textwidth]{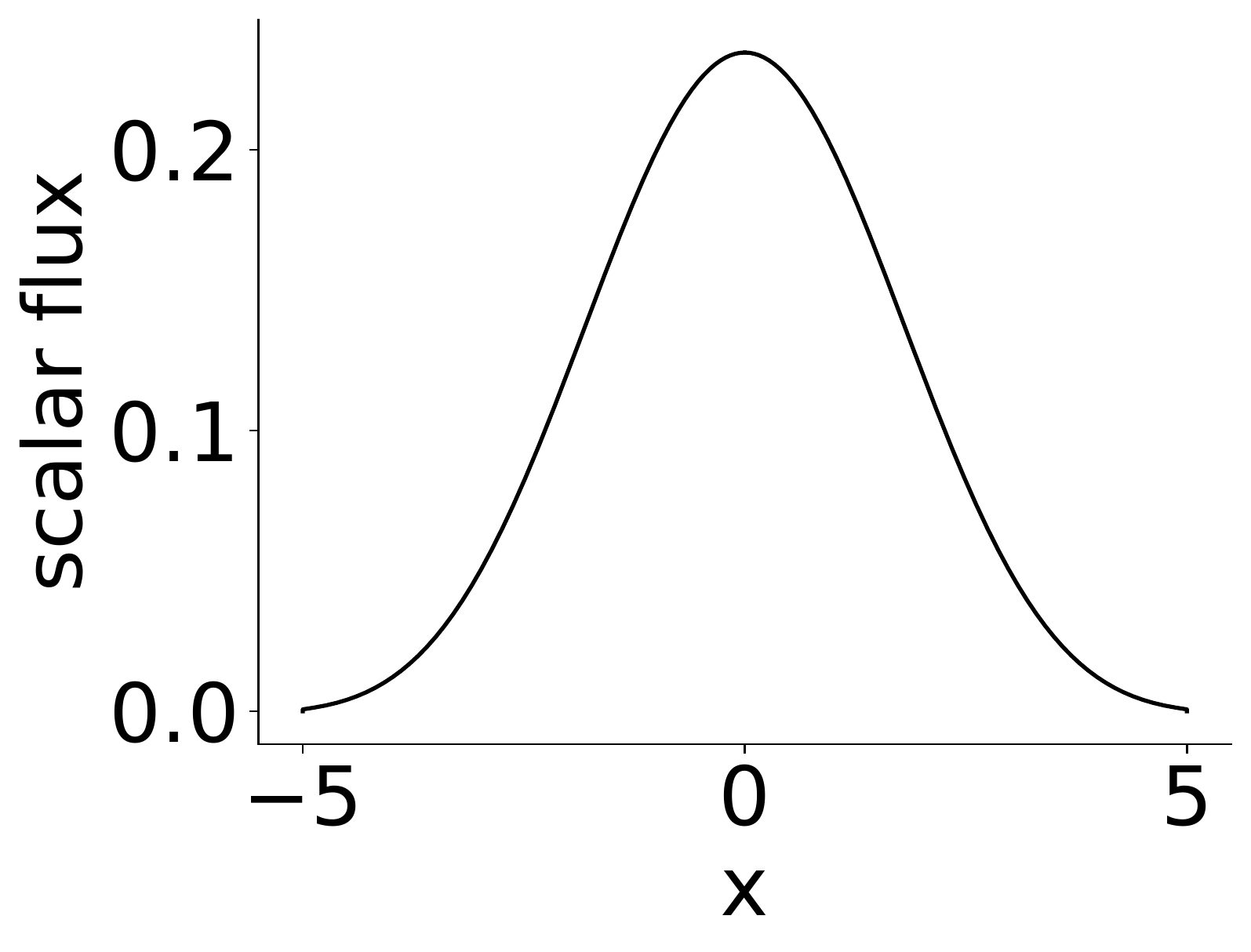}
         \caption{$t=5$}
         \label{fig:fig:pl_IC_5}
     \end{subfigure}
     \hfill
     \begin{subfigure}[b]{0.3\textwidth}
         \centering
         \includegraphics[width=\textwidth]{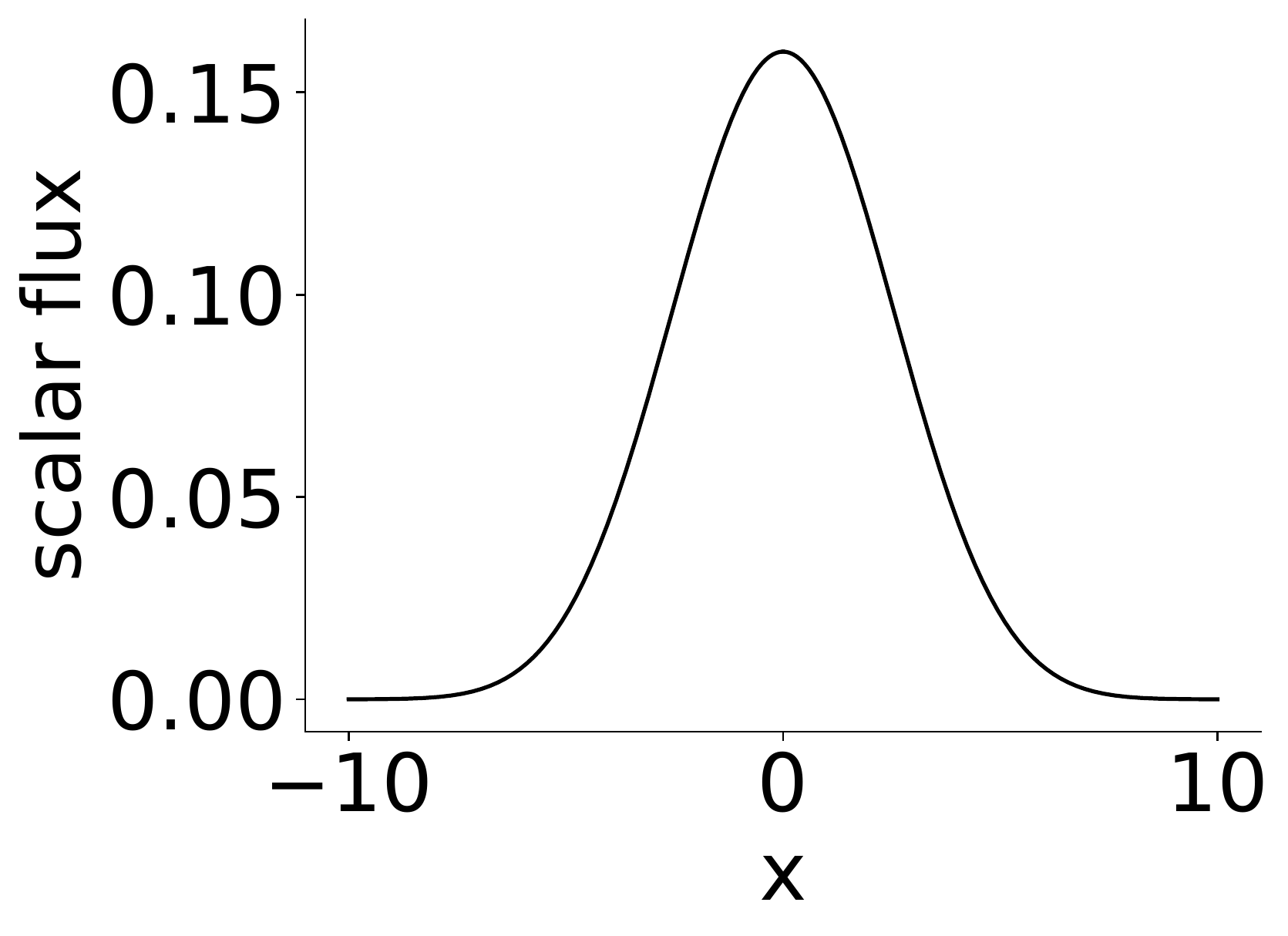}
         \caption{$t=10$}
         \label{fig:fig:pl_IC_10}
     \end{subfigure}
        \caption{Plane pulse semi-analytic solution, $\phi$ (solid) and $\phiu$ (dashed) with $c=1$. The uncollided flux is not shown for times where it is negligible.}
        \label{fig:pl_IC}
\end{figure}

Ganapol's frequently used benchmark solution for an infinite plane source is a Green's function for the source,
\begin{equation}
    S_{\mathrm{pl}} = \delta(x)\delta(t).
\end{equation}
The full solution is plotted for three times in Figure \ref{fig:pl_IC}. This solution is shows how a discontinuous uncollided scalar flux can cause discontinuities in the first derivative of the full solution. In this case, the discontinuity is manifested as traveling wavefront at early time. The solution smooths to be redolent of a Gaussian  at later times. 

The uncollided solution for this configuration, also given by \cite{ganapol}, is
\begin{equation}\label{eq:uncollided_plane_IC}
    \phi_{\mathrm{u}}^{\mathrm{pl}}(x,t) = \frac{\exp\left(-t\right)}{2t}\Theta\left(1-\left|\frac{x}{t}\right|\right).
\end{equation}
Equation~\eqref{eq:uncollided_plane_IC} is substituted for $S$ in Eq.~\eqref{eq:S} for the uncollided source treatment. For the two cases where an uncollided source is not used, it is necessary to approximate the delta function initial condition. For the static mesh case, an initial condition of,
\begin{equation}\label{eq:pl_IC}
    \psi^{\mathrm{pl}}(x,t=0) =  \frac{1}{2}\frac{\Theta(x_0-|x|)}{2x_0},
\end{equation}
where $x_0$ is a very small number, is used to approximate the initial condition in Eq.~\eqref{eq:icbcmms}. The moving mesh, no uncollided source case did not converge due to how we represent the initial condition and is not included here.

The moving mesh is governed by Eq.~\eqref{eq:mesh_edges} with $x_0$ set to approximately zero, which ensures that the step function in Eq.~\eqref{eq:uncollided_plane_IC} is always one. The static mesh for this source spans $[-t_{\mathrm{final}}, t_{\mathrm{final}}]$ with evenly spaced cells. 

For $j>0$, our $B_j$ basis are orthogonal to functions that are constant in the range $x_L$ to $x_R$. The uncollided solution always satisfies this criterion since the wavefronts are never inside any of the cells. Therefore, it is simple to integrate the uncollided solution for the moving mesh over a cell in the source term. The source becomes,
\begin{equation}
    Q_0(x,t) = \int^{x_R}_{x_L}\!dx\,B_{0,k}(z)\,\frac{\exp\left(-t\right)}{2t} = \sqrt{x_R-x_L}\,\frac{\exp\left(-t\right)}{2t},
\end{equation}
This simplification is not possible in the static mesh case due to the step function, which requires integration at the moving wavefront. 
 
For the finite width sources, the plane pulse and the square source and pulse, spectral convergence is unrealistic with our methods. The discontinuities in the scalar flux can be easily matched with mesh edges, resulting in a smooth problem, but the angular flux poses a more difficult problem. In the angular flux of the plane pulse for example, the angular collided flux has a discontinuity in the first derivative travelling outward from the origin at speed $\mu t$ that is caused by the uncollided angular flux. This means that there is as many discontinuities in the angular fluxes as there are discrete angles. 

\begin{figure}
    \centering
    \begin{subfigure}[b]{0.48\textwidth}
    \includegraphics[width=\textwidth]{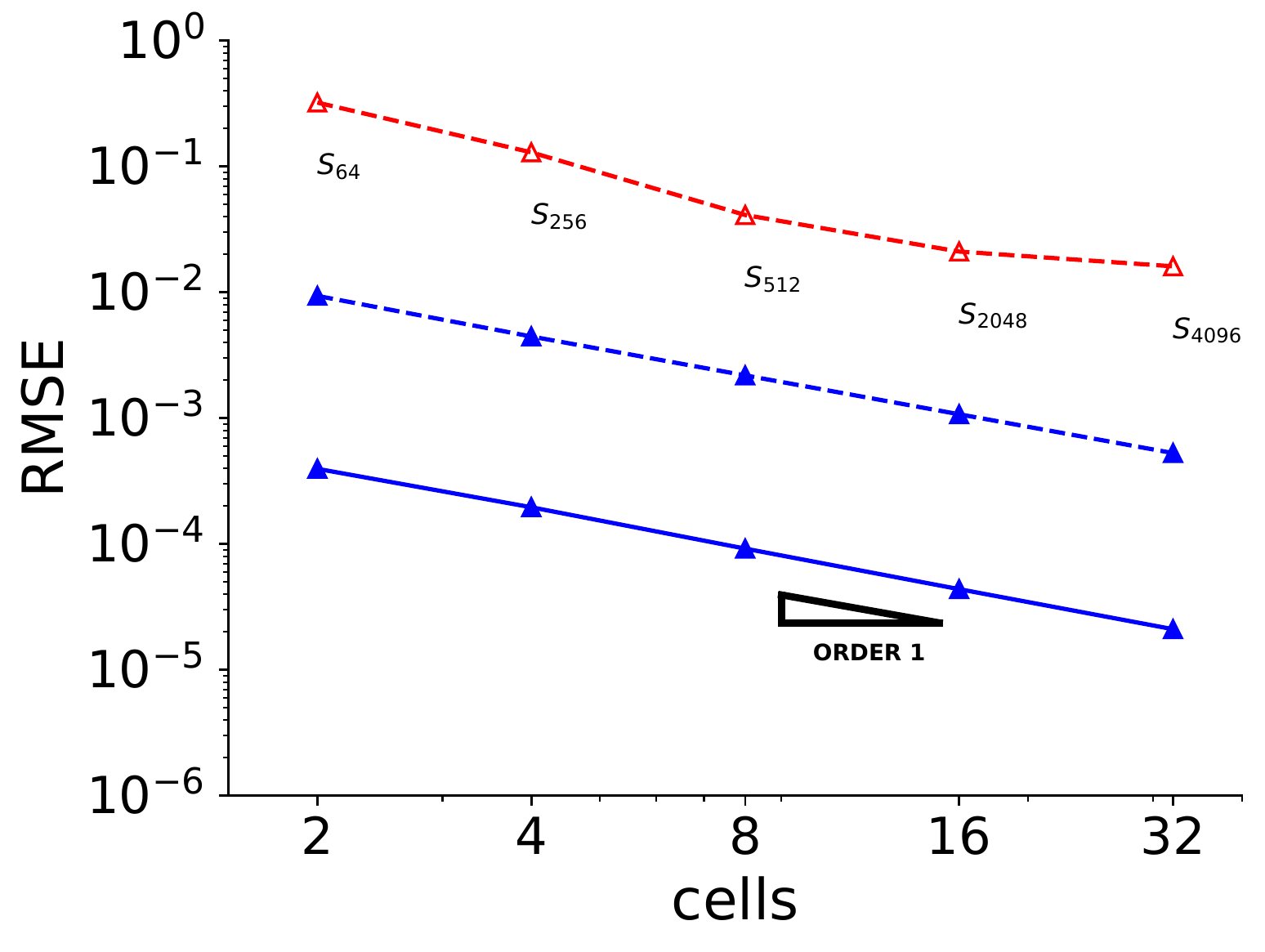}
    \caption{$M=4$}
    \end{subfigure}
    \centering
    \begin{subfigure}[b]{0.48\textwidth}
        \includegraphics[width=\textwidth]{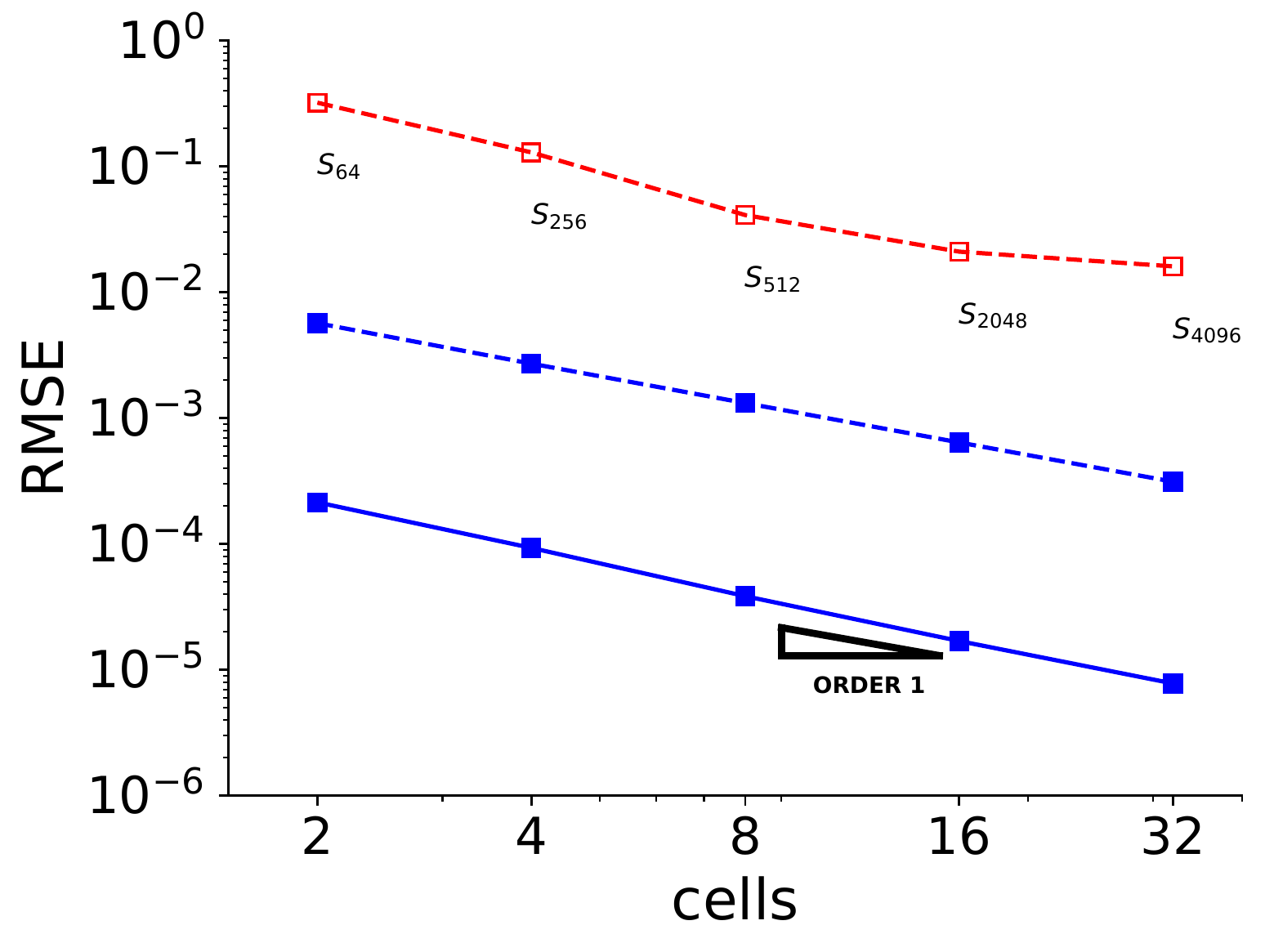}
        \caption{$M=6$}
    \end{subfigure}
    \caption{Plane pulse convergence results on a logarithmic scale with $c=1$ at $t=1$. Blue lines indicate the uncollided solution is used, red that no uncollided source is used.  Dashed lines are for a static mesh and solid lines are for the moving mesh.}
    \label{fig:plane_IC_rms}
\end{figure}

Though the highly nonsmooth nature of the solution to the plane pulse at early time restricts all methods in Figure \ref{fig:plane_IC_rms} to first order convergence, there is a substantial difference in the accuracy of each method. Table \ref{table:C_plane_IC} shows that in the $M=6$ case, there is an almost $900$ times reduction in the intercept of the uncollided source, moving mesh case compared to the standard DG implementation. These results show that using an uncollided source is effective in reducing the error, in this case $36$ times, but using the uncollided source and the moving mesh is a significantly more effective method.

\begin{figure}

    \includegraphics[width=\textwidth]{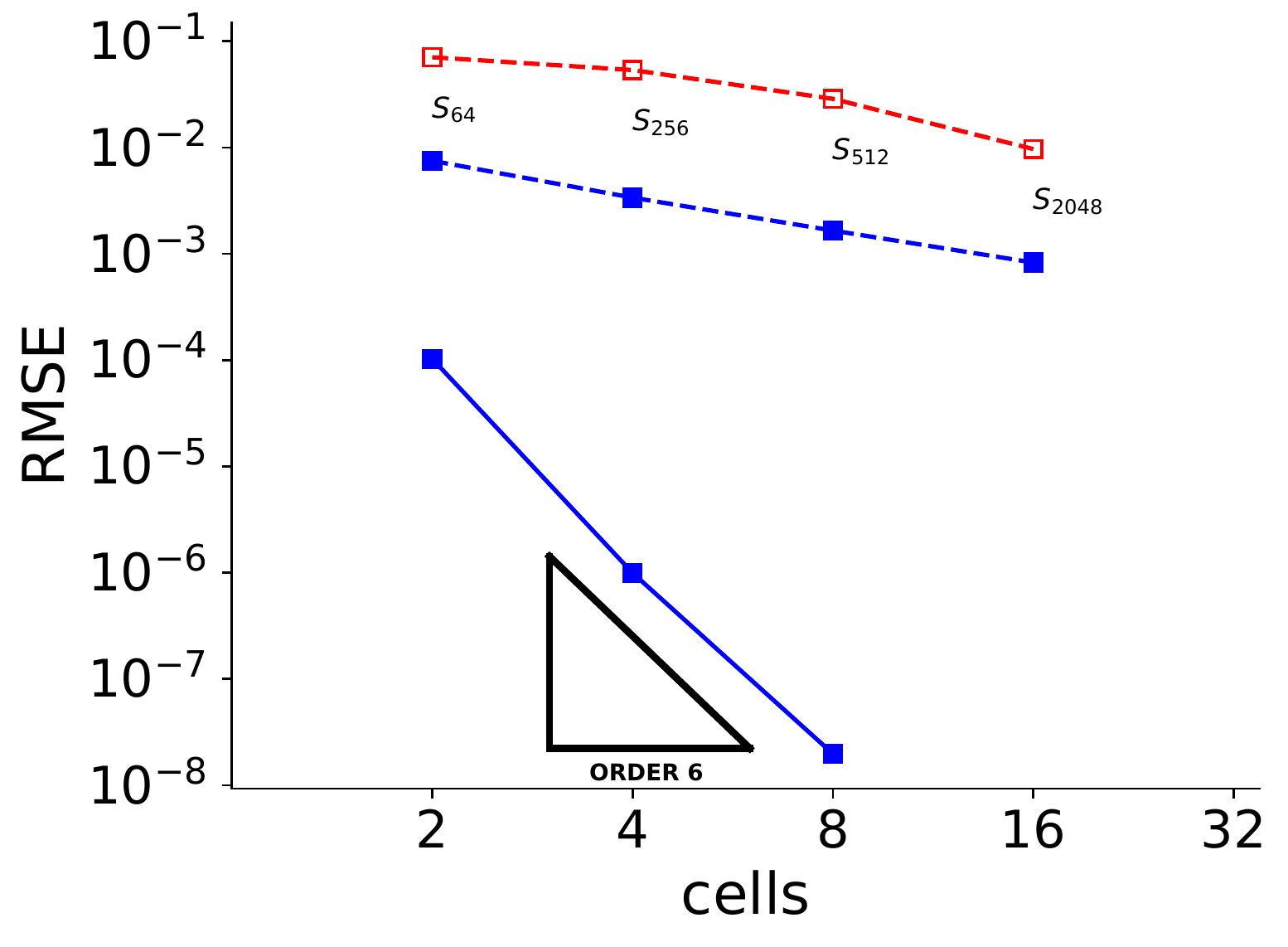}

    \caption{Plane pulse convergence results on a logarithmic scale with $c=1$ at $t=10$ with $M=6$ . Blue lines indicate the uncollided solution is used, dashed that the mesh is static, and solid that the mesh is moving.}
    \label{fig:plane_IC_RMSE_10}
\end{figure}

The Gaussian nature of the plane pulse solution at later times allows for higher order convergence, as shown in Figure \ref{fig:plane_IC_RMSE_10}. The no-uncollided solution method in this case is restricted by the inherent error in approximating a delta function and the uncollided source, static mesh method is restricted by the necessity to resolve the discontinuous wavefront at early times with a polynomial. The uncollided, moving mesh method shows nearly optimal, sixth-order convergence with a significantly lower error level. 

The plane pulse solution requires more angular resolution than all of the other source configurations tested, with $S_{4096}$ required for errors less than $10^{-4}$. This angular error is reduced as the solution smooths at later times.

\begin{table}[]
\caption{Plane pulse intercepts from Figure \eqref{fig:plane_IC_rms} where the intercept, is found from the curve fit, $\mathrm{RMSE} = C \,K^{-A}$, where $K$ is the number of mesh subdivisions and $A$ is the order of convergence. The improvement from the baseline is found by dividing the intercept from the no uncollided solution, static mesh case by the intercept from the given case.}

\begin{tabular}{|l|ll|ll|}
\hline
                              & \multicolumn{2}{c|}{M = 4}                                 & \multicolumn{2}{c|}{M = 6}                                 \\ \cline{2-5} 
                              & \multicolumn{1}{l|}{intercept} & improvement &
                              \multicolumn{1}{l|}{intercept} & improvement \\
                               \hline
no uncollided $+$ static mesh & \multicolumn{1}{l|}{0.41393}   & --                         & \multicolumn{1}{l|}{0.4140}    & --                         \\
uncollided $+$ static mesh    & \multicolumn{1}{l|}{0.01848}   & 22                        & \multicolumn{1}{l|}{0.01145}   & 36                        \\
uncollided $+$ moving mesh    & \multicolumn{1}{l|}{0.0008613} & 481                       & \multicolumn{1}{l|}{0.0004731} & 875                       \\ \hline
\end{tabular}
\label{table:C_plane_IC}
\end{table}

\afterpage{\clearpage}
\subsection{Square pulse}\label{sec:sq_ic}
\begin{figure}
     \centering
     \begin{subfigure}[b]{0.3\textwidth}
         \centering
         \includegraphics[width=\textwidth]{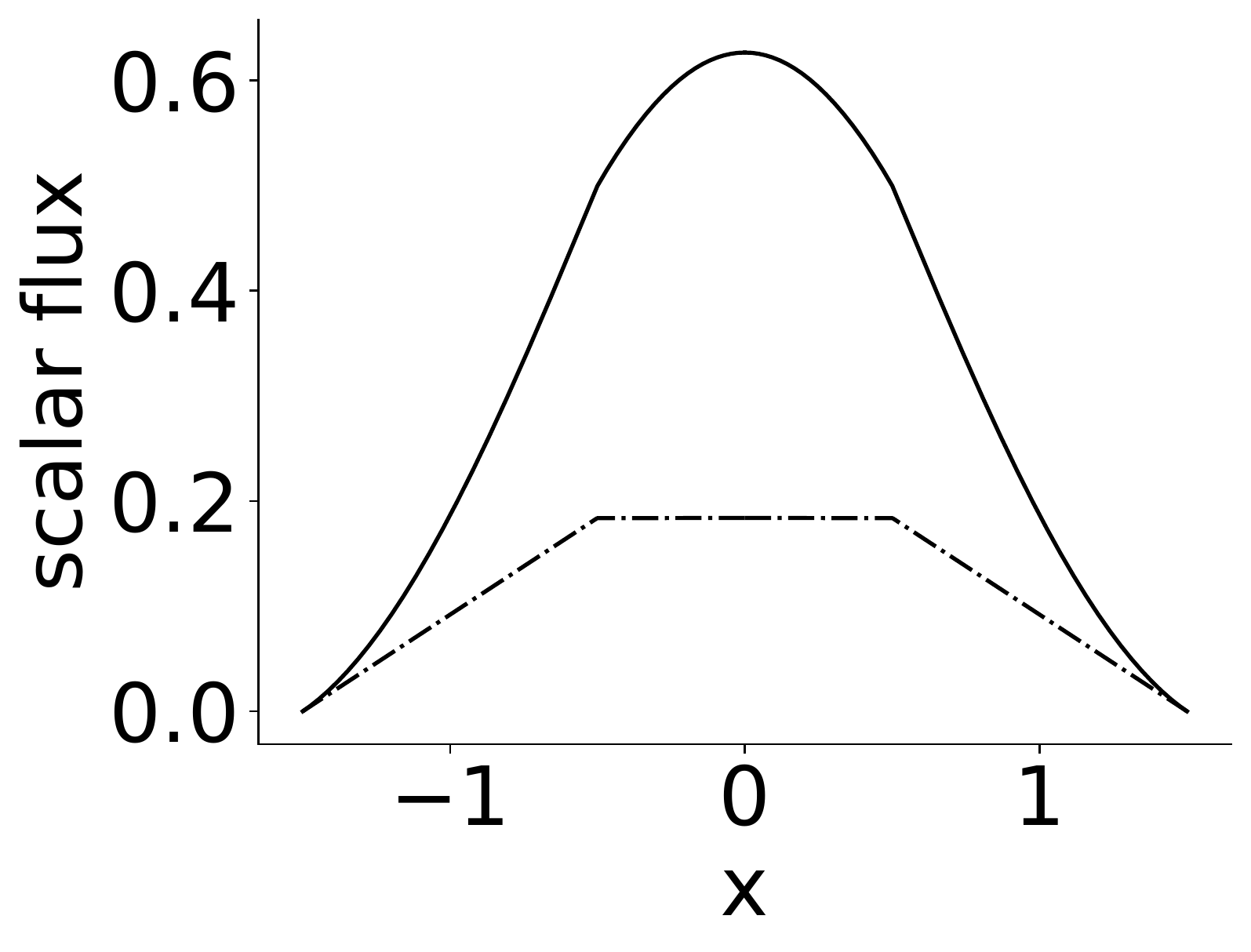}
         \caption{$t=1$}
         \label{fig:sq_IC_1}
     \end{subfigure}
     \hfill
     \begin{subfigure}[b]{0.3\textwidth}
         \centering
         \includegraphics[width=\textwidth]{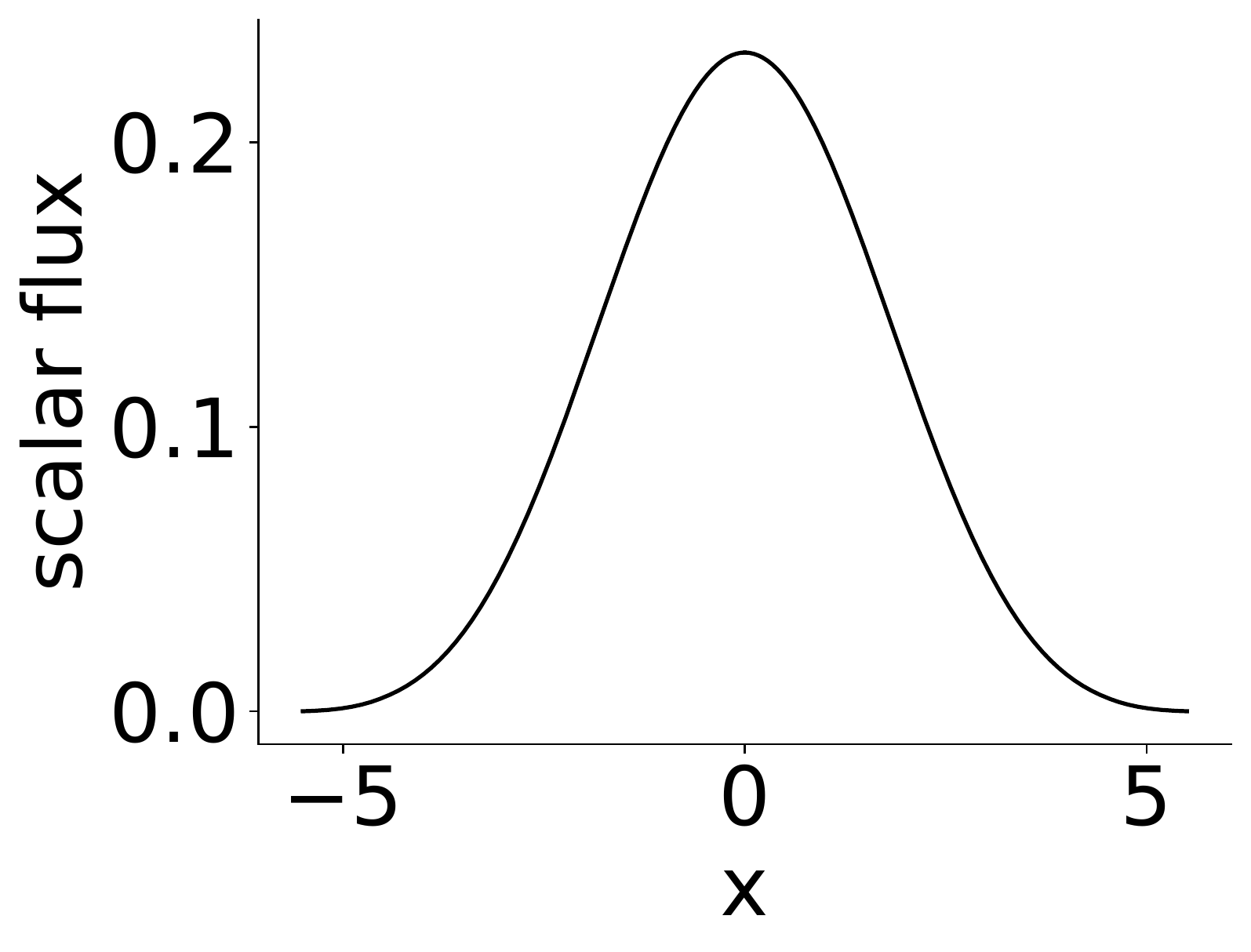}
         \caption{$t=5$}
         \label{fig:sq_IC_5}
     \end{subfigure}
     \hfill
     \begin{subfigure}[b]{0.3\textwidth}
         \centering
         \includegraphics[width=\textwidth]{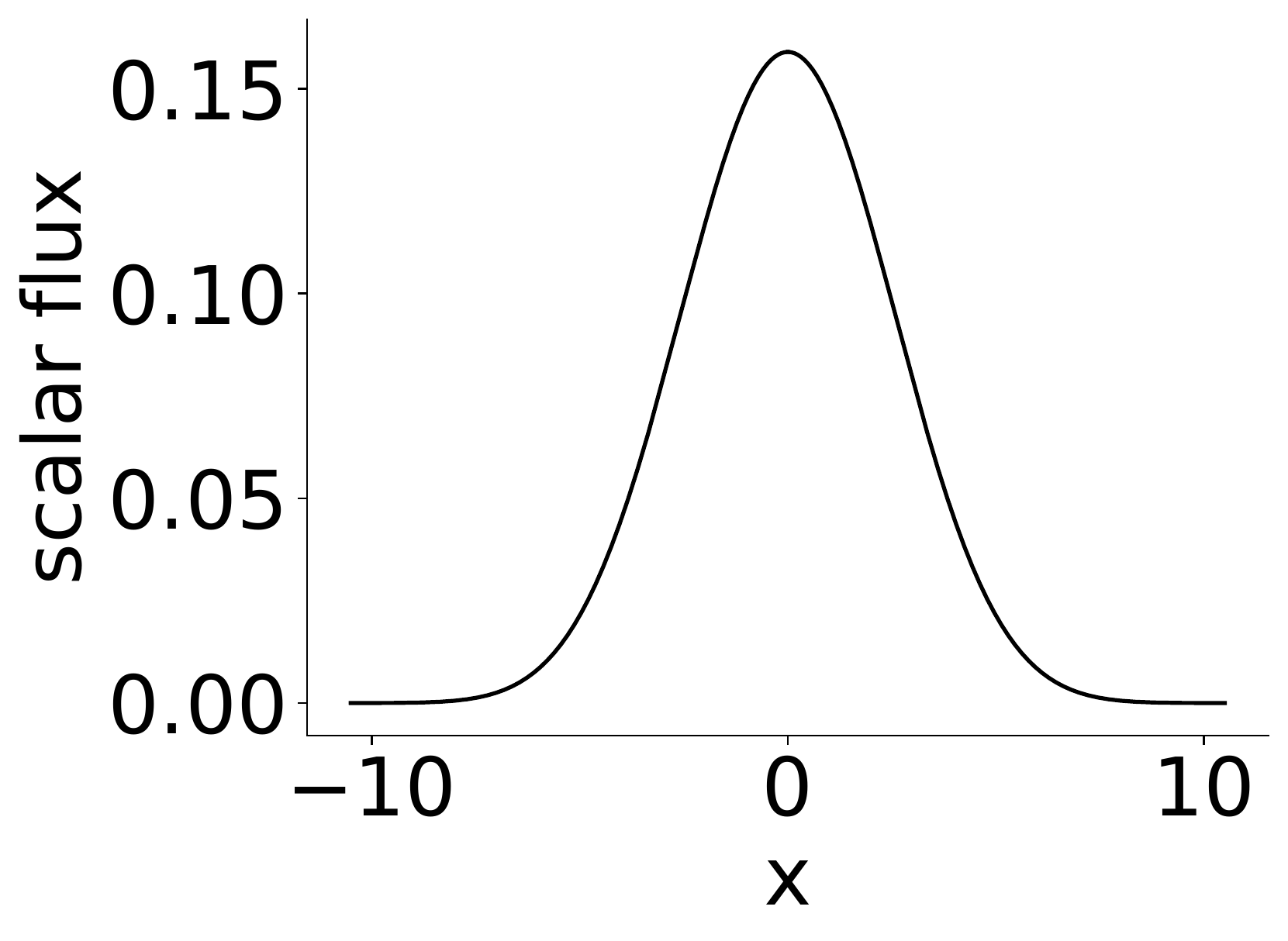}
         \caption{$t=10$}
         \label{fig:sq_IC_10}
     \end{subfigure}
        \caption{Square pulse semi-analytic solution, $\phi$ (solid) and $\phiu$ (dashed) with $x_0= 0.5$ and $c=1$. The uncollided flux is not shown for times where it is negligible. }
        \label{fig:sq_IC}
\end{figure}

We next consider a finite width square pulse source of the form,
\begin{equation}
    S_{sp} = \Theta(x_0-|x|)\delta(t).
\end{equation}
\cite{bennett2022benchmarks} gives the piecewise linear uncollided solution for this problem,
\begin{equation}\label{eq:sq_ic_cases}
    \phi^\mathrm{sp}_u(x,t) = 
    \begin{cases}
     0 & |x| - t >   x_0\\
    x_0 \frac{e^{-t}}{t} &   t > x_0 \: \& \:   x_0-t \leq x \leq t - x_0  \\
    e^{-t} & t \leq x_0 \: \& \:   t-x_0 \leq x \leq x_0-t\\
    \frac{e^{-t} (t+x+x_0)}{2 t} & -t-x_0 < x < t + x_0 \:\&\: x_0+x \leq t \leq x_0 -x \\
    \frac{e^{-t} (t-x+x_0)}{2 t} &  -t-x_0 < x < t + x_0 \: \& \: x_0-x \leq t \leq x_0 + x.
    \end{cases}
\end{equation}
The full solution and the uncollided piecewise solution for an initial width $x_0=0.5$ at $t=1$ is plotted in Figure \ref{fig:sq_IC}. The uncollided flux is discontinuous in the first derivative at $x=\pm x_0$. These discontinuities travel towards the origin at the wavespeed, meet at $x = 0$, then travel outwards again. Tracking this discontinuity involves moving edges towards the origin, which preformed poorly in test implementation, so a hybrid static-moving mesh was adopted for the moving mesh implementation. This method requires the number of cells to multiples of two greater than or equal to four. Half of the zones span the source and never move, and half begin at the edges with width zero and move outwards. Therefore, edges on the left edges of the source obey
\begin{equation}\label{eq:mesh_edges_square1}
    x_k(t) = x_{ko}  -\frac{x_k}{x_{0}} \times v t \qquad  \mathrm{for} \: k < \frac{K}{4}.
\end{equation}
Mesh edges from $k=\frac{K}{4}$ to  $k=\frac{3K}{4}$ stay at their initial values. For the rest,
\begin{equation}\label{eq:mesh_edges_square2}
        x_k(t) = x_{ko}  + \frac{x_k}{x_{K}} \times v t \qquad  \mathrm{for} \: k > \frac{3K}{4}.
\end{equation}
While this mesh method does not track the interior discontinuity, it ensures better resolution of the solution inside the source where the uncollided solution presents difficulty by allocating two-thirds of the cells to the source region. The static mesh in this case spans $[-t_{\mathrm{final}}-x_0, t_{\mathrm{final}}+x_0]$ with an edge at the initial source width. 

For our uncollided source treatment, Eq.~\eqref{eq:sq_ic_cases} is substituted into Eq.~\eqref{eq:S} and the initial condition is set to zero. For the standard source treatment, the initial condition is found by substituting $\psi(x,t=0) = \frac{1}{2}\Theta\left(x_0-|x|\right)$ into Eq.~\eqref{eq:icbcmms} and setting $S$ in Eq.~\eqref{eq:S} to zero.

\begin{figure}
     \hfill
     \begin{subfigure}[b]{0.48\textwidth}
         \centering
         \includegraphics[width=\textwidth]{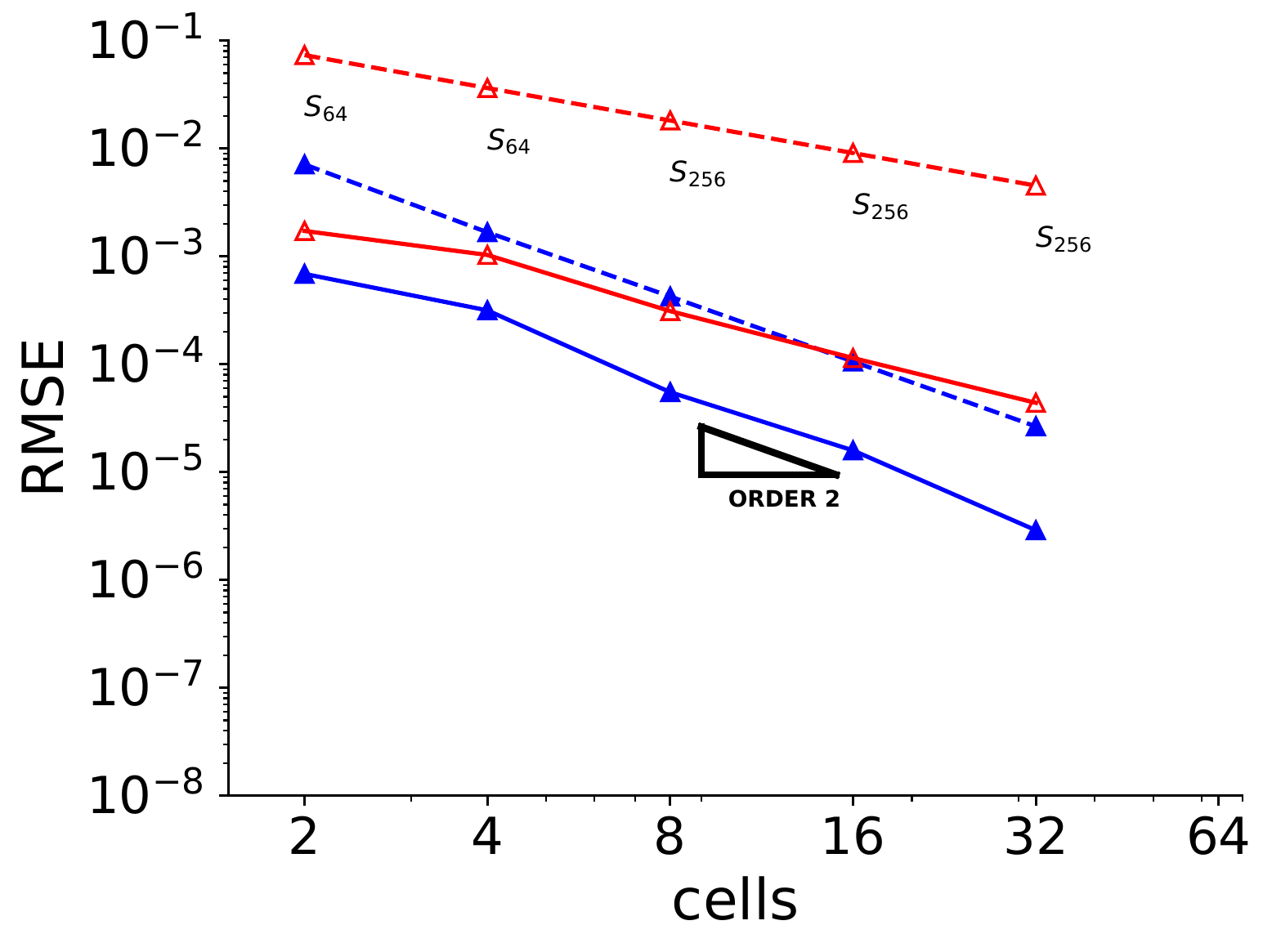}
         \caption{$M=4$}
         \label{fig:sq_IC_rms_4}
     \end{subfigure}
     \centering
     \begin{subfigure}[b]{0.48\textwidth}
         \centering
         \includegraphics[width=\textwidth]{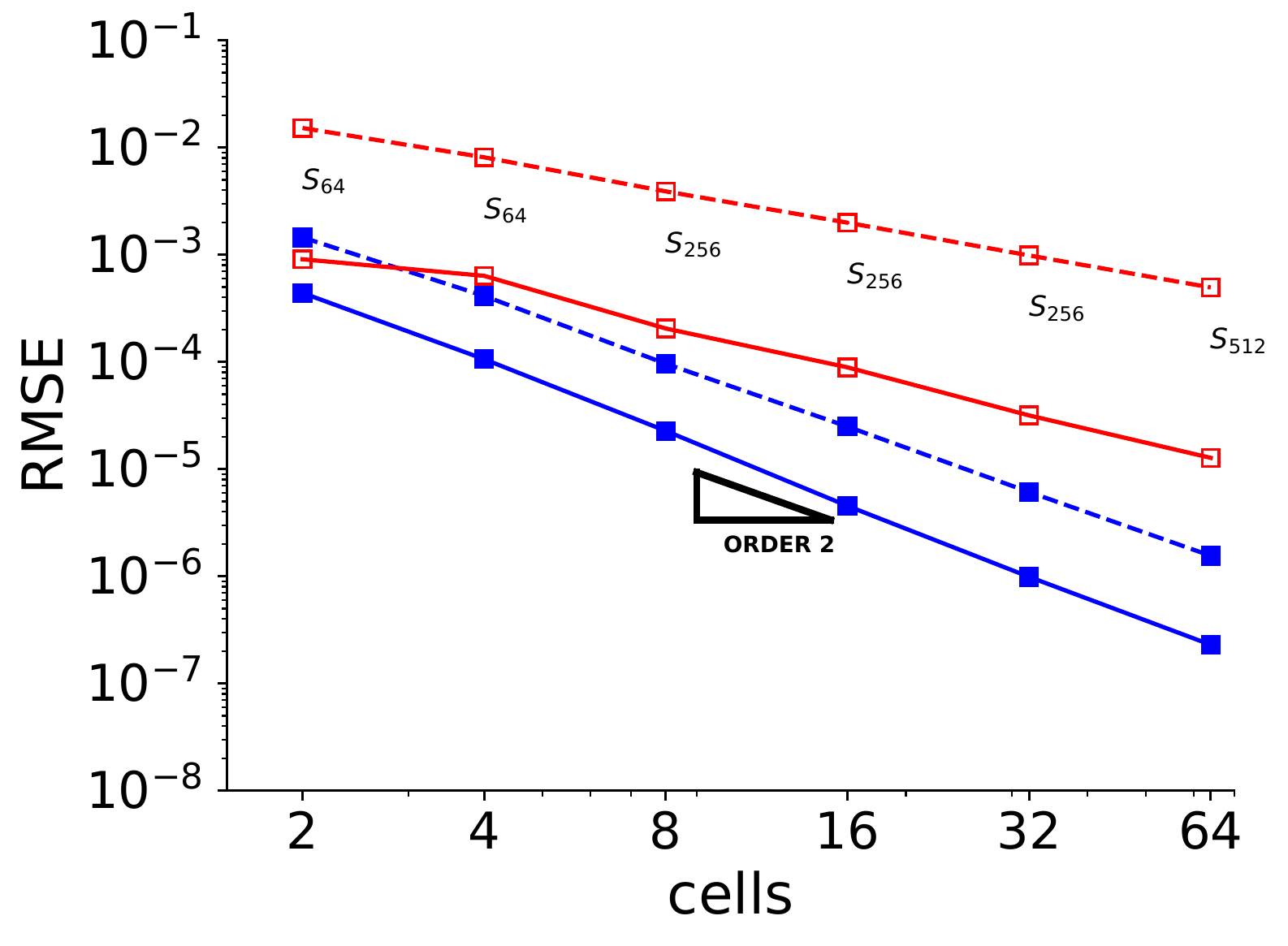}
         \caption{$M =6$}
         \label{fig:sq_IC_rms_6}
     \end{subfigure}
        \caption{Square pulse convergence results on a logarithmic scale  with $c=1$ at $t=1$. Blue lines indicate the uncollided solution is used, red that no uncollided source is used.  Dashed lines are for a static mesh and solid lines are for the moving mesh. }
        \label{fig:sq_IC_RMSE}
\end{figure}

The square pulse source is significantly smoother than the plane pulse. The former has an uncollided angular flux that is piecewise smooth, where the latter has an uncollided angular flux made of of travelling delta functions. Figure \ref{fig:sq_IC_RMSE} shows that the uncollided methods are able to achieve second order convergence at early time, with the standard source treatments doing slightly worse. The square pulse requires similar angular error resolution as do the smooth Gaussian problems, with $S_{512}$ required for RMSE less than $10^{-6}$.

As with the plane pulse, the intercept value is important. For six basis functions, the uncollided, moving mesh and the uncollided static mesh have the same order of convergence, but the former has a 10 times intercept reduction from the baseline and the latter is reduced 5 times (Table \ref{tab:sq_IC}). Table \ref{tab:sq_IC} also shows a feature that is less discernible in Figure \ref{fig:sq_IC_RMSE}, the intercept for $M=6$ is approximately 3 times smaller than for $M=4$.

\begin{table}
\caption{Square pulse intercepts from Figure \ref{fig:plane_IC_rms} where the intercept, $C$ is found from the curve fit, $\mathrm{RMSE} = C \,\Delta x^{-A},$ where $\Delta x$ is the number of mesh subdivisions and $A$ is the order of convergence. The improvement from the baseline is found by dividing the intercept from the no uncollided solutions, static mesh case by the intercept from the given case}
\begin{tabular}{|l|ll|ll|}
\hline
                              & \multicolumn{2}{c|}{M = 4}                                 & \multicolumn{2}{c|}{M = 6}                                 \\ \cline{2-5} 
                              & \multicolumn{1}{l|}{intercept} & improvement &
                              \multicolumn{1}{l|}{intercept} & improvement \\
                              \hline
no uncollided $+$ static mesh & \multicolumn{1}{l|}{0.14594}   & -                         & \multicolumn{1}{l|}{0.03220}   & -                         \\
uncollided $+$ static mesh    & \multicolumn{1}{l|}{0.02664}   & 6                         & \multicolumn{1}{l|}{0.006501}  & 5                         \\
no ucollided $+$ moving mesh  & \multicolumn{1}{l|}{0.0078136} & 19                        & \multicolumn{1}{l|}{0.00414}   & 8                         \\
uncollided $+$ moving mesh    & \multicolumn{1}{l|}{0.006379}  & 23                        & \multicolumn{1}{l|}{0.002266}  & 10                        \\ \hline
\end{tabular}
\label{tab:sq_IC}
\end{table}

\subsection{Square source}\label{sec:square_s}
\begin{figure}
     \centering
     \begin{subfigure}[b]{0.3\textwidth}
         \centering
         \includegraphics[width=\textwidth]{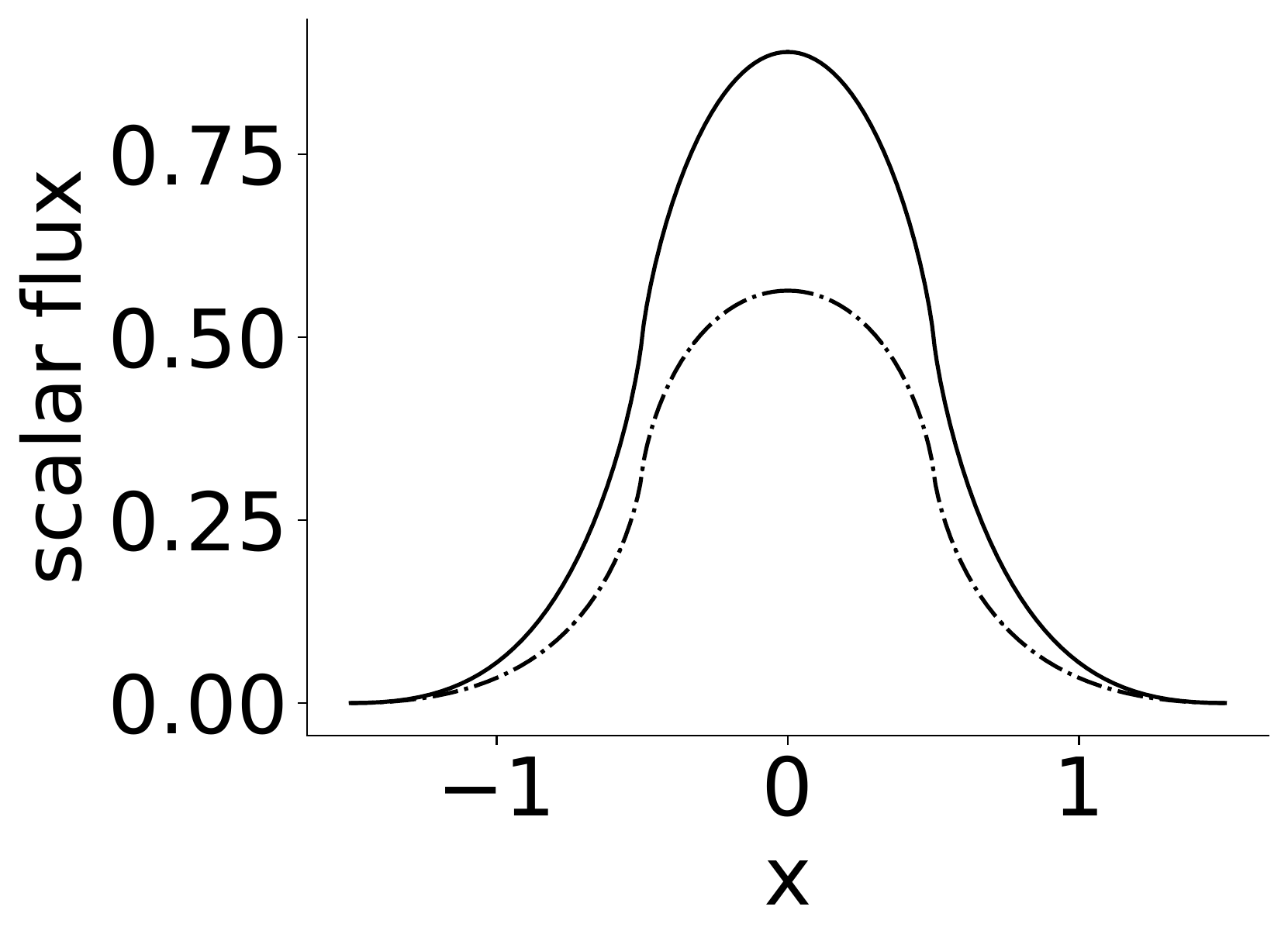}
         \caption{$t=1$}
         \label{fig:sq_s_1}
     \end{subfigure}
     \hfill
     \begin{subfigure}[b]{0.3\textwidth}
         \centering
         \includegraphics[width=\textwidth]{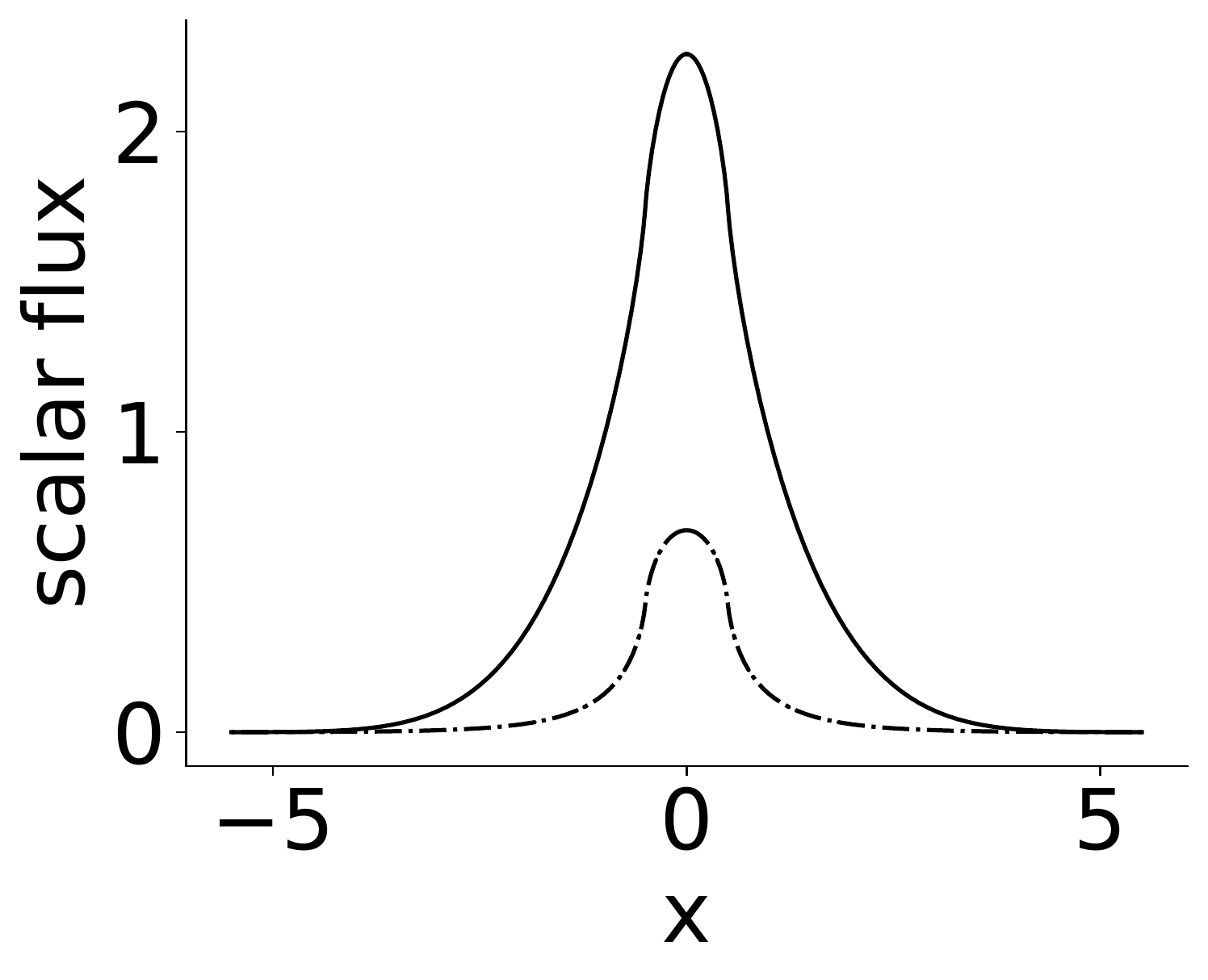}
         \caption{$t=5$}
         \label{fig:sq_s_5}
     \end{subfigure}
     \hfill
     \begin{subfigure}[b]{0.3\textwidth}
         \centering
         \includegraphics[width=\textwidth]{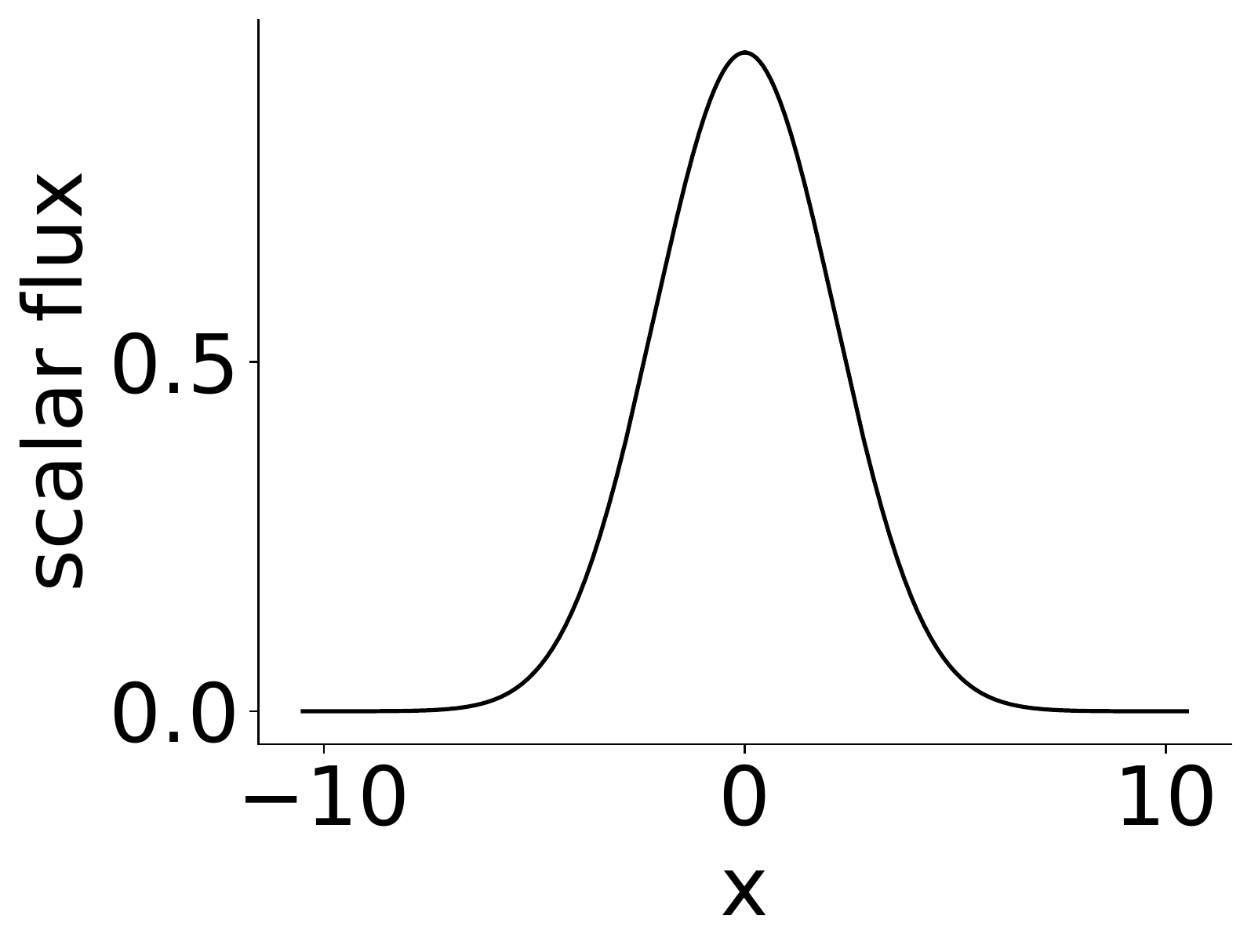}
         \caption{$t=10$}
         \label{fig:sq_s_10}
     \end{subfigure}
        \caption{Square source semi-analytic solution, $\phi$ (solid) and $\phiu$ (dashed) with $x_0 = 0.5$, $t_0= 5$, and $c=1$. The uncollided flux is not shown for times where it is negligible.}
        \label{fig:sq_s}
\end{figure}

We  define a square source which is a superposition of square pulses while $t<t_0$,
\begin{equation}\label{eq:sq_s}
    S_{ss} = \Theta(x_0-|x|)\Theta(t_0-t).
\end{equation}
The solution for this configuration is shown in Figure \eqref{fig:sq_s} where the uncollided solution is continuous, but not everywhere smooth and is still significant by the time the source is turned off, $t=5$. The uncollided solution for this source, given by \cite{bennett2022benchmarks}, is 
\begin{equation}\label{eq:sq_s_uncollided}
    \phi_u^\mathrm{sp}(x,t) = \left[-x_0\mathrm{E_i}(\tau-t)\right]\bigg{|}_0^b + \frac{1}{2} \left[(| x| -x_0) \text{Ei}(\tau -t)+e^{\tau -t}\right]\bigg{|}_b^c + \left[e^{-(t-\tau)}\right]\bigg{|}_c^d.
\end{equation}
Where the intervals are defined by
\begin{equation}
    b = \left[\mathrm{min}\left(d, t - |x| - x_0\right)\right]_+,
\end{equation}
\begin{equation}
    c = \left[\mathrm{min}\left(d, t + |x| - x_0\right)\right]_+,
\end{equation}
\begin{equation}
    d = \left[\mathrm{min}\left(t_0, t, t - |x| + x_0\right)\right]_+.
\end{equation}
$[\cdot]_+$ returns the positive part of its argument and
$\mathrm{E_i}$ is the exponential integral.

To find the source term in the weak formulation, Eq.~\eqref{eq:sq_s_uncollided} is inserted into Eq.~\eqref{eq:S}. For the standard source evaluation, Eq.~\eqref{eq:sq_s} is used as the source in Eq.~\eqref{eq:S}.

The square source uncollided flux creates a discontinuity in the first derivative at $x=\pm x_0$ while the source is on. The moving mesh defined in Section \ref{sec:sq_ic} is ideal for handling this discontinuity since a mesh edge is always at $\pm x_0$. Also, a two-thirds of the cells available cells are inside of the source where the solution is more difficult to resolve. The static mesh in this case spans $[-t_{\mathrm{final}}-x_0, t_{\mathrm{final}}+x_0]$ with evenly spaced cells. An edge is always located at the source edge. 

\begin{figure}
     \hfill
     \begin{subfigure}[b]{0.48\textwidth}
         \centering
         \includegraphics[width=\textwidth]{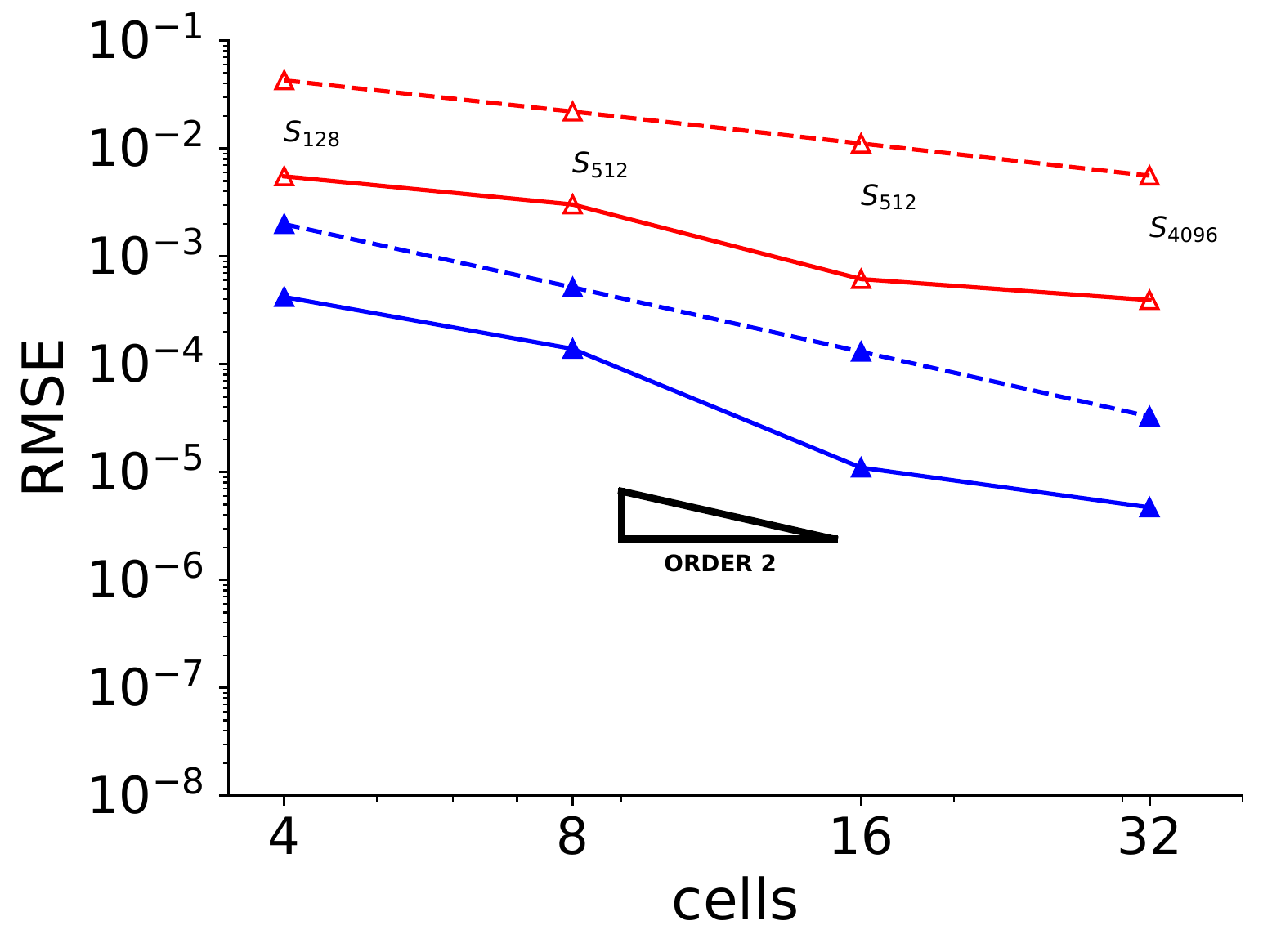}
         \caption{$M=4$}
         \label{fig:sq_s_rms_4}
     \end{subfigure}
     \centering
     \begin{subfigure}[b]{0.48\textwidth}
         \centering
         \includegraphics[width=\textwidth]{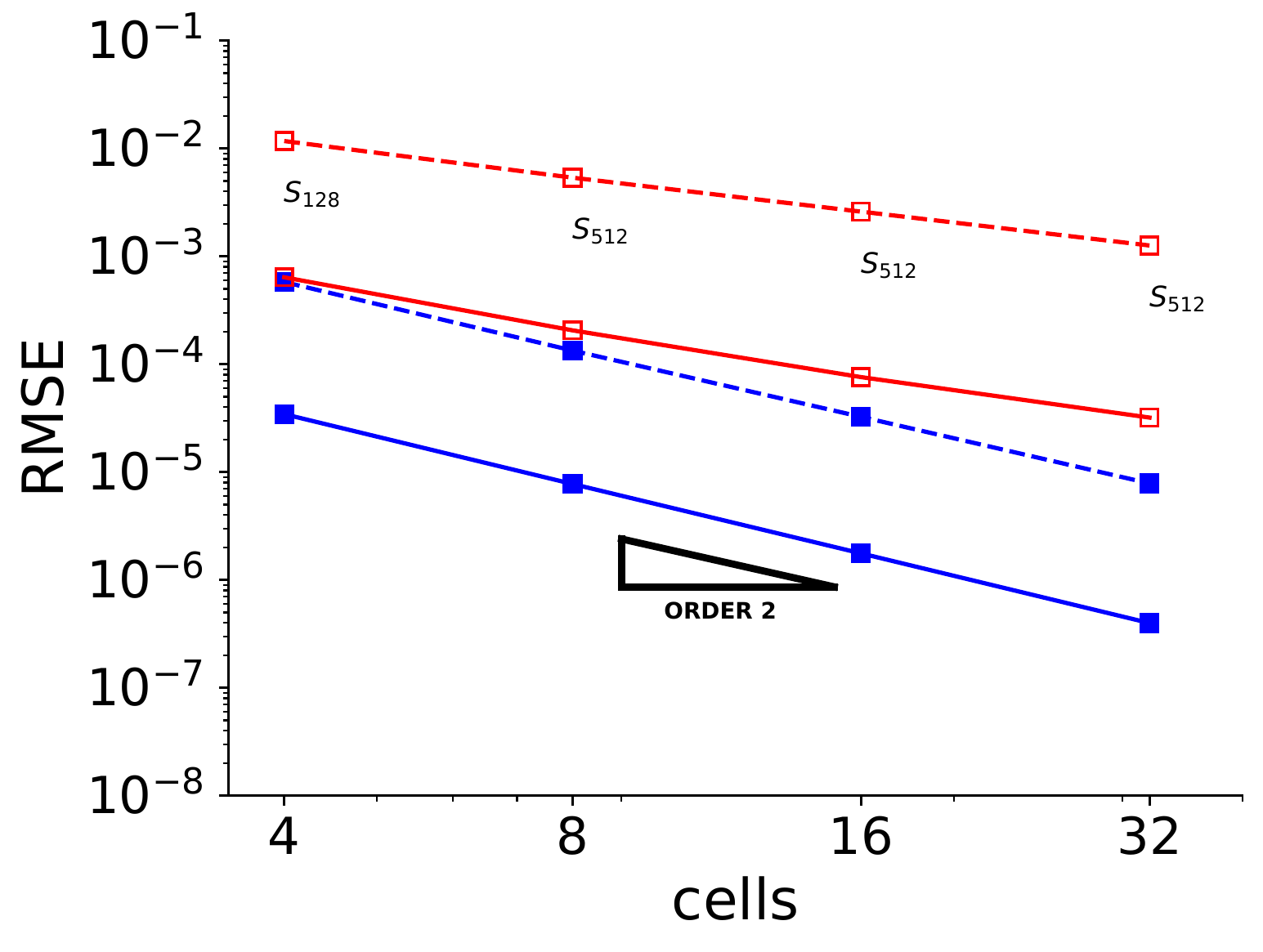}
         \caption{$M =6$}
         \label{fig:sq_s_rms_6}
     \end{subfigure}
        \caption{Square source convergence results at $t=1$. Blue lines indicate the uncollided solution is used, red that no uncollided source is used.  Dashed lines are for a static mesh and solid lines are for the moving mesh. }
        \label{fig:sq_s_RMSE}
\end{figure}

The uncollided source methods for the square source achieves second order convergence at early times and the standard source treatments do slightly worse, shown in Figure \ref{fig:sq_s_RMSE}. Like the square pulse, discontinuities in the uncollided angular flux restrict the order of convergence of the uncollided source methods. The angular error dependence in this case is similar to the square pulse, with $S_{512}$ required for errors less than $10^{-6}$. 

Since the uncollided solution in this case lasts longer than the square pulse case before decaying to zero, using the uncollided source, moving mesh treatment creates a more significant intercept reduction from the standard method. Table \ref{tab:sq_s} shows that for the $M=6$ case, the intercept is reduced 70 times from the standard method. It is interesting to note that, according to Eq.~\eqref{eq:algebraic}, for $M=6$, the standard DG method would require 1300 hundred spatial subdivisions to achieve an error that the uncollided, moving mesh method returns with 32 cells. 

\begin{figure}
    \centering
    \begin{subfigure}[b]{0.48\textwidth}
    \includegraphics[width=\textwidth]{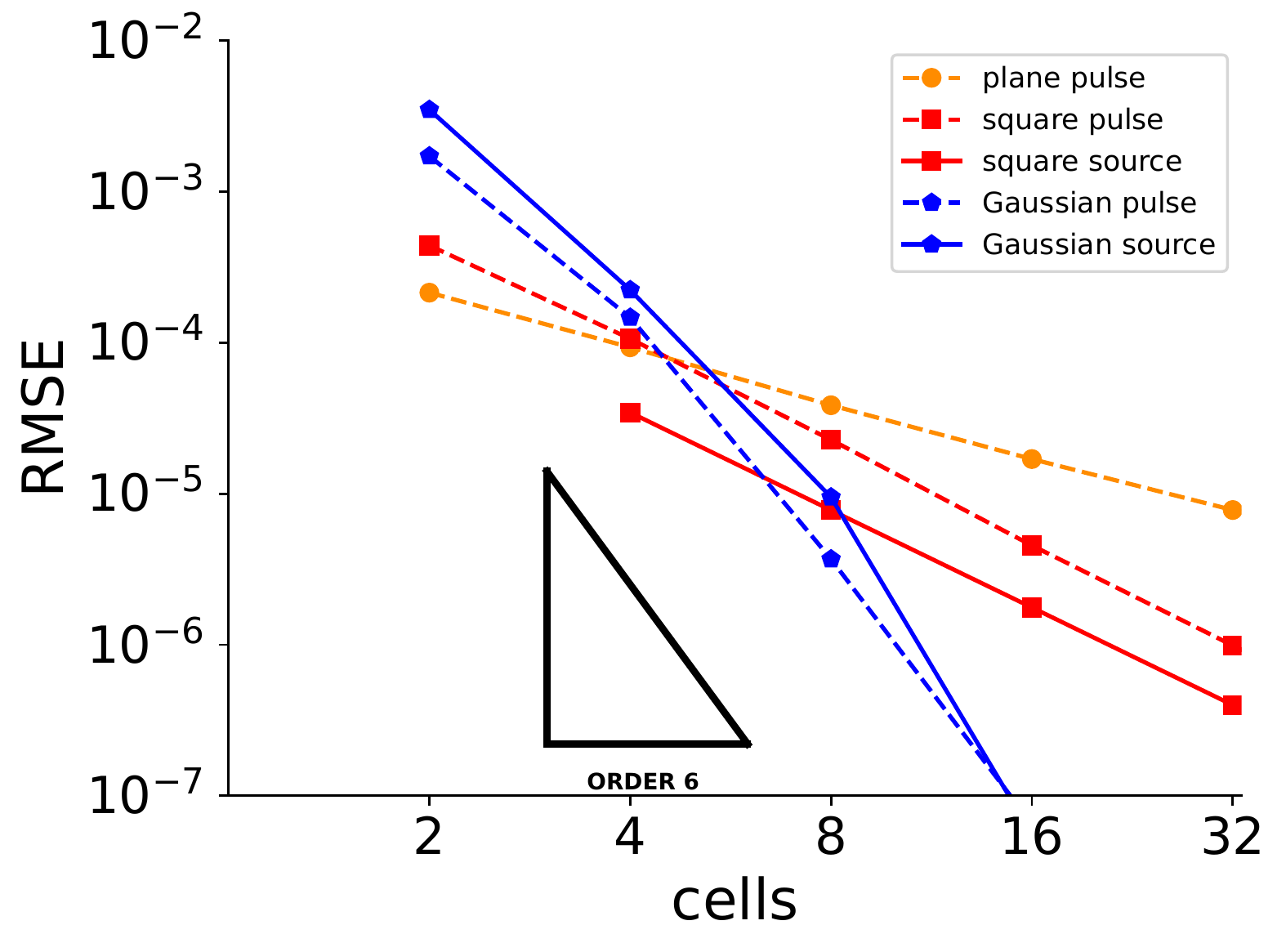}
    \caption{Moving mesh}
    \label{subfig:all}
    \end{subfigure}
    \centering
    \begin{subfigure}[b]{0.48\textwidth}
        \includegraphics[width=\textwidth]{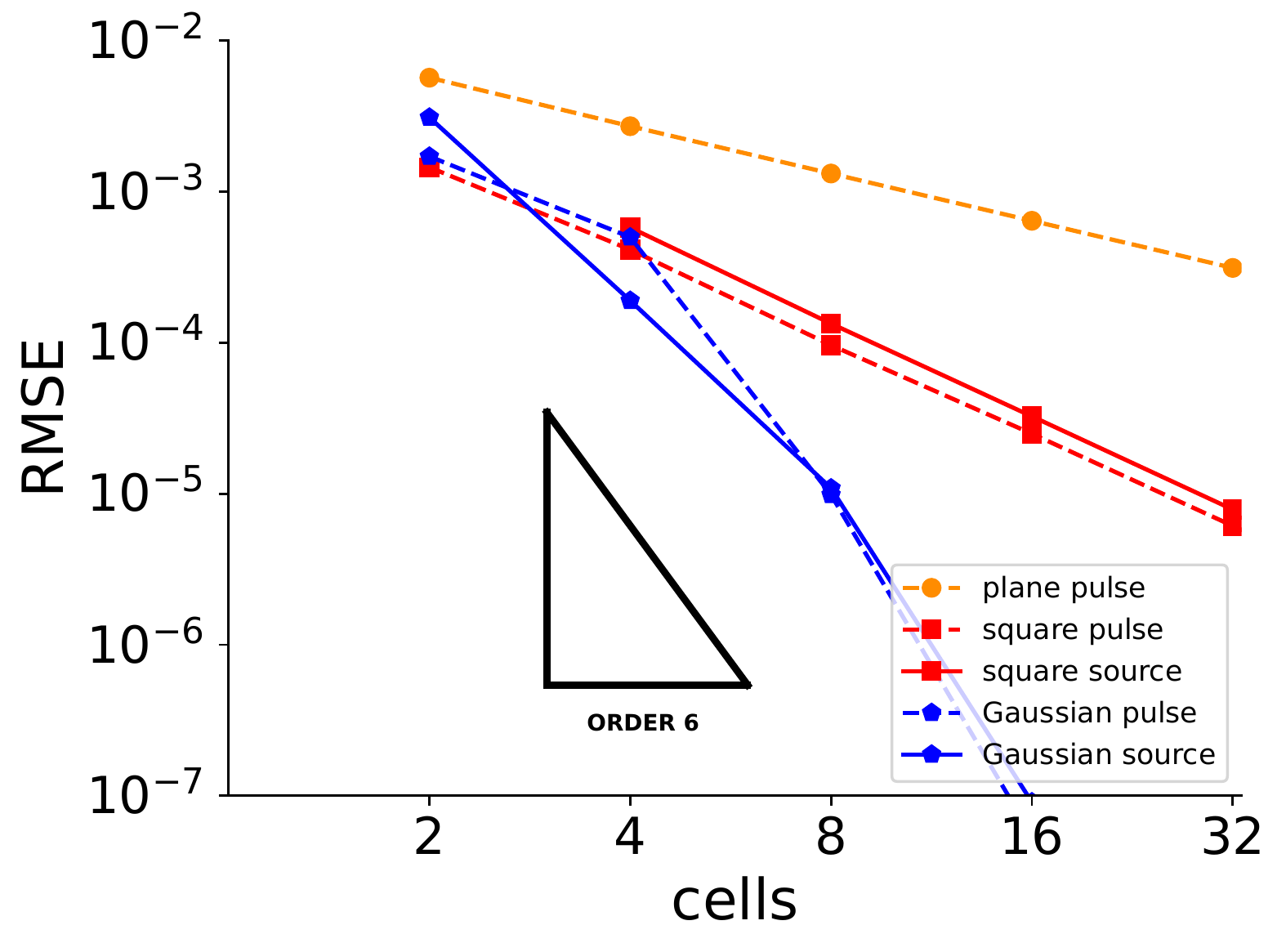}
        \caption{Static mesh}
    \label{subfig:all_static}
    \end{subfigure}
    \caption{Convergence results for an increasing number of cell divisions, $K$, for $t=1$ for $M=6$ for the uncollided source moving mesh (\ref{subfig:all}) and static mesh (\ref{subfig:all_static}) cases for every test problem except the MMS problem.}
    \label{fig:ALL_RMSE}
\end{figure}

The $t=1$ convergence results for every case except the MMS problem for the uncollided source, moving mesh and the standard source, moving mesh are shown side by side in Figure~\ref{fig:ALL_RMSE}. This figure is a good illustration of the difference in convergence between sources within the same method and the difference in accuracy between the uncollided, moving mesh and the standard moving mesh methods.

\begin{table}
\caption{Square source intercepts from Figure \eqref{fig:sq_s_RMSE} where the intercept, $C$ is found from the curve fit $\mathrm{RMSE} = C \,\Delta x^{-A}$, where $\Delta x$ is the number of mesh subdivisions and $A$ is the order of convergence. The improvement from the baseline is found by dividing the intercept from the no uncollided solutions, static mesh case by the intercept from the given case}
\begin{tabular}{|l|ll|ll|}
\hline
                              & \multicolumn{2}{c|}{M = 4}                                 & \multicolumn{2}{c|}{M = 6}                                 \\ \cline{2-5} 
                              & \multicolumn{1}{l|}{intercept} & improvement &
                              \multicolumn{1}{l|}{intercept} & improvement \\
                              \hline
no uncollided $+$ static mesh & \multicolumn{1}{l|}{0.1713}    & -                         & \multicolumn{1}{l|}{0.04705}   & -                         \\
uncollided $+$ static mesh    & \multicolumn{1}{l|}{0.03235}   & 5                         & \multicolumn{1}{l|}{0.009397}  & 5                         \\
no ucollided $+$ moving mesh  & \multicolumn{1}{l|}{0.05331}   & 3                         & \multicolumn{1}{l|}{0.003389}  & 14                        \\
uncollided $+$ moving mesh    & \multicolumn{1}{l|}{0.01674}   & 10                        & \multicolumn{1}{l|}{0.0006706} & 70                        \\ \hline
\end{tabular}
\label{tab:sq_s}
\end{table}


\subsection{Results for non-purely scattering problems}\label{sec:c_not_one}
All of the test problems considered so far have been pure scattering problems ($c=1$). In this section, we briefly show that for the Gaussian pulse and square pulse problems, our method performs well with partially absorbing and multiplying scattering ratios. For these problems, we are able to avoid recalculating a benchmark by rescaling already calculated results. In an underappreciated footnote, Case and Zweifel (\cite[p.~175]{case1967linear}) presented a clever scaling to relate the solution for $c=1$ ($\psi^1$) to solutions for any other nonzero scattering ratio ($\psi^c$),
\begin{equation}\label{eq:scaling_psi}
    \psi^c(x,\mu,t) = c \exp\left(-(1-c)t\right)\psi^1(cx,\mu, ct).
\end{equation}
This scaling was derived for initial value problems with no source term, hence the choice of the pulse-type problems in this section.

It is also necessary to scale the initial condition of each source in the code implementation so that our code is running the same problem as the benchmark. This is done by scaling the parameters, $x_0$ and $\sigma$ for the square pulse and Gaussian pulse respectively, in the initial condition and the uncollided source,
\begin{equation}
    x_0' = \frac{x_0}{c} \qquad \sigma' =  \frac{\sigma}{c},
\end{equation}
where $x_0'$ and $\sigma'$ are the new parameters for our source term. Also, the scaled evaluation time becomes,
\begin{equation}
    t_\mathrm{final} = \frac{t}{c},
\end{equation}
where $t$ is the evaluation time of the benchmark problem. 

\begin{figure}
     \hfill
     \begin{subfigure}[b]{0.48\textwidth}
         \centering
         \includegraphics[width=\textwidth]{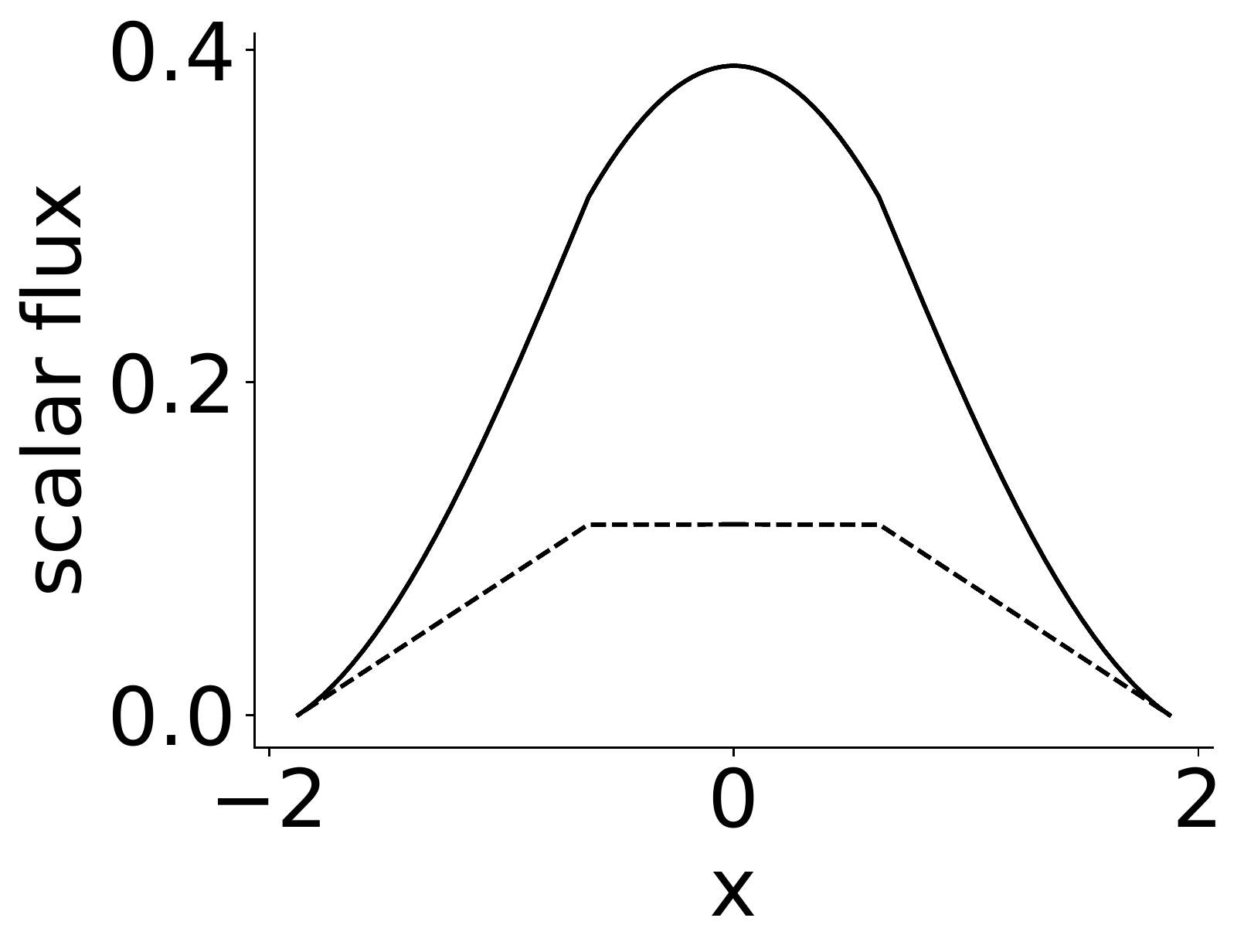}
         \caption{$c=0.8$, $t = 1.25$}
         \label{fig:sq_IC_c=0.8}
     \end{subfigure}
     \centering
     \begin{subfigure}[b]{0.48\textwidth}
         \centering
         \includegraphics[width=\textwidth]{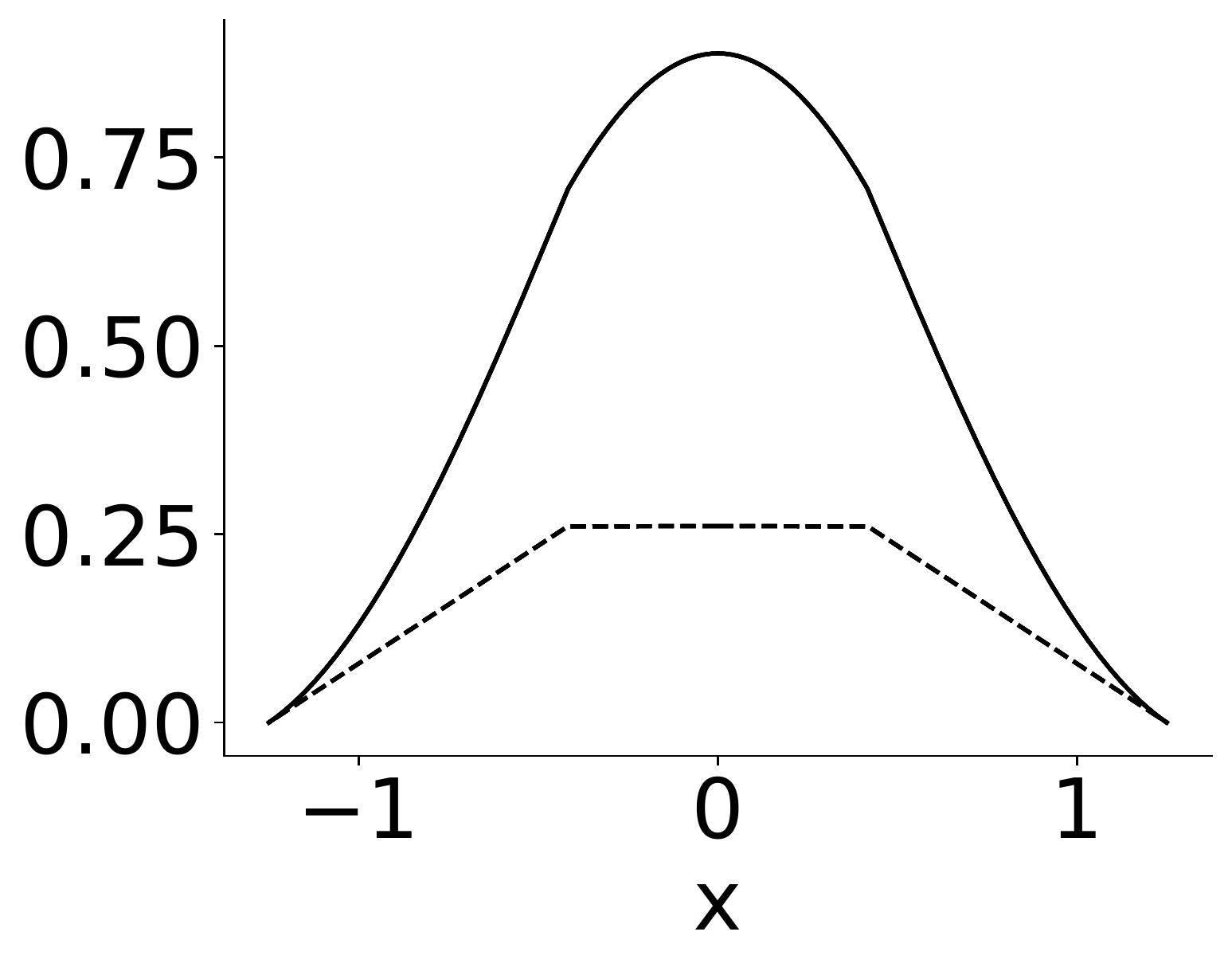}
         \caption{$c=1.2$, $t=0.83$}
         \label{fig:sq_IC_c=1.2}
     \end{subfigure}
        \caption{Square pulse semi-analytic solutions, $\phi$ (solid) and $\phiu$ (dashed), at two times for different values of $c$, scaled from $t = 1$, $c= 1$ solution (\ref{fig:sq_IC_1}). }
        \label{fig:sq_IC_not_1}
\end{figure}

\begin{figure}
     \hfill
     \begin{subfigure}[b]{0.48\textwidth}
         \centering
         \includegraphics[width=\textwidth]{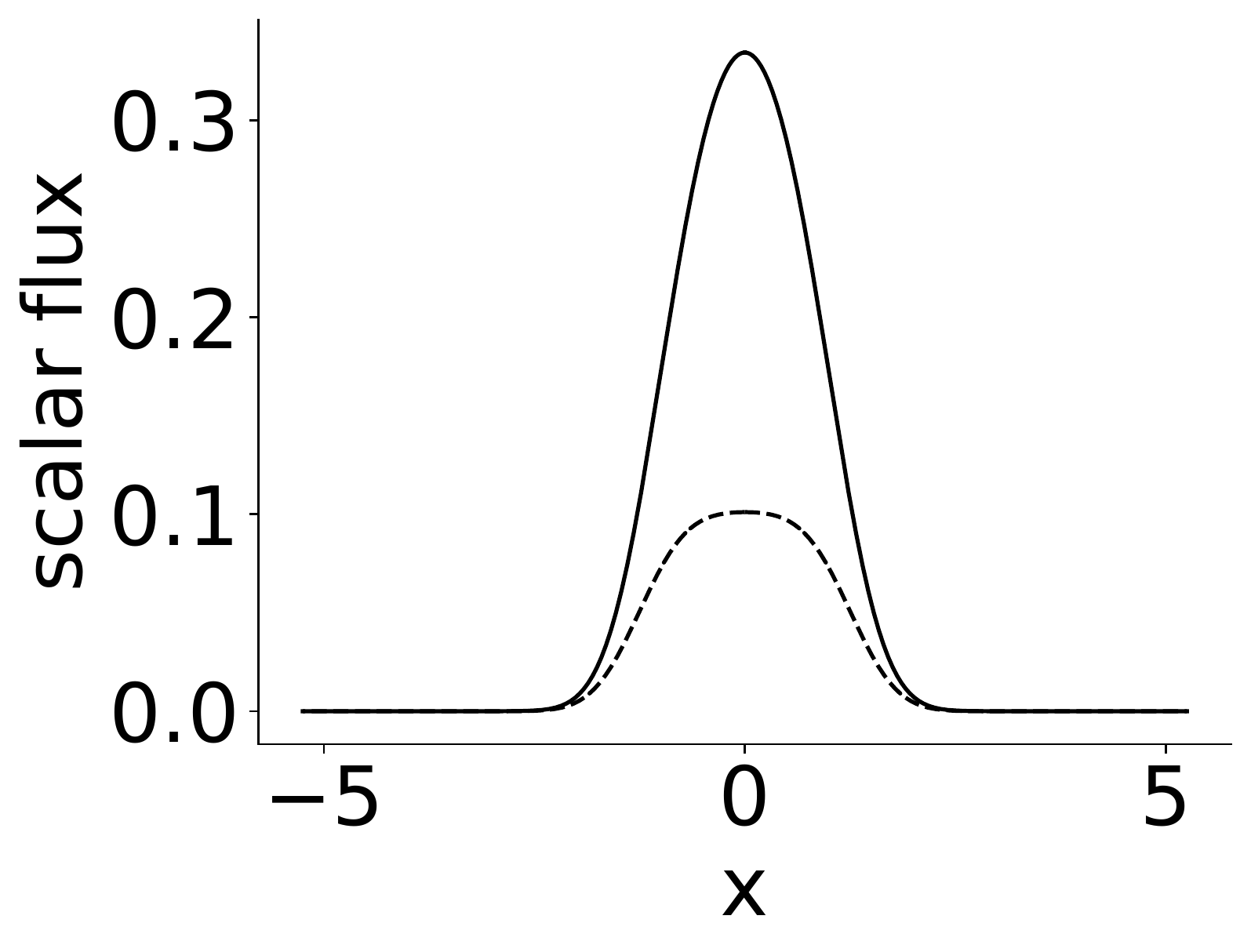}
         \caption{$c=0.8$, $t = 1.25$}
         \label{fig:gs_IC_c=0.8}
     \end{subfigure}
     \centering
     \begin{subfigure}[b]{0.48\textwidth}
         \centering
         \includegraphics[width=\textwidth]{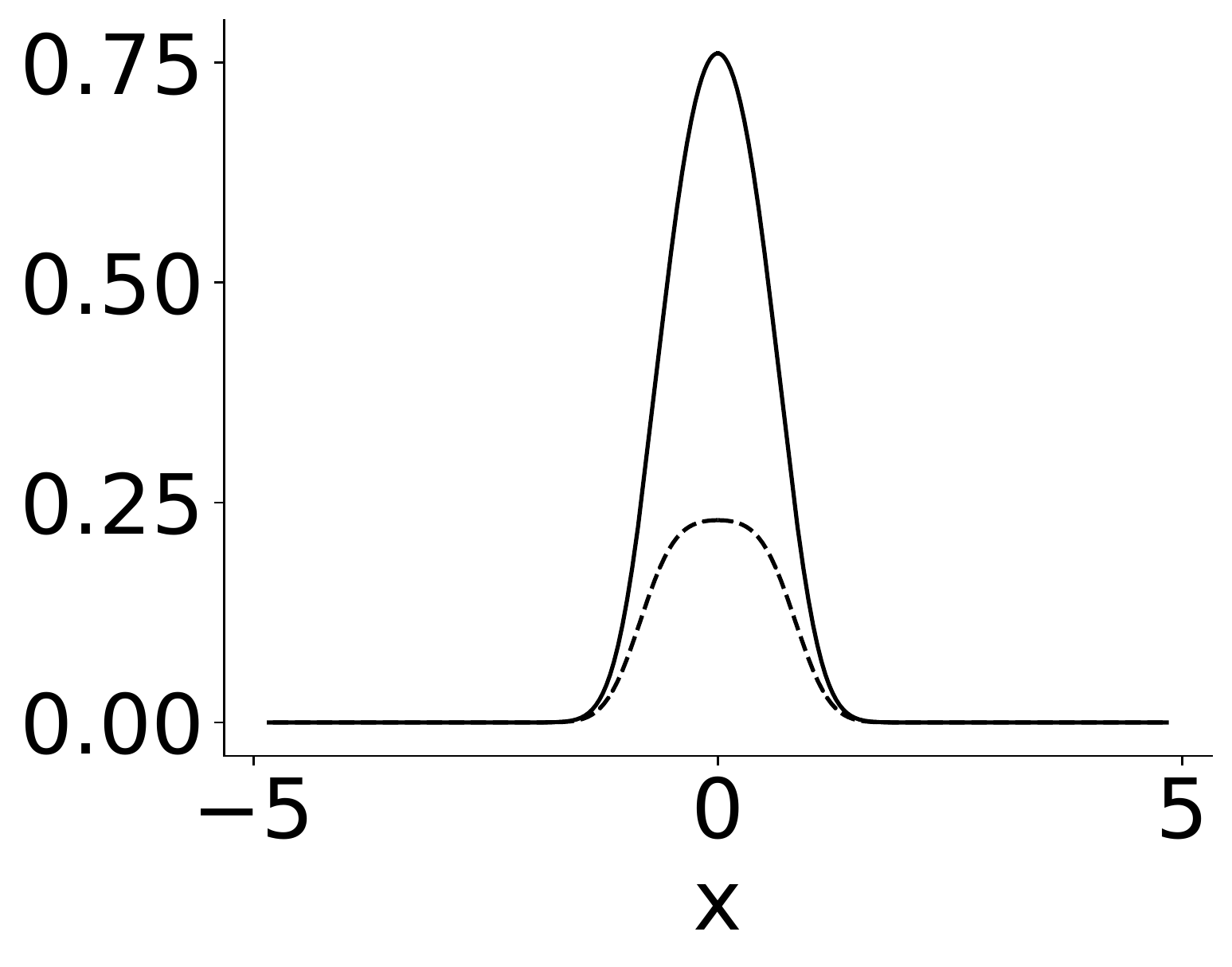}
         \caption{$c=1.2$, $t=0.83$}
         \label{fig:gs_IC_c=1.2}
     \end{subfigure}
        \caption{Gaussian pulse semi-analytic solutions, $\phi$ (solid) and $\phiu$ (dashed), at two times for different values of $c$, scaled from $t = 1$, $c= 1$ solution (\ref{fig:gs_IC_1})}
        \label{fig:gs_IC_not_1}
\end{figure}

For $c=0.8$ and $c=1.2$, the benchmarks for the square pulse and Gaussian pulse are plotted in Figures~\ref{fig:sq_IC_not_1} and \ref{fig:gs_IC_not_1} respectively. We choose the evaluation times $t_{\mathrm{final}}=1.25$ and $t_{\mathrm{final}}=0.83$ so we may use the $t=1$ benchmark.

\begin{figure}
    \centering
    \begin{subfigure}[b]{0.48\textwidth}
    \includegraphics[width=\textwidth]{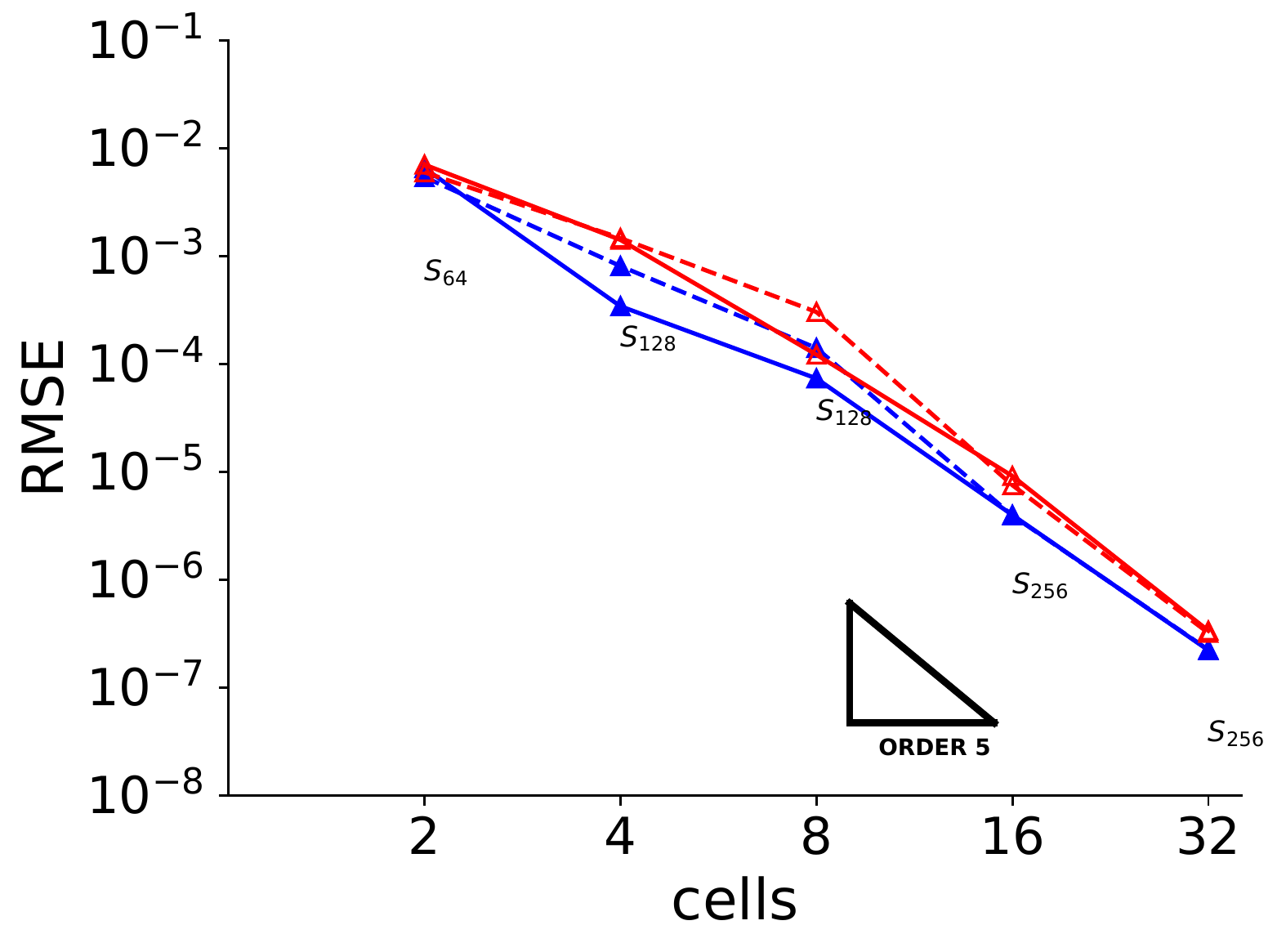}
    \caption{$M=4$, $c=0.8$, $t=1.25$}
    \label{subfig:gauss_M4_rms_c=0.8}
    \end{subfigure}
    \centering
    \begin{subfigure}[b]{0.48\textwidth}
        \includegraphics[width=\textwidth]{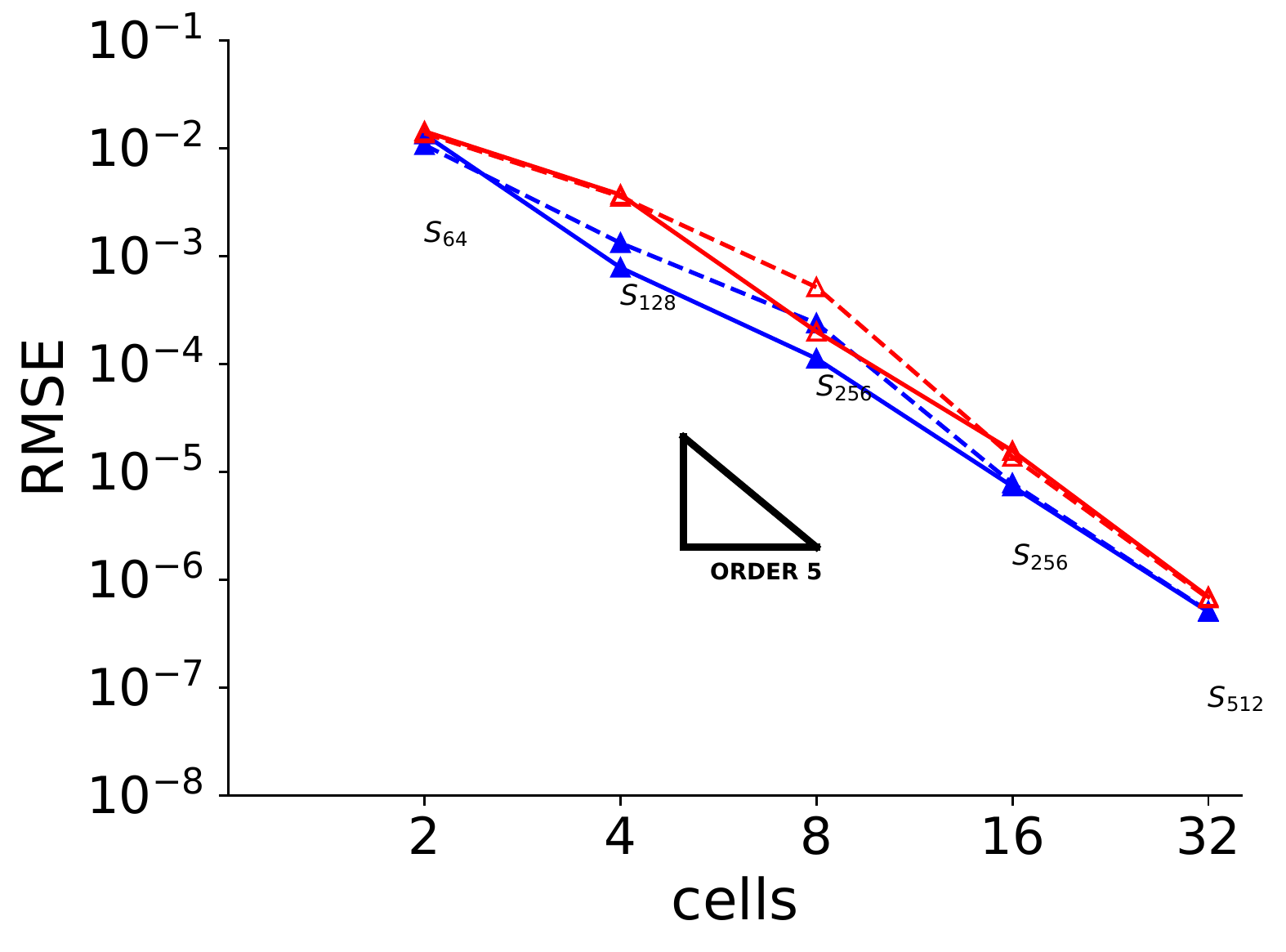}
        \caption{$M=4$, $c=1.2$, $t=0.83$}
        \label{subfig:gauss_M4_rms_c=1.2}
    \end{subfigure}
    \caption{Gaussian pulse convergence results on a logarithmic scale with $c=0.8$ evaluated at $t=1.25$ and $c=1.2$ evaluated at $t=0.83$. The standard deviation of the initial condition, $\sigma$ is $0.625$ in the first case and $0.417$ in the second. Blue lines indicate the uncollided solution is used, red that no uncollided source is used.  Dashed lines are for a static mesh and solid lines are for the moving mesh.}
    \label{fig:gauss_IC_RMS_c_not_one}
\end{figure}

\begin{figure}
     \hfill
     \begin{subfigure}[b]{0.48\textwidth}
         \centering
         \includegraphics[width=\textwidth]{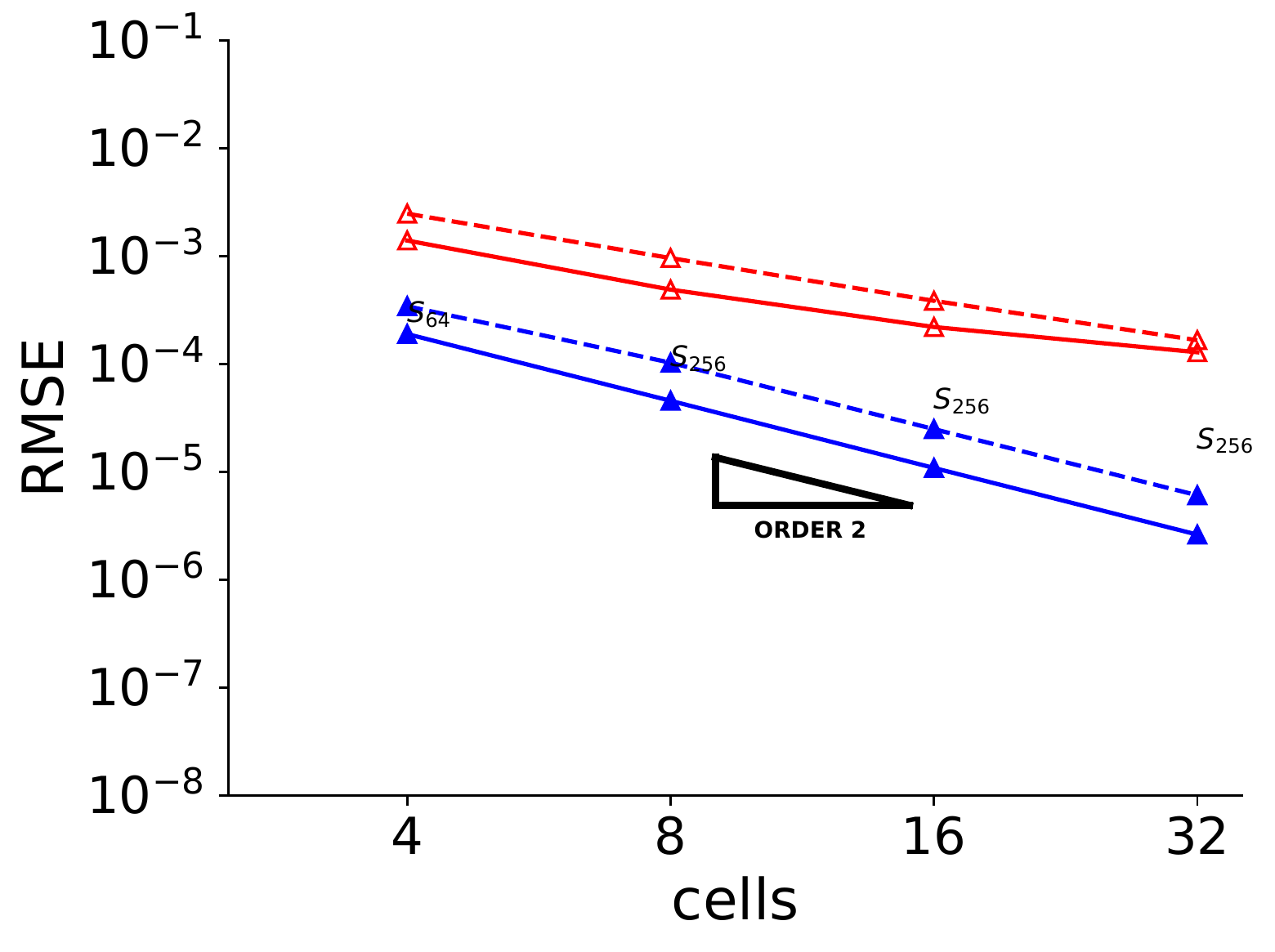}
         \caption{$M=4$, $c=0.8$, $t=1.25$}
         \label{fig:sq_ic_rms_4_c=0.8}
     \end{subfigure}
     \centering
     \begin{subfigure}[b]{0.48\textwidth}
         \centering
         \includegraphics[width=\textwidth]{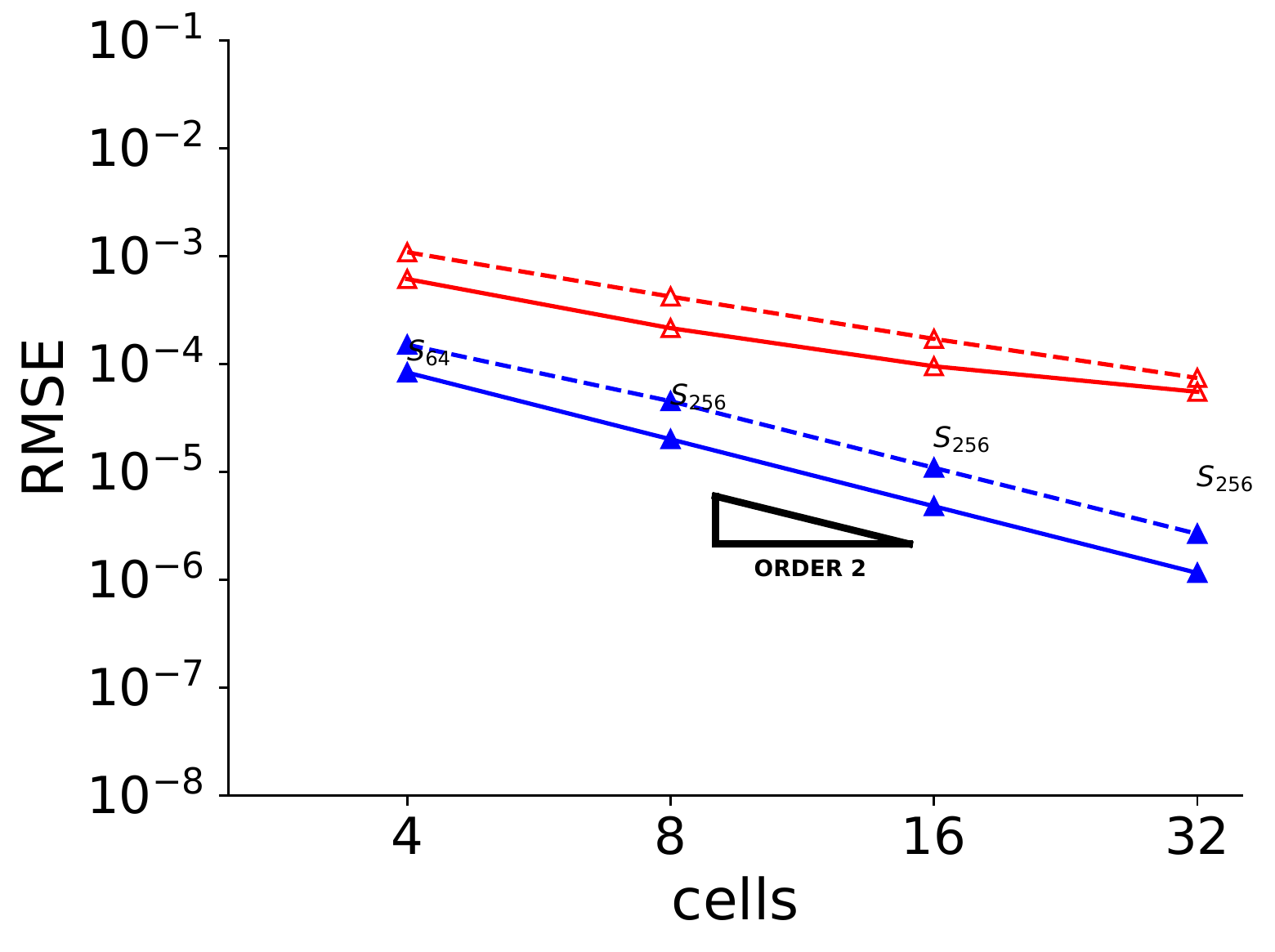}
         \caption{$M=4$, $c=1.2$, $t=0.83$}
         \label{fig:sq_ic_rms_4_c=1.2}
     \end{subfigure}
        \caption{Square pulse convergence results on a logarithmic scale with $c=0.8$ evaluated at $t=1.25$ and $c=1.2$ evaluated at $t=0.83$. The standard deviation of the initial condition, $\sigma$ is $0.625$ in the first case and $0.417$ in the second. Blue lines indicate the uncollided solution is used, red that no uncollided source is used.  Dashed lines are for a static mesh and solid lines are for the moving mesh. }
            \label{fig:sq_IC_RMS_c_not_one}
\end{figure}

Convergence tests showed that our moving mesh, uncollided solutions method handles these partially absorbing or multiplying problems just as well as purely scattering problems. The results for the plane pulse behave similarly to $c=1$ results, with second order convergence at early times and a significant difference in magnitude between the uncollided, moving mesh case and the standard DG implementation. These results are shown in Figure~\ref{fig:sq_IC_RMS_c_not_one}. Changing the scattering ratio also did not impact the convergence behavior of the method on the Gaussian pulse problem, with the method achieving fifth order convergence with $M=4$ for the tested values of $c$ (Figure~\ref{fig:gauss_IC_RMS_c_not_one}). In both tests, the multiplying scattering ratio showed similar convergence characteristics to the absorbing problem, but with a higher relative error. This is not necessarily an effect of the larger scattering ratio, but more likely of earlier evaluation time, when the solution is more challenging.

\afterpage{\clearpage}
\subsection{Computational efficiency}
\begin{figure}
    \centering
    \begin{subfigure}[b]{0.48\textwidth}
    \includegraphics[width=\textwidth]{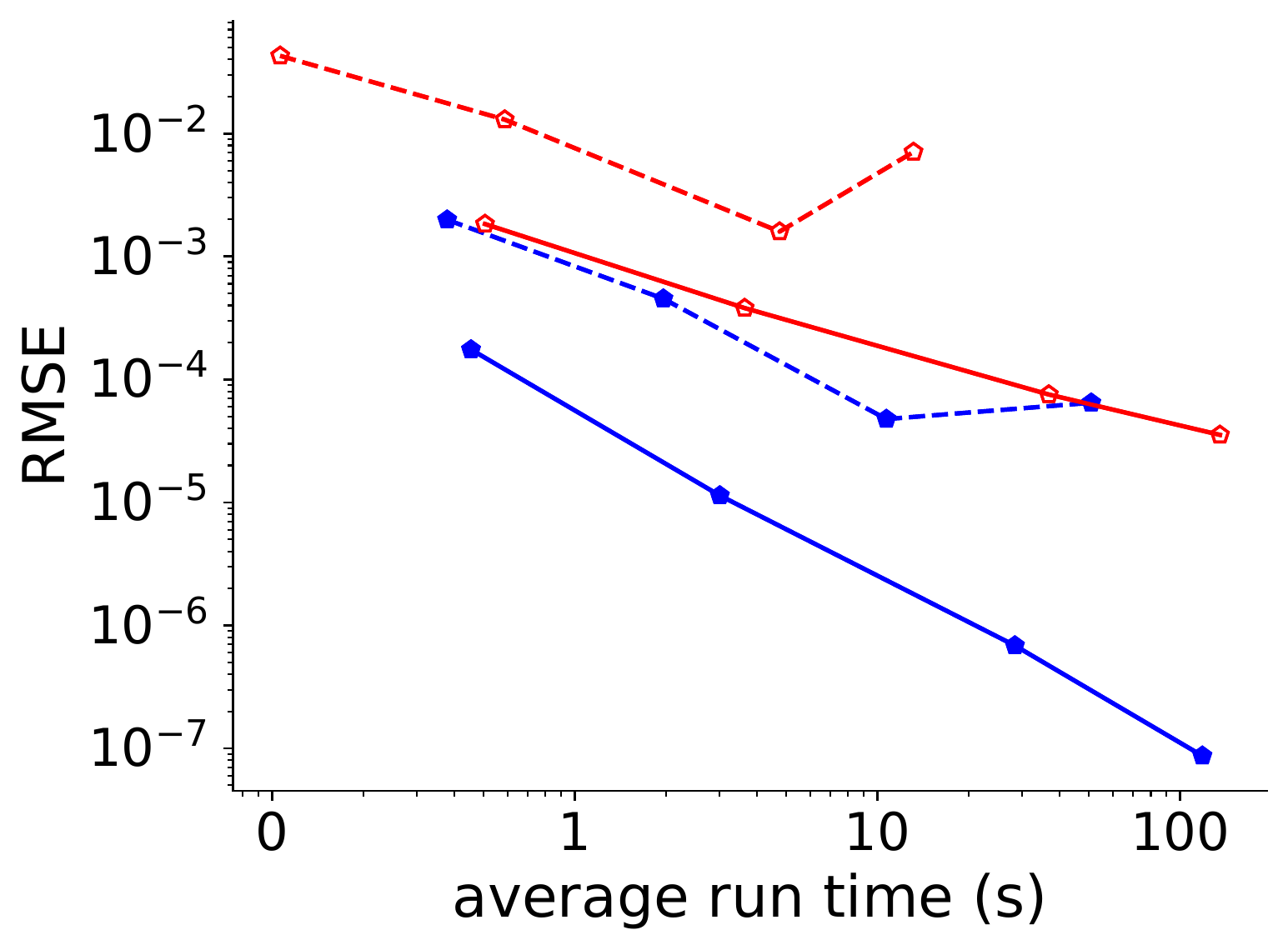}
    \caption{$4$ cells,  $M=2,4,8$ and $16$}
    \label{subfig:Ms_time}
    \end{subfigure}
    \centering
    \begin{subfigure}[b]{0.48\textwidth}
        \includegraphics[width=\textwidth]{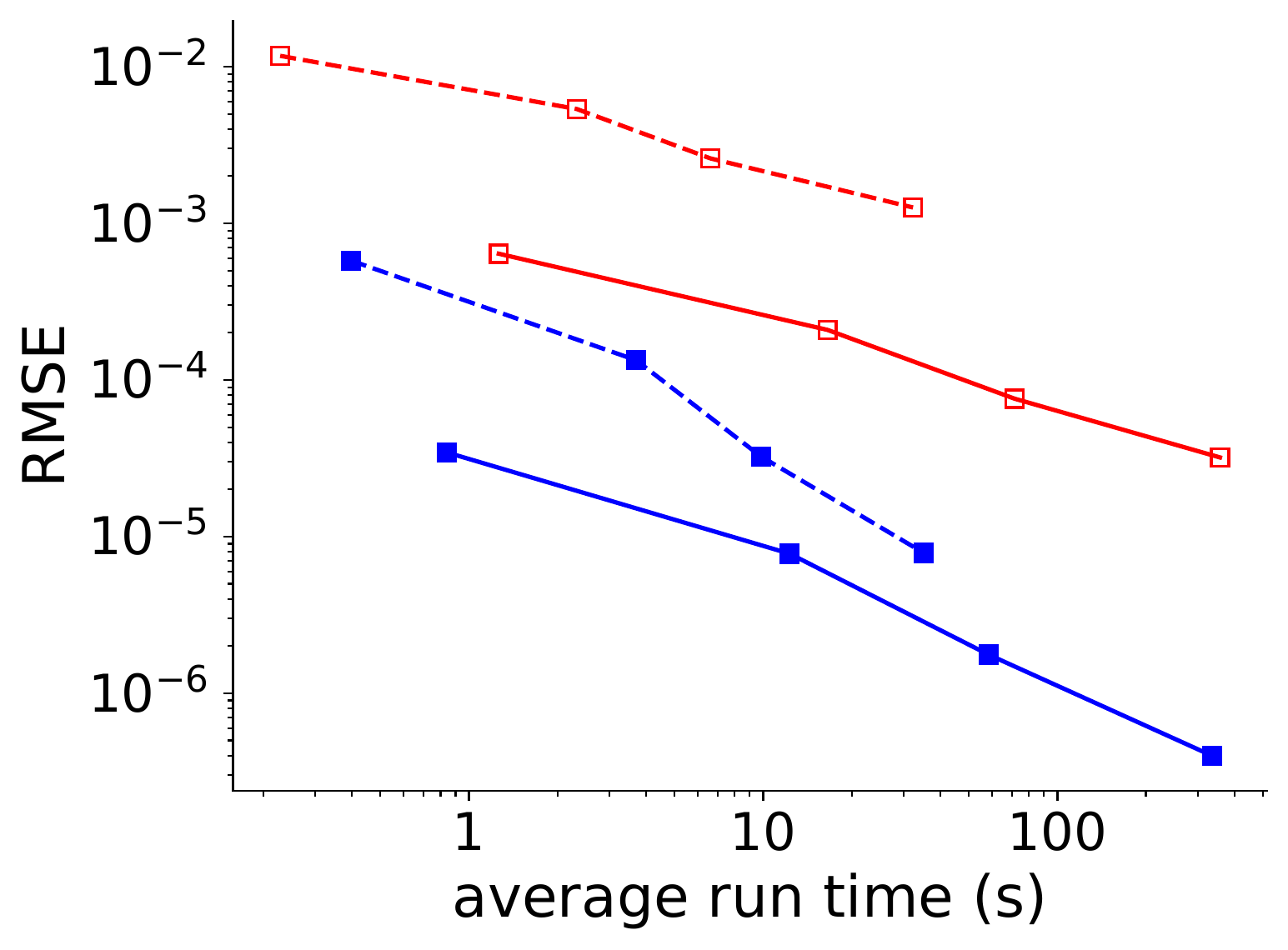}
        \caption{$M=6$, $2,4,8$ and $16$ cells}
    \label{subfig:cells_time}
    \end{subfigure}
    \caption{Calculation time vs RMSE for five runs of the square source problem. Blue lines indicate the uncollided solution is used, dashed that the mesh is static, and solid that the mesh is moving. Panel (\ref{subfig:Ms_time}) contains data from calculations holding the number of mesh subdivisions constant and increasing the number of basis functions and  Panel (\ref{subfig:cells_time}) contains data from calculations holding $M$ constant and increasing the number of mesh subdivisions. }
    \label{fig:time_calc_square}
\end{figure}

For the square source problem of Section \ref{sec:square_s}, benchmarks were created storing the average computational time over $5$ runs holding $M=6$ constant and increasing number of cell divisions and holding the number of cells constant ($4$) while increasing the number of basis functions, Figure \ref{fig:time_calc_square}. The calculations were performed on a 2020 MacBook Pro Laptop with an 8 core M1 CPU and 16GB of RAM. For both cases, the uncollided source, moving mesh method returned more accurate solutions for less calculation time. The results also show that the most efficient way to obtain accurate solutions with this method is to use less mesh subdivisions and a higher order polynomial. The most accurate solution in Figure \ref{subfig:Ms_time} takes about three times less time than the most accurate solution in Figure \eqref{subfig:cells_time}. This can also be understood in light of  the trend apparent in Figures \ref{fig:plane_IC_rms}, \ref{fig:sq_IC_RMSE}, and \ref{fig:sq_s_RMSE} where increasing the number of basis functions from 5 to 7 returns the same convergence but starting from a smaller intercept value. 
\FloatBarrier
\section{Conclusions}\label{sec:conclusions}
For smooth problems, like the MMS problem, the Gaussian pulse, the Gaussian source, and the nonsmooth sources at later times, the method of uncollided solutions with a moving mesh achieves spectral convergence and produces highly accurate solutions for relatively few degrees of freedom. In the Gaussian source case, there is a definite benefit to using the uncollided solution, but it does not seem that using a moving mesh gives a large added benefit since the solution does not contain discontinuous wavefronts. The moving mesh, uncollided source treatment was most useful in problems with finite width, nonsmooth sources. For the plane pulse problem, the highly nonsmooth angular flux prohibited high order convergence, but also amplified the difference in efficiency between the uncollided, moving mesh case and the standard DG implementation. The comparatively smooth square pulse and square source achieved second order convergence at early times, also with the uncollided and moving mesh performing the best by a significant margin. These cases also showed an improvement of the order of convergence from using the uncollided solutions. For the square source case, we also achieved second order convergence with significant difference in solution accuracy achieved with a moving mesh and the uncollided source. Also for this case, we showed that using a higher order polynomial to interpolate the solution and a lower number of mesh subdivisions returned accurate solutions with low computational expense compared to a method with a lower order polynomial and more mesh subdivisions. Additionally, we showed for that the method preforms well for scattering ratios not equal to one ($c\neq1$) with tests of the square and Gaussian pulse source problems. 

Given the encouraging results from this study, we plan for future work to utilize the moving mesh and uncollided flux treatments to solve problems of nonlinear radiative transfer to create novel verification-quality benchmark solutions. Since there are no inherent challenges to applying the Discontinuous Galerkin method (and spectral methods in general) to nonlinear problems, we do not anticipate any special difficulties other than those introduced by the allowance of negative solutions in this undertaking. Systems with anisotropic scattering, finite medium problems, and problems in spherical and cylindrical geometry are also of interest. We also suggest that these approaches could be used for other, transport problems not amenable to (semi-)analytic solution technique such as neutronics problems with thermal feedback or rarefied gas dynamics. 

\section*{Acknowledgement}
\noindent This work was supported by a Los Alamos National Laboratory
under contract \#599008, “Verification for Radiation Hydrodynamics Simulations”.
\FloatBarrier
\bibliographystyle{elsarticle-num}
\bibliography{references.bib}





\end{document}